\pdfoutput=1

\documentclass[11pt,twoside,a4paper,cmspaper,final,collab]{cms-tdr}
\def\svnVersion{493532P}\def\cmsCernNoTag{CERN-EP-2018-254}\def\cmsCernDate{\today}\def\cmsMessage{Published in the Journal of High Energy Physics as \href{http://dx.doi.org/10.1007/JHEP04(2019)031}{\doi{10.1007/JHEP04(2019)031.}}}

\begin{document}\cmsNoteHeader{B2G-17-017}

\hyphenation{had-ron-i-za-tion}
\hyphenation{cal-or-i-me-ter}
\hyphenation{de-vices}
\RCS$Revision: 491530 $
\RCS$HeadURL: svn+ssh://svn.cern.ch/reps/tdr2/papers/B2G-17-017/trunk/B2G-17-017.tex $
\RCS$Id: B2G-17-017.tex 491530 2019-03-11 16:05:54Z drberry $

\newcommand{\ck}{\ensuremath{\checkmark}\xspace}
\newcommand{\gKK}{\ensuremath{\Pg_{\text{KK}}}\xspace}
\newcommand{\GKK}{\ensuremath{\text{G}_{\text{KK}}}\xspace}
\newcommand{\mjet}{\ensuremath{m_{\text{jet}}}\xspace}
\newcommand{\mttbar}{\ensuremath{m_{\ttbar}}\xspace}
\newcommand{\msd}{\ensuremath{m_{\text{SD}}}\xspace}
\newcommand{\ST}{\ensuremath{S_{\text{T}}}\xspace}
\newcommand{\htlep}{\ensuremath{\HT^{\ell}}\xspace}
\newcommand{\ttagging}{\ensuremath{\cPqt}\text{ tagging}\xspace}
\newcommand{\btagging}{\ensuremath{\cPqb}\text{ tagging}\xspace}
\newcommand{\tmistag}{\ensuremath{\cPqt}\text{ mistag}\xspace}
\newcommand{\wjets}{\ensuremath{\PW}\text{+jets}\xspace}
\newcommand{\wjetsbdt}{\ensuremath{\PW}\text{+jets BDT}\xspace}
\newcommand{\zjets}{\ensuremath{\PZ}\text{+jets}\xspace}
\newcommand{\pb}{\unit{pb}}

\cmsNoteHeader{B2G-17-017}
\title{Search for resonant \ttbar production in proton-proton collisions at $\sqrt{s}=13\TeV$}

\date{\today}

\abstract{
A search for a heavy resonance decaying into a top quark and antiquark
(\ttbar) pair is performed using proton-proton collisions at
$\sqrt{s}=13\TeV$. The search uses the data set collected
with the CMS detector in 2016, which corresponds to an integrated
luminosity of 35.9\fbinv. The analysis considers
three exclusive final states and uses reconstruction techniques that
are optimized for top quarks with high Lorentz boosts, which requires
the use of nonisolated leptons and jet substructure techniques. No
significant excess of events relative to the expected yield from
standard model processes is observed. Upper limits on the production
cross section of heavy resonances decaying to a \ttbar pair are
calculated. Limits are derived for a leptophobic topcolor
\PZpr\ resonance with widths of 1, 10, and
30\%, relative to the mass of the resonance, and exclude masses up
to 3.80, 5.25, and 6.65\TeV, respectively. Kaluza--Klein excitations of
the gluon in the Randall--Sundrum model are excluded up to
4.55\TeV. To date, these are the most stringent limits on \ttbar
resonances.
}

\hypersetup{%
pdfauthor={CMS Collaboration},%
pdftitle={Search for resonant ttbar production in proton-proton collisions at sqrt(s)=13 TeV},%
pdfsubject={CMS},%
pdfkeywords={CMS, zprime, top quark, resonance}}

\maketitle

\section{Introduction}
\label{sec:introduction}
The top quark (\PQt) is the most massive known fundamental
particle~\cite{Abe:1995hr,D0:1995jca} in the standard model. It has a
Yukawa coupling to the Higgs field that is near unity. It is also
closely connected to the hierarchy problem, where the largest
corrections to the Higgs mass arise from top quark loops.
Furthermore, studies of the top quark may provide insight into the
mechanism of electroweak (EW) symmetry breaking.

Many theories beyond the standard model (SM) predict
heavy resonances at the TeV scale, which would decay to top quark and
antiquark (\ttbar) pairs. These resonances can present themselves as
peaks on top of the falling \ttbar invariant mass spectrum or as a
distortion of the \ttbar spectrum if the resonance has a large width
and a mass above the center-of-mass energy of the colliding
partons. Resonances decaying to \ttbar pairs can be found in models
that contain TeV scale color singlet \PZpr\
bosons~\cite{theory_rosner,theory_lynch,theory_carena}, a pseudoscalar
Higgs boson that may couple strongly to \ttbar
pairs~\cite{theory_dicus},
axigluons~\cite{theory_frampton_glashow,theory_choudhury,Godbole:2008qw},
or colorons~\cite{theory_hill1,theory_hill_parke,theory_hill2,
Harris:1999ya}, and especially models that contain a leptophobic
topcolor \PZpr~\cite{theory_jainharris}. Additionally, extensions of
the Randall--Sundrum model~\cite{theory_randall_sundrum,
Randall:1999vf} with extra dimensions predict Kaluza--Klein (KK)
excitations of the gluons \gKK~\cite{theory_agashe} or gravitons
\GKK~\cite{theory_davoudiasl}, which can have large branching
fractions to \ttbar pairs. This analysis searches for spin-1
resonances that do not interfere with SM \ttbar production. Previous
searches at the Fermilab Tevatron have excluded a leptophobic \PZpr\
boson up to 900\GeV~\cite{cdf_ttbar_resonance1,
cdf_ttbar_resonance2, cdf_ttbar_resonance3, d0_ttbar_resonance1,
cdf_ttbar_resonance4, d0_ttbar_resonance2} at 95\% confidence level
(\CL). Experiments at the CERN LHC have excluded various \PZpr\ and
\gKK models at 95\% \CL in the 1--4\TeV mass
range~\cite{cms_ttbar_resonance1,atlas_ttbar_resonance1,
atlas_ttbar_resonance2, cms_ttbar_resonance2, cms_ttbar_resonance3,
cms_ttbar_resonance4,Sirunyan:2017uhk,atlas_ttbar_resonance3}.
The results presented here represent a significant improvement on
the previous searches for \ttbar resonances.

This paper presents a model-independent search for
\ttbar resonances. Since no excess is seen, limits are calculated on several
spin-1 resonance models of varying widths. The \ttbar system, and all
its daughter particles, decay as described by the SM. The top quark
predominately decays to a \PW\ boson and a bottom quark (\PQb). Each of the two
\PW\ bosons in the event can decay to either a lepton and its
corresponding neutrino or to hadrons. The analysis considers three
subanalyses based on the decay modes of the two \PW\ bosons: dilepton,
single-lepton, and fully hadronic decay modes of the \ttbar system. In
the fully hadronic channel, both \PW\ bosons decay to hadrons. In the
single-lepton channel, one \PW\ boson decays to an electron (\Pe)
or muon (\Pgm) and its neutrino ($\nu$) counterpart, while the other
\PW\ boson decays to hadrons. In the dilepton channel, both \PW\
bosons decay to an \Pe\ or \Pgm\ and a $\nu$. The leptonic selections
are not optimized to identify electrons or muons originating from
leptonically decaying tau leptons; however, such particles are not excluded by
the event selections. The search is based on $\sqrt{s} = 13\TeV$
proton-proton (\Pp\Pp) collision data collected in 2016 by the CMS
experiment at the LHC, corresponding to an integrated luminosity of
35.9\fbinv.

The dilepton final state consists of two leptons (\Pgm\Pgm,
\Pe\Pe, or \Pgm\Pe), two jets originating from bottom quarks
(\PQb jets) with high transverse momentum (\pt), and missing
transverse momentum (\ptvecmiss). The large mass of the
resonance causes the resulting top quarks to have a significant
Lorentz boost, which leads to a collimated system consisting of
a lepton and a \PQb jet. To account for the overlap between
the lepton and the \PQb jet, special reconstruction and
selection criteria are used to increase lepton selection efficiency
and reduce the SM background. The dominant irreducible SM
background arises from \ttbar nonresonant production.
Smaller contributions are due to a \PZ\ boson produced in
association with jets (\zjets), single top quark, and diboson
processes. Events that have a large separation between the
lepton and \PQb jet are allocated to control regions (CR),
which are used to validate the modeling of the SM backgrounds.

The single-lepton final state consists of one lepton (\Pgm\ or
\Pe), at least two high-\pt jets, and \ptvecmiss. In this channel also,
the final state particles from the decay of the \ttbar
pairs have a large Lorentz boost because of the mass of the
resonance. Leptons from the decay of the \PW\ boson are found in near
proximity to the \PQb jet from the top quark decay. The same lepton
reconstruction and selection criteria used in the dilepton
channel are used in the single-lepton channel. In addition
to those techniques, a special triggering technique is used
 to select events with a single nonisolated lepton and an additional jet.
A \ttagging algorithm is used to identify top quarks where
the daughter \PW\ boson decays hadronically
($\PQt \to \PW\ \PQb \to \cPq \cPaq^{\prime} \PQb$). Events with a jet
that passes the \ttagging criteria are classified into a category with
higher sensitivity. The largest irreducible background is the \ttbar
continuum production, while the largest reducible background is from \PW\
bosons produced in association with jets (\wjets). The latter
background is separated from the signal using a multivariate analysis
technique.

The fully hadronic channel contains events with a dijet topology,
where both large radius jets are required to pass t tagging criteria that
select Lorentz-boosted hadronically decaying top quarks. Because of the dijet
topology of the search region, the largest reducible background arises
from dijet events produced from quantum chromodynamic (QCD)
interactions between the colliding protons. This background, referred
to as QCD multijet production, can be reduced considerably by
requiring one of the subjets in each of the two large radius
jets, which are selected by the \ttagging algorithm,
to be consistent with the fragmentation of a bottom
quark~\cite{Sirunyan:2017ezt}. A subjet is defined as a smaller radius
jet reconstructed within a larger radius jet. The use of subjet \btagging
for categorization nearly eliminates the QCD multijet
background leaving only the \ttbar continuum in the highest
sensitivity category.

Except for the QCD multijet background in the fully hadronic
channel, the shapes of all SM backgrounds are estimated from
simulation. The total normalization of each simulated sample is
obtained from a simultaneous binned maximum likelihood fit to the
reconstructed \ttbar invariant mass (\mttbar) distribution for the
single-lepton and fully hadronic analyses and \ST for the dilepton
analysis, where \ST is defined as
\begin{linenomath*}
\begin{equation}
	\ST = \sum_{i=1}^{N_{\text{jet}}}p_{\text{T}_{i}}^{\text{jet}}+\sum_{i=1}^{2}p_{\text{T}_{i}}^{\ell}+\ptmiss.
\end{equation}
\end{linenomath*}
The variable \ST is used because it has a greater sensitivity to
signal than \mttbar, in the dilepton final state. A limit on the
production cross section of heavy resonances is extracted by
performing a template-based statistical evaluation of
the \mttbar (single-lepton and fully hadronic) and \ST (dilepton)
distributions simultaneously in all of the channels.

This paper is organized the following way. Section~\ref{sec:detector}
provides a description of the CMS detector. The reconstruction and
identification of electrons, muons, and jets are described in
Section~\ref{sec:reco}. Section~\ref{sec:reco} also gives an overview
of the \ttagging algorithms used. The data sets and triggering
techniques are described in Section~\ref{sec:trigger}. The simulated
Monte Carlo (MC) samples used in the analysis are discussed in
Section~\ref{sec:simulation}. Section~\ref{sec:selection} describes
the event selection for the three different channels.
Section~\ref{sec:background} describes the evaluation of the SM
background processes. Systematic uncertainties affecting the signal
and background shapes and normalization are discussed in
Section~\ref{sec:systematics}. The statistical analysis and the
results are given in Sections~\ref{sec:stats} and~\ref{sec:results},
respectively, and a summary is presented in Section~\ref{sec:summary}.

\section{The CMS detector}
\label{sec:detector}

The central feature of the CMS detector is a superconducting solenoid
of 6\unit{m} internal diameter, providing a magnetic field of
3.8\unit{T}. Within the solenoid volume are a silicon pixel and strip
tracker, a lead tungstate crystal electromagnetic calorimeter (ECAL),
and a brass and scintillator hadron calorimeter (HCAL), each composed of a
barrel and two endcap sections. In addition to the barrel and endcap
detectors, CMS has extensive forward calorimetry. Muons are detected
by four layers of gas-ionization detectors embedded in the steel
flux-return yoke of the magnet. The inner tracker measures
charged-particle trajectories within the pseudorapidity range
$\abs{\eta} < 2.5$, and provides an impact parameter resolution of
approximately 15\mum. A two-stage trigger
system~\cite{Khachatryan:2016bia} selects \Pp\Pp\ collision events of
interest for use in physics analyses. A more detailed description of
the CMS detector, together with a definition of the coordinate system
used and the relevant kinematic variables, can be found in
Ref.~\cite{Chatrchyan:2008zzk}.

\section{Event reconstruction}
\label{sec:reco}

The CMS event reconstruction uses a particle-flow (PF) technique that
aggregates input from all subdetectors for event
reconstruction~\cite{CMS-PRF-14-001}. Typical examples of PF inputs
are charged-particle tracks from the tracking system and energy
deposits from the ECAL and HCAL. The PF approach enables the global event
description to take advantage of the excellent granularity of the CMS
detector. Clusters of tracks and energy deposits are iteratively
classified as muons, electrons, photons, charged hadrons, and neutral
hadrons. Vertices are reconstructed from tracks using a deterministic
annealing filter algorithm~\cite{Chatrchyan:2014fea}. The
reconstructed vertex with the largest value of summed physics-object
$\pt^2$ is taken as the primary \Pp\Pp\ interaction vertex (PV). For the
PV reconstruction, the physics objects are jets, clustered with the
jet finding algorithm~\cite{Cacciari:2008gp,Cacciari:2011ma} using
only tracking information, with the tracks assigned to the PV as inputs.
The reconstructed leptons and photons in the event are included as
inputs to the jet clustering algorithm.

The \ptvecmiss is defined as the projection onto the plane
perpendicular to the beam axis of the negative vector sum of the
momenta of all reconstructed PF candidates in an
event~\cite{Chatrchyan:2011ds}. Its magnitude is referred to
as \ptmiss. Corrections to the jet energy scale and jet energy
resolution are propagated to the measurement of \ptmiss.

Muons are reconstructed in the pseudorapidity range $\abs{\eta} < 2.4$
using the information from the tracker and muon
chambers~\cite{Chatrchyan:2014fea}. Tracks associated with muon
candidates must be consistent with a muon originating from the PV, and
tracks must satisfy fit quality requirements.

Electrons are detected and measured in the pseudorapidity range
$\abs{\eta} < 2.5$, by combining tracking information with energy
deposits in the ECAL~\cite{electronreco,
Chatrchyan:2013dga}. Candidate electrons are required to originate
from the PV. The track quality, electromagnetic shower shape,
displacement between the track and electromagnetic shower, and ratio
of energy between the HCAL and ECAL are used to identify
electrons. Reconstructed electrons that originate from photon
conversions are rejected.

No isolation requirements are placed on the leptons at the trigger or
analysis level. This is because the lepton, bottom quark, and neutrino
from the top quark decay are highly collimated, and the lepton is not
well separated from the products of fragmentation of the bottom quark.
Additionally, jets that contain an electron are reclustered and
corrected with the track and calorimeter deposit of the electron
removed. Kinematic restrictions are placed on the electron and on the overall
event to reduce the contribution from electrons not originating from t decays.
Details on these requirements can be found in Section~\ref{sec:selection}.

The PF candidates are clustered into jets using the \FASTJET software
package~\cite{Cacciari:2011ma}. Charged hadrons that are not
associated with the PV in the event are excluded from the jet
clustering procedure via charged hadron subtraction
(CHS)~\cite{CMS-PRF-14-001}. All jets are required to have
$\abs{\eta} < 2.4$. Jets are clustered using the anti-\kt jet
clustering algorithm~\cite{Cacciari:2008gp} with a distance parameter
of 0.4 (AK4 jets). If a lepton is found with ${\Delta R < 0.4}$ of
an AK4 jet, its four-momentum is subtracted from that of the jet. The
single-lepton and fully hadronic analyses also use anti-\kt clustered
jets with a distance parameter of 0.8 (AK8 jets). These larger-radius
jets are used to tag the hadronic decay of top quarks. A high-mass
resonance decay creates daughter particles with significant Lorentz
boost. The three jets from the top quark decay merge into a
single-larger AK8 jet. Jets in all three channels are contaminated
with neutral particles that are generated from additional \Pp\Pp\
collisions within the same or a neighboring bunch crossing (pileup).
The extra energy in each jet is
corrected based on the average expectation of the pileup within the
jet footprint~\cite{Cacciari:2008gn}. The expected energy offset due
to pileup is modeled as a function of the number of primary vertices
in the event~\cite{Chatrchyan:2011ds}. Jets that are produced from the
decay of charm and bottom quarks are identified using the combined
secondary vertex algorithm (CSV)~\cite{CMS-PAS-JME-15-002}. Loose,
medium, and tight operating points are used in this analysis. They
have a probability of 10, 1, and 0.1\%, respectively, of
misidentifying a light-parton jet as heavy flavor, where the
light-flavor jet has $\pt > 30\GeV$ and is determined from a
simulated multijet sample with a center-of-mass energy between 80 and
120\GeV~\cite{Sirunyan:2017ezt}. They correspond to a \btagging
efficiency of 81, 63, and 41\%, respectively, for \PQb\ jets ($\pt >
20\GeV$) in simulated \ttbar events. All jets are required to pass a
minimal set of criteria to separate them from calorimeter noise and
other sources of jets that do not originate from the
PV~\cite{CMS-PAS-JME-10-003}. Events are also required to pass a set
of selections that remove spurious \ptmiss that is generated from
calorimeter noise~\cite{Chatrchyan:2011tn}.

The \ttagging algorithm~\cite{JME-09-001,JME-13-007}, which is based
on the algorithm described in Ref.~\cite{Kaplan:2008ie}, is applied to
AK8 jets that use pileup per particle identification (PUPPI)
corrections~\cite{Bertolini:2014bba}, referred to as PUPPI jets, in
order to separate hadronically decaying top quarks from light quark or
gluon jets. While CHS only removes charged particles originating from
pileup, PUPPI corrects for both charged and neutral pileup
particles. PUPPI jets, as opposed to CHS jets, are therefore used
for \ttagging because of their better performance as a function of
pileup. The CMS \ttagging algorithm only considers jets with $\pt >
400\GeV$, as lower-momentum top quarks frequently decay into resolved
jets. The algorithm iteratively reverses the jet clustering procedure in
order to remove soft radiation. First, it reclusters the AK8 PUPPI jet
with the Cambridge-Aachen jet clustering algorithm~\cite{Dokshitzer:1997in}. It then
separates the jet ($j$) into two subjets, $j_{1}$ and $j_{2}$, which must
satisfy the ``soft drop''  (SD) criterion
\begin{linenomath*}
\begin{equation}
	\frac{\min(p_{\text{T}1},p_{\text{T}2})}{p_{\text{T}1}+p_{\text{T}2}} > z_{\text{cut}} \left(\frac{\Delta R_{12}}{R_0}\right)^{\beta},
\end{equation}
\end{linenomath*}
where $p_{\text{T}1}$ and $p_{\text{T}2}$ are the transverse momenta
of the two subjets and $\Delta R_{12}$ is the distance between them.
The implementation of the SD algorithm used in this analysis
has an angular exponent $\beta=0$, making it equivalent to the
``modified mass drop tagger'' algorithm~\cite{mmdt}. Additionally,
a soft cutoff threshold of $z_{\text{cut}} = 0.1$ and a characteristic
radius $R_0=0.8$~\cite{Larkoski:2014wba} are used. If the SD criterion
is met, the procedure ends with $j$ as the resulting jet. If not, the
lower-$\pt$ subjet is discarded and the declustering procedure
continues with the higher-$\pt$ subjet. The SD mass (\msd)
of the jet pair is required to be near the mass of the top quark
($105 < \msd < 210\GeV$). The CMS \ttagging algorithm also requires
that the $N$-subjettiness~\cite{Thaler:2010tr,Thaler:2011gf} ratio
($\tau_{32} \equiv \tau_3/\tau_2$) must be less than 0.65. The
$N$-subjettiness ($\tau_N$) is a measure of the consistency of an AK8
PUPPI jet with $N$ or fewer subjets, and is defined as
\begin{linenomath*}
\begin{equation}
	\tau_N = \frac{1}{d_0} \sum_i p_{\text{T},i} \min\left[\Delta
	R_{1,i}, \Delta R_{2,i}, \cdots, \Delta R_{N,i}\right],
\end{equation}
\end{linenomath*}
where $i$ is a summation over all jet constituents, $d_0$ is a
normalization constant, and $\Delta R$ is the distance between a given
jet constituent $i$ and a candidate subjet axis $N$.

\section{Triggers and data set}
\label{sec:trigger}

The events in the dilepton channel are triggered by single-lepton and
dilepton triggers without isolation requirements. The triggers for
\Pgm\Pgm\ and \Pe\Pgm\ events require one muon with $\pt > 50\GeV$
and with $\abs{\eta} < 2.4$ that is seeded by hits in either the muon
chambers or the inner tracker. The \Pe\Pe\ events are selected using a
dielectron trigger that requires the presence of two electrons with
$\pt > 33\GeV$ and $\abs{\eta} < 2.5$.

Events used in the single-lepton channel pass either a single electron
or a single muon trigger. The single-lepton muon channel uses the
same triggers as the dilepton \Pgm\Pgm\ and \Pe\Pgm\ channels. The
triggers for the electron channel require one electron with $\pt > 115\GeV$
or an electron with $\pt > 55\GeV$ and a PF jet with $\pt > 165\GeV$.
Both triggers require electrons within $\abs{\eta} < 2.5$, and the
electron-jet combination trigger requires
the jet to be within $\abs{\eta} < 2.4$. In the combination trigger, if
the electron lies within the jet footprint, the four-vector of the
electron is subtracted from the uncorrected four-vector of the jet,
and then the jet energy corrections are reapplied. Neither the
muon or electron triggers have isolation requirements.

The fully hadronic analysis uses events that are selected by a logical
`OR' of five different triggers. The first trigger requires a single AK8
jet with $\pt > 450\GeV$, a second trigger requires an AK4 jet with
$\pt > 360\GeV$ and mass $(\mjet) > 30\GeV$. A third trigger
requires $\HT > 800\GeV$, where the \HT is the scalar sum of the
\pt of every AK4 PF jet above 30\GeV in the event. A fourth trigger requires $\HT
> 900\GeV$, and remains un-prescaled during the acquisition of
data. The final trigger requires that the $\HT > 700\GeV$, but also
requires a jet with $\mjet > 50\GeV$.

Small differences in trigger efficiency between data and simulation in the
dilepton and single-lepton channels are accounted for with corrections
determined from events selected by triggers with different conditions.

\section{Simulated events}
\label{sec:simulation}

The $\PZpr\to\ttbar$ process is simulated using the
\MGvATNLO~v5.2.2.2~\cite{Alwall:2014hca} event
generator, which produces a resonance with the same spin and left- and
right-handed couplings to fermions as the SM \PZ\ boson. Matrix
element calculations are done at tree level and include up to three additional partons
for the \gKK and most \PZpr\ models, \PZpr\ bosons above 5\TeV are
simulated with only up to two additional partons in their final state. The
$\PZpr\to\ttbar$ process is simulated at masses between
500\GeV and 7\TeV for resonances with a relative decay width
($\Gamma/m$) of 1\% (narrow), 10\% (wide), and 30\% (extra-wide). Matching
between the hard matrix element interactions and the lower energy
parton showers is done using the MLM algorithm~\cite{Alwall:2007fs}.
The KK gluon excitation is simulated using
\PYTHIA~8.212~\cite{Sjostrand:2014zea} with the couplings
described in Ref.~\cite{Ask:2011zs}. The $\Gamma/m$ of the \gKK
resonance lies between the wide and extra-wide \PZpr\ resonances,
depending on its coupling to the top quark. The expected \PZpr\
production cross section is calculated at NLO accuracy, and the \gKK
production cross section is calculated at LO. A multiplicative factor of 1.3
is applied to the \gKK cross section as an NLO $K$ factor~\cite{Bonciani:2015hgv}.
Both the \PZpr\ and \gKK processes are simulated without interference from
SM \ttbar production.

\begin{figure}[!htbp]
\centering
\includegraphics[width=0.495\textwidth]{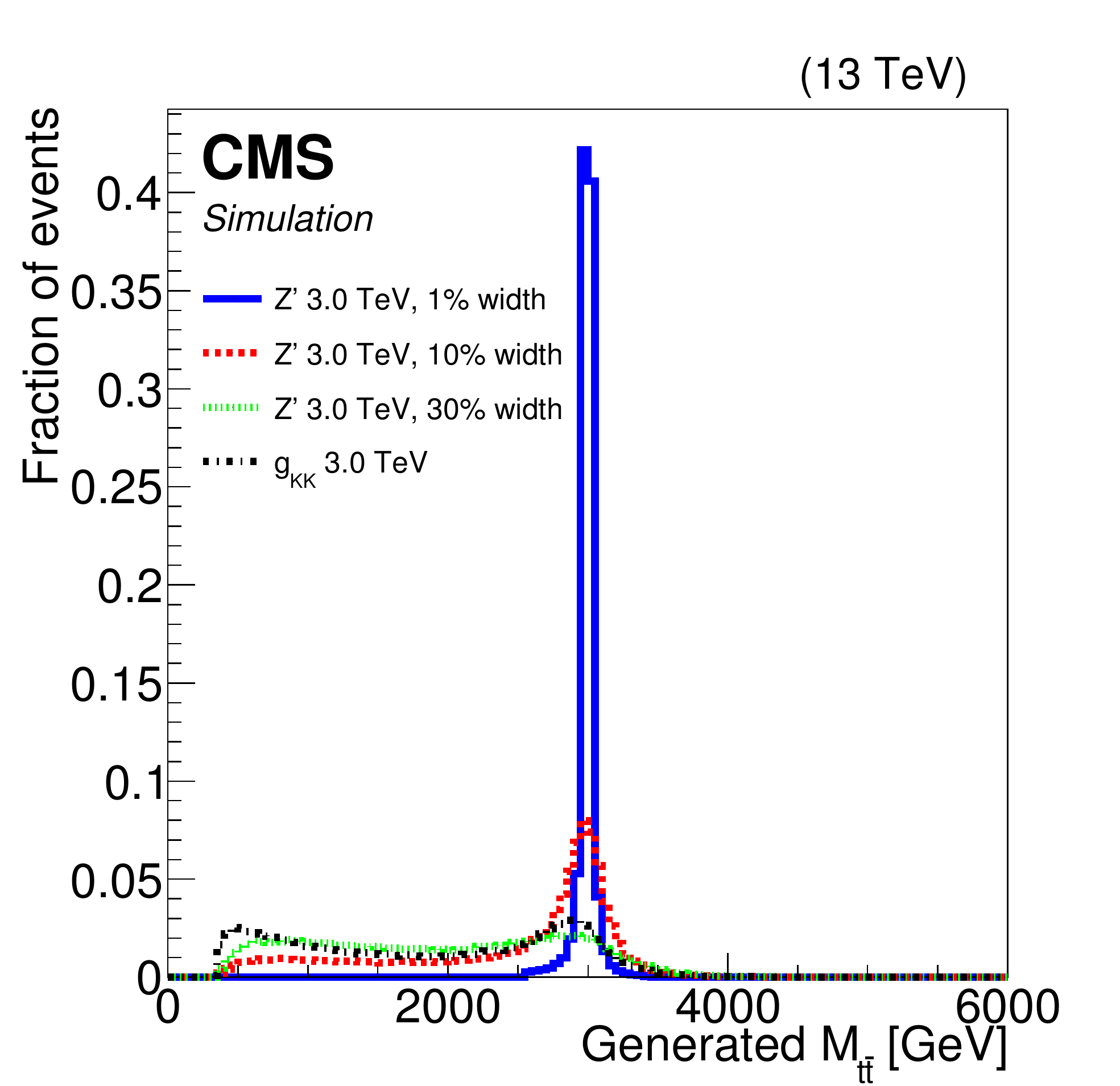}
\includegraphics[width=0.495\textwidth]{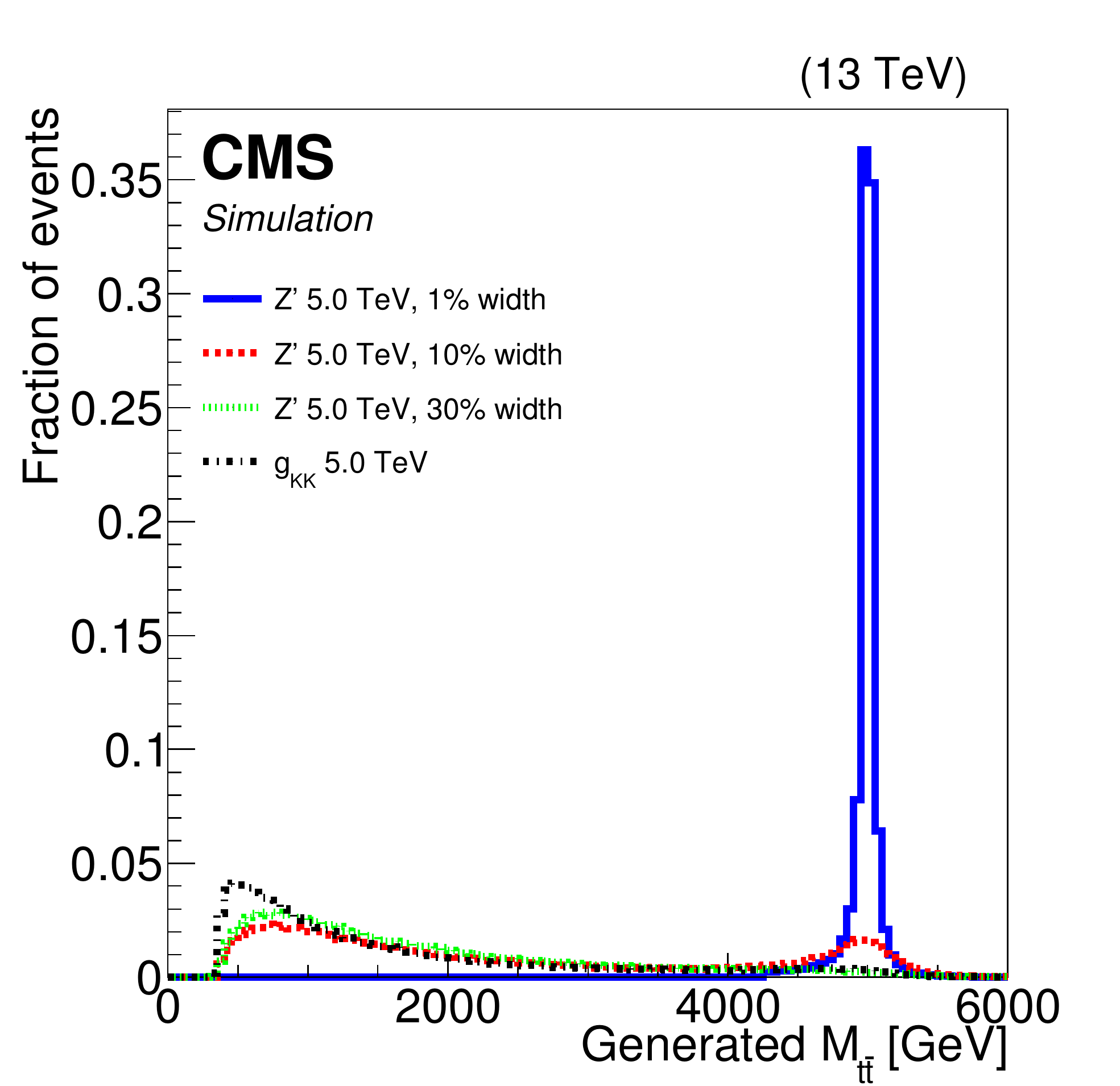}
\caption{The \ttbar invariant mass distributions for four signal models with resonance masses of 3\TeV (left) and 5\TeV (right). The \gKK resonance has a relative width $\Gamma/m \approx \text{15--20}\%$, which is between those of the wide and extra-wide \PZpr\ boson signal models.}
\label{fig:sig_model}
\end{figure}

The invariant mass distribution of the \ttbar system at the parton
level for \PZpr\ resonances with three different widths and a \gKK resonance
can be seen in Fig.~\ref{fig:sig_model}. The plots are normalized such
that the total integral of each signal model is 1. A resonant
structure is manifest at 3\TeV, but at 5\TeV the off-shell
component of the signal is strongly enhanced by the available parton
luminosity at lower masses. This effect is not noticeable for the narrow
\PZpr\ signal, but becomes more apparent for the wider \PZpr\ resonances. Such
behavior is expected for resonant \ttbar production in general.

{\tolerance=800 The \ttbar pair production background is simulated at next-to-leading
order (NLO) with the \POWHEG~v2
generator~\cite{Nason:2004rx,Frixione:2007nw,Frixione:2007vw,Alioli:2010xd}.
The \POWHEG generator is also used to simulate single top quark
production via EW interactions at NLO~\cite{Alioli:2009je,Re:2010bp}.
The \wjets background is simulated with the \MGvATNLO
generator with the FxFx matching prescription between matrix element
calculations and parton shower simulations~\cite{Frederix:2012ps}. The
Drell--Yan (DY) process with an invariant mass between 10 and 50\GeV is
simulated at NLO with the same generator, while for an invariant mass
above 50\GeV, leading order (LO) precision is used. Diboson and QCD multijet
production are simulated at LO with \PYTHIA. It should be noted that
simulated multijet events are only used for the background estimate
when QCD multijet production is a secondary background. In the case of
the fully hadronic analysis, the multijet background is estimated from
a CR in data, as described in Section~\ref{sec:bkg_allhad}. For all
simulated events, \PYTHIA with the CUETP8M1
tune~\cite{CMS-PAS-GEN-14-001} is used to describe the fragmentation and
hadronization. All the samples are generated with the NNPDF~3.0 parton
distribution functions (PDFs)~\cite{Ball:2014uwa}. All sample cross
sections are normalized to the latest theoretical calculations,
usually at next-to-NLO
precision~\cite{Czakon:2011xx,Li:2012wna,Kant:2014oha,Kidonakis:2012rm}. \par}

All samples are processed through a \GEANTfour-based
simulation~\cite{bib:GEANT}, which models the propagation of the
particles through the CMS apparatus and the corresponding detector
response. For all samples, the pileup distributions are weighted to
have an average of 23 pileup interactions per event, as measured in
data. The same event reconstruction software is used for data and
simulated events. Differences of a few percent in the resolution and
reconstruction efficiency are corrected to match those
measured in data using dedicated samples from data~\cite{Hildreth:2015kps}.

\section{Reconstruction and categorization of \texorpdfstring{\ttbar}{ttbar} events}
\label{sec:selection}

\subsection{Dilepton channel}
\label{sec:preselection_dilepton}

Events in the dilepton channel are selected by requiring oppositely
charged high-\pt lepton pairs: \Pgm\Pgm\, \Pe\Pe\ or \Pe\Pgm. Leptons
with $\pt > 53$ and 25 (45 and 36)\GeV in the \Pgm\Pgm\ (ee) channel
are selected. In the \Pe\Pgm\ channel, muons are required to have
$\pt > 53\GeV$ and electrons are required to have $\pt > 25\GeV$. Muons
(electrons) are required to be within $\abs{\eta} < 2.4$ (2.5). To
remove contributions from low-mass resonances and
$\PZ/\gamma(\to\ell\ell)$+jets production in events with same-flavor lepton
pairs, the dilepton invariant mass is required
to be above 20\GeV and outside of the \PZ\ boson mass window 76 to
106\GeV. Contamination from QCD multijet background is reduced by
applying a two-dimensional (2D) selection for both leptons: $\Delta
R_{\min}(\ell, j) > 0.4$ or $p_{\text{T},\text{rel}}(\ell, j) > 15\GeV$, where $\Delta
R_{\min}(\ell, j)$ is the minimum $\Delta R$-distance between the
lepton candidate and any AK4 jet with $\pt > 15\GeV$ and $\abs{\eta} < 3$ and
$p_{\text{T},\text{rel}}(\ell, j)$ is the \pt of the lepton with
respect to the axis of the $\Delta R$-nearest AK4 jet. The 2D
selection reduces the QCD multijet background by a factor of ${\approx}100$.
Events are further required to contain at least two AK4 jets
with $\abs{\eta} < 2.4$ and $\pt > 100$ and 50\GeV for the leading and
subleading jets, respectively. It is required that at least one of
the two leading jets must be \PQb\ tagged as determined by the loose
CSV tagger operating point. Finally, \ptmiss is required to be larger
than 30\GeV. The resulting sample is dominated by the
irreducible \ttbar background, which amounts to $ > $90\% of the total
background.

\begin{figure}[!htbp]
\centering
\includegraphics[width=0.495\textwidth]{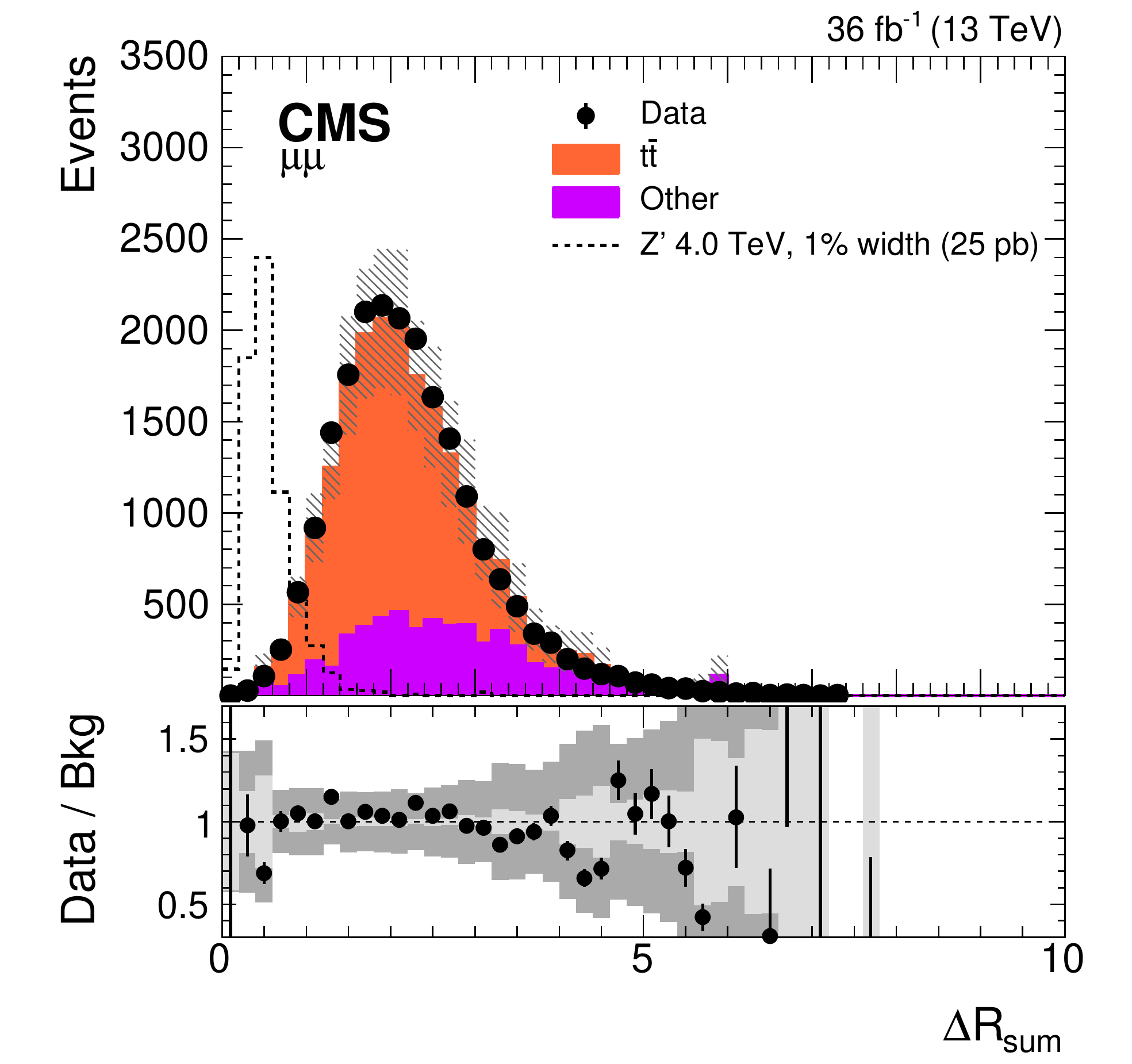}
\includegraphics[width=0.495\textwidth]{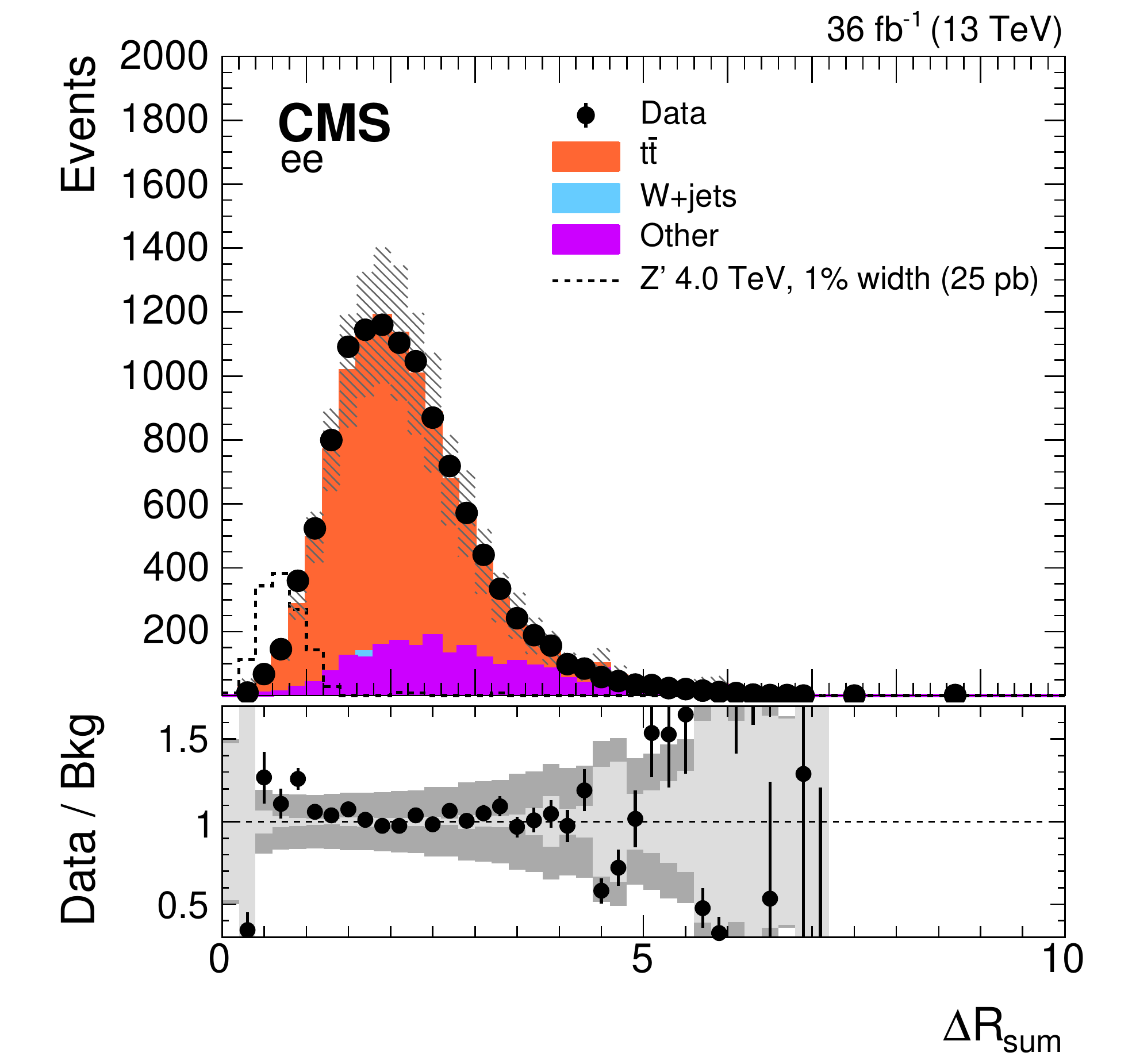}\\
\includegraphics[width=0.495\textwidth]{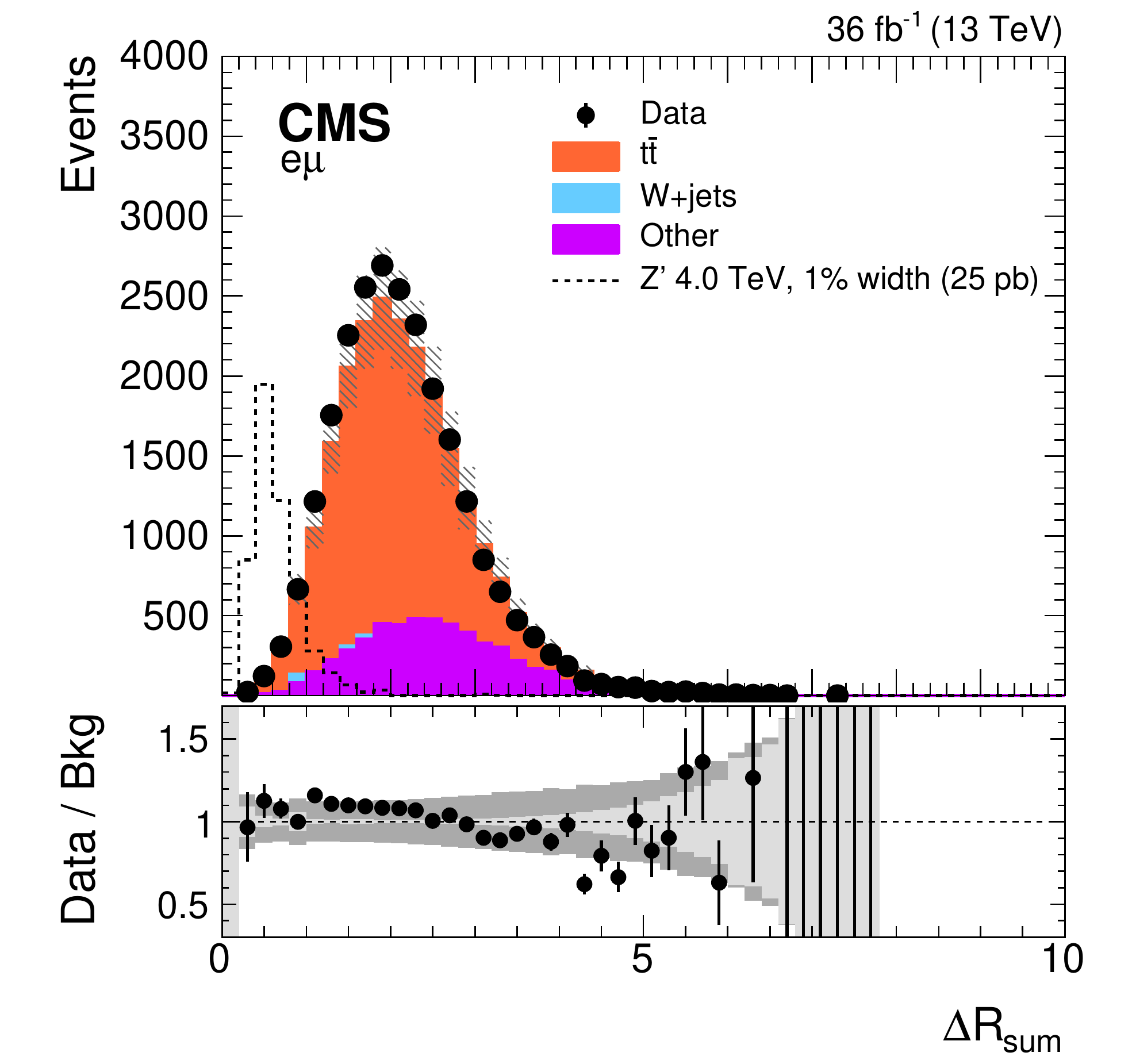}
\caption{Distributions of $\Delta R_{\text{sum}}$ in \Pgm\Pgm\ (upper left), \Pe\Pe\ (upper right), and \Pe\Pgm\ (lower) events. The contribution expected from a 4\TeV \PZpr\ boson, with a relative width of 1\%, is shown normalized to a cross section of 25\pb. The hatched band on the simulated distribution represents the statistical and systematic uncertainties. The lower panels in each plot show the ratio of data to the SM background prediction and the light (dark) gray band represents statistical (systematic) uncertainty. The error bars on the data points indicate the Poisson statistical uncertainties.}
\label{fig:DRsum}
\end{figure}

Figure~\ref{fig:DRsum} shows the distributions of $\Delta
R_{\text{sum}}= \Delta R(\ell_{1}, j) + \Delta R(\ell_{2}, j)$ in
\Pgm\Pgm, \Pe\Pe, and \Pe\Pgm\ subchannels, where $\Delta
R(\ell_{1}, j)$ and $\Delta R(\ell_{2}, j)$ are the
$\Delta \text{R}$ variables between the leading and subleading lepton and the
nearest jet. The lepton-jet pairs from \PZpr\ boson decays are expected
to be collimated and populate the low-$\Delta R_{\text{sum}}$
region. The $\Delta R_{\text{sum}}$ variable is used to separate events into signal- and
background-enriched samples: $\Delta R_{\text{sum}} < 1$ and $1 < \Delta
R_{\text{sum}} < 2$ defines the boosted and nonboosted signal regions (SRs), respectively, whereas $\Delta R_{\text{sum}} > 2$
defines the background-enriched region. The shape and normalization
are in agreement between data and simulation at low $\Delta
R_{\text{sum}}$, which is the region of interest for separating
boosted and resolved events.

\subsection{Single-lepton channel}
\label{sec:preselection_singlelepton}

The selection for events used in the single-lepton analysis requires
the presence of a muon with ${\pt > 55\GeV}$ and ${\abs{\eta} < 2.4}$ or
an electron with ${\pt > 80\GeV}$ and ${\abs{\eta} < 2.5}$. Neither
lepton has an isolation requirement other than passing the lepton 2D
selection, which requires the $\Delta R_{\min}(\ell, j) > 0.4 $ or
the $p_{\text{T},\text{rel}}(\ell, j) > 25\GeV$, where both
quantities are calculated with respect to all AK4 jets with
$\pt > 15\GeV$. Events with a second lepton are removed from the sample
to avoid any overlap with the dilepton channel. Events are also
required to contain at least two AK4 jets with $\abs{\eta} < 2.4$ and a
minimum \pt of 150 (185)\GeV for the leading jet in the muon
(electron) channel, and 50\GeV for the subleading jet. To reduce the
contributions to the sample from QCD multijet events, additional
requirements are imposed. In the muon channel, \ptmiss and
\htlep are required to be greater than 50 and
150\GeV, respectively, where $\htlep \equiv \ptmiss + \pt^{\ell}$. In the
electron channel, it is required that $\ptmiss > 120\GeV$. The
electron channel has a higher \ptvecmiss requirement because of the
larger QCD multijet background. As a result of this requirement,
an additional selection on \htlep would not increase performance.
In order to suppress the contamination from events originating from
\wjets events, a boosted decision tree~\cite{Roe:2004na} (\wjetsbdt)
was trained using the \textsc{tmva} software
package~\cite{Hocker:2007ht} on the jet-related variables listed below,
in order of importance.

\begin{enumerate}
\item $\Delta R_{\text{min}}(\ell,j)$, \ie, the separation between the lepton and its closest jet.
\item The CSV score of the subleading and leading AK4 jets.
\item The number of jets.
\item $p_{\text{T},\text{rel}}(\ell,j)$, \ie, the relative momentum between the jet and nearby lepton.
\item The reconstructed mass of the leading AK4 jet.
\item $\Delta R_{\text{min}}(\ell,j) \,\pt(j)$, \ie, the $\Delta R$ separation between the jet and nearby lepton scaled by the \pt of the jet.
\item The reconstructed mass of the subleading AK4 jet.
\item The shape variable $S^{33}$ of the sphericity tensor $S^{\alpha\beta} = (\sum_{i}p^{\alpha}_{i}p^{\beta}_{i})/(\sum_{i}\abs{p_{i}}^{2})$, where $\alpha,\beta$ correspond to the $x$, $y$, and $z$ components of the momentum vectors of the jets~\cite{Bjorken:1969wi,Hanson:1975fe}.
\item \HT+ \htlep, \ie, the summation of the hadronic, leptonic, and \ptmiss in the event.
\end{enumerate}

\begin{figure}[!htbp]
\centering
\begin{tabular}{cc}
\includegraphics[width=0.495\textwidth]{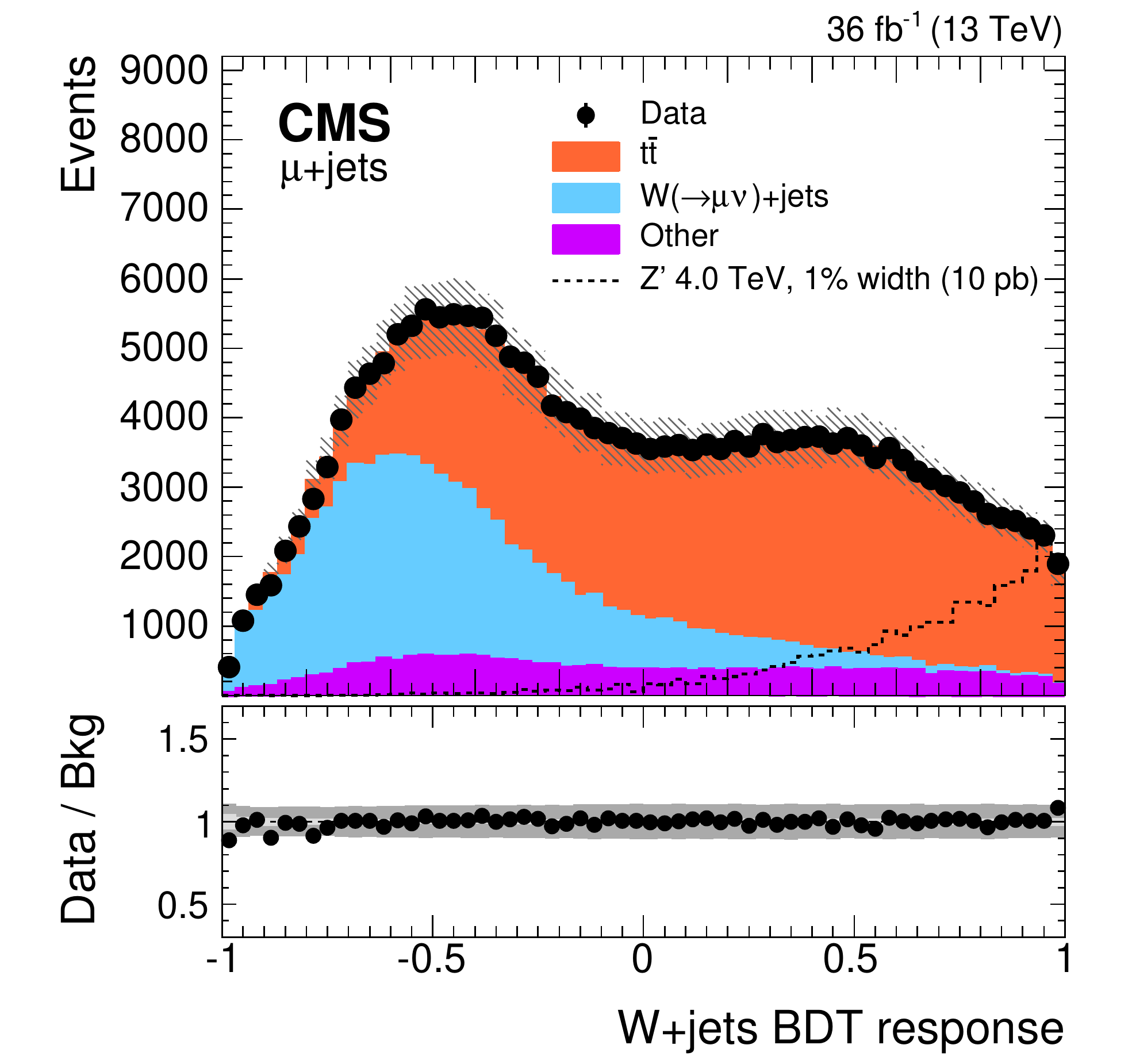}
\includegraphics[width=0.495\textwidth]{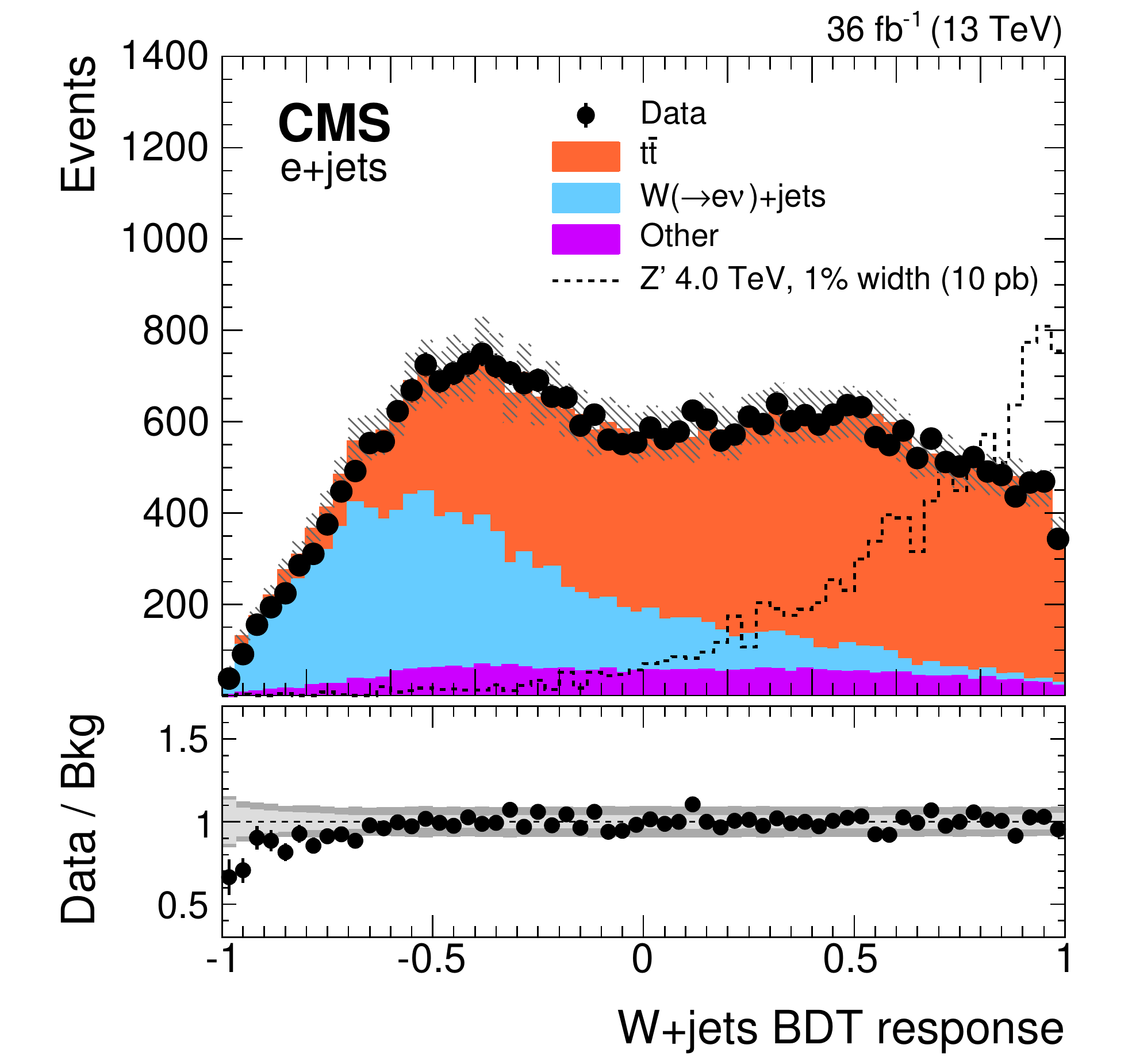}
\end{tabular}
\caption{\wjetsbdt distributions in the muon (left) and electron (right) single-lepton channel. The SR is defined as events with $\wjetsbdt \ge 0.5$. The contribution expected from a 4\TeV \PZpr\ boson, with a relative width of 1\%, is shown normalized to a cross section of 10\pb. The hatched band on the simulation represents the statistical and systematic uncertainties. The lower panels in each plot shows the ratio of data to the SM background prediction and the light (dark) gray band represents statistical (systematic) uncertainty. The error bars on the data points indicate Poisson statistical uncertainty.}
\label{fig:WBDT}
\end{figure}

Figure~\ref{fig:WBDT} shows the \wjetsbdt distribution in the muon and
electron channels. The requirement $\wjetsbdt \ge 0.5$ is applied to
the events in the SR, which is further separated in two regions,
depending on the presence of a \cPqt-tagged AK8 jet with $\pt >
400\GeV$ and rapidity $\abs{y} < 2.4$. Events with no \cPqt-tagged AK8 jet and
$\wjetsbdt < -0.75$ or $0 < \wjetsbdt < 0.5$ are dominated
by \wjets and \ttbar events, respectively, and constitute the
background enriched CRs.

The \ttbar system is reconstructed by assigning the four-vectors of
the reconstructed final-state objects (charged lepton, \ptmiss, and
jets) to the leptonic or hadronic legs of the \ttbar decay. For
events without an AK8 jet, several hypotheses are built based on possible
assignments of each AK4 jet to either the leptonic \PQt\ decay, the
hadronic \PQt\ decay, or neither. For events with an AK8 jet, that
jet is associated with the hadronic \PQt\ decay, and the leptonic
\PQt\ decay hypotheses only consider AK4 jets that are separated
from the AK8 jet by $\Delta R > 1.2$. In both cases, the combination
chosen is the one that minimizes the $\chi^{2}$ discriminator, where
\begin{linenomath*}
\begin{equation}
 \chi^{2} = \chi^{2}_{\text{lep}} + \chi^{2}_{\text{had}} = \left [ \frac{m_{\text{lep}} - \overline{m}_{\text{lep}}}{\sigma_{m_{\text{lep}}}} \right ]^2
      + \left [ \frac{m_{\text{had}} - \overline{m}_{\text{had}}}{\sigma_{m_{\text{had}}}} \right ]^2.
 \label{eq:chi2}
\end{equation}
\end{linenomath*}

In this equation, $m_{\text{lep}}$ and $m_{\text{had}}$ are the
invariant masses of the reconstructed leptonically and hadronically
decaying top quarks, respectively. The parameters
$\overline{m}_{\text{lep}}$, $\sigma_{m_{\text{lep}}}$,
$\overline{m}_{\text{had}}$, and $\sigma_{m_{\text{had}}}$ in the $\chi^2$
discriminator are determined from simulation by matching reconstructed
final-state objects of the hypothesis to the corresponding generator-level
particles from the \ttbar decay. Events in signal- and
background-enriched regions are all required to have
$\chi^{2} < 30$. Events with two \cPqt-tagged AK8 jets are removed
from the sample in order to avoid any overlap with the fully hadronic
channel.

\subsection{Fully hadronic channel}
\label{sec:preselection_allhad}

All events used in the fully hadronic analysis are required to fulfill
the following kinematic and \ttagging criteria. In order to reach a trigger
efficiency of ${\approx}100$\%, each event must have $\HT >
950\GeV$. Events are reconstructed using the two \pt-leading AK8
jets, both of which are required to have $\pt > 400\GeV$ and
$\abs{y} < 2.4$. In order to ensure a back-to-back topology, the two jets
must have an azimuthal separation $\abs{\Delta \phi} > 2.1$. These
kinematic requirements are later referred to as the fully hadronic
preselection. Both AK8 jets are required to be \PQt\ tagged for
events to enter the SR. These events are then separated into six SRs
based on two criteria: the rapidity difference between the two jets
($\abs{\Delta y} < 1.0$ or $\abs{\Delta y} > 1.0$) and the number of jets with
a \cPqb-tagged subjet (0, 1, or 2).

\begin{figure}[!htbp]
\centering
\begin{tabular}{cc}
\includegraphics[width=0.495\textwidth]{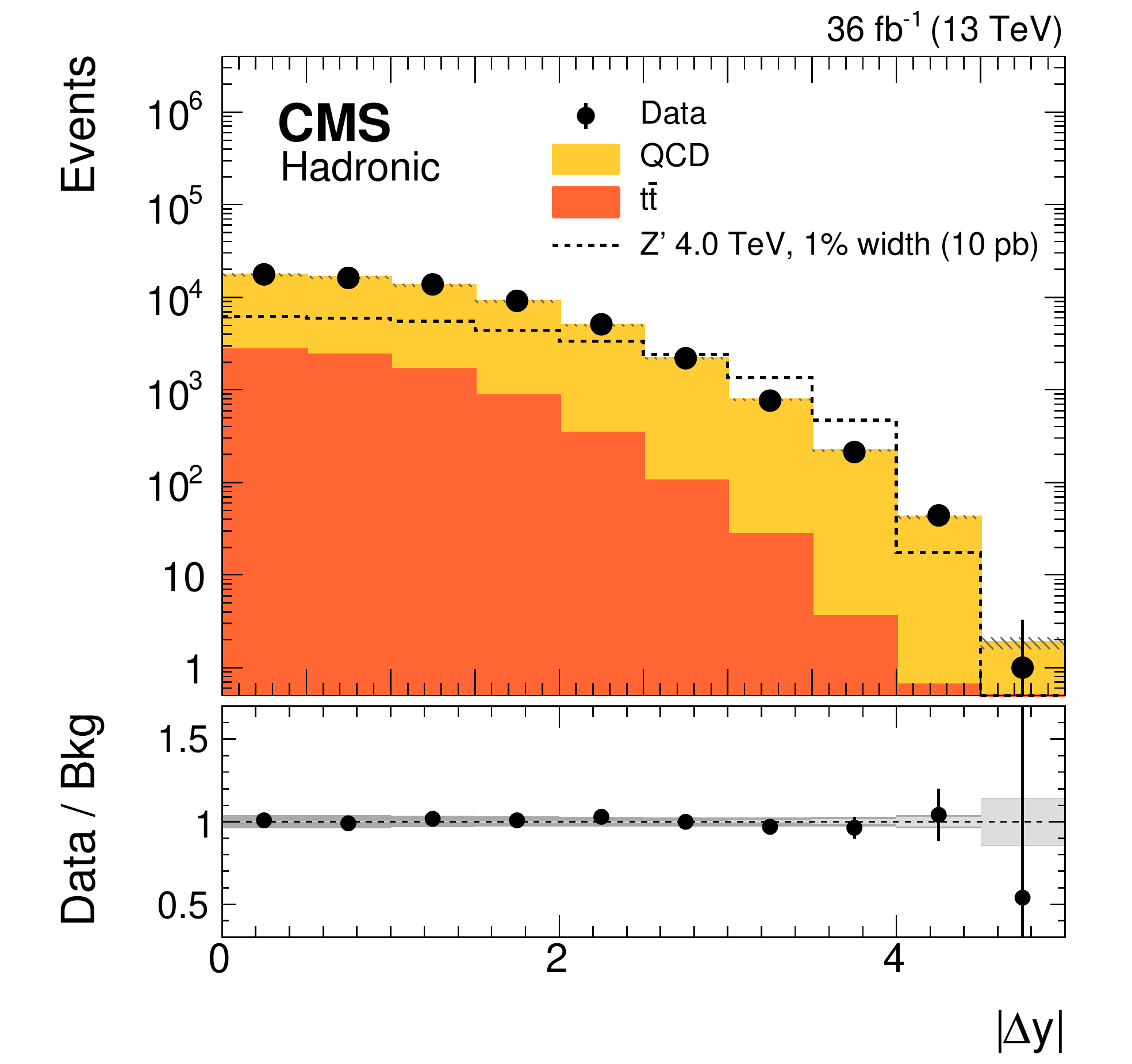}
\includegraphics[width=0.495\textwidth]{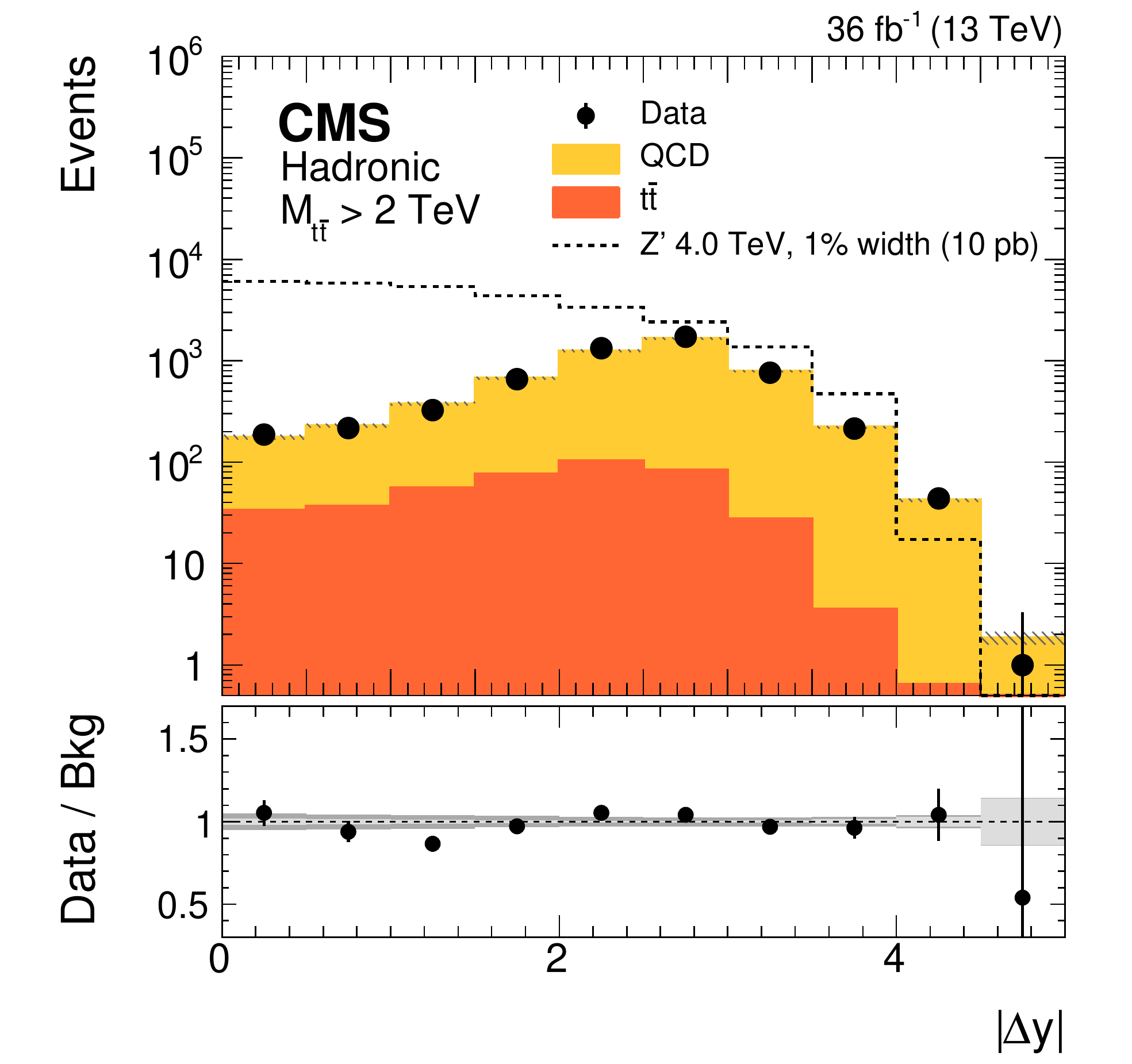}
\end{tabular}
\caption{Dijet rapidity difference $(\Delta y)$ for events passing the fully hadronic event selection for all \mttbar (left) and for events with an $\mttbar > 2\TeV$ (right). The contribution expected from a 4\TeV \PZpr\ boson, with a relative width of 1\%, is shown normalized to a cross section of 10\pb. The hatched band around the simulated distribution represents the statistical and systematic uncertainties. The lower panels in each plot show the ratio of data to the SM background prediction and the light (dark) gray band represents statistical (systematic) uncertainty.}
\label{fig:kinem_DeltaRap_2ttag}
\end{figure}

The categories with a greater number of jets with a \cPqb-tagged
subjet are expected to provide higher sensitivity, while those with
fewer \cPqb-tagged subjets are included to provide better
constraints on the backgrounds and additional sensitivity to the
analysis. The low-$\abs{\Delta y}$ region is expected to be more
sensitive than the high-$\abs{\Delta y}$ region. At high values
of \mttbar, QCD multijet events will have jets with greater $y$ separation, as
compared to those from a massive particle decay, in order to achieve
such high invariant masses. This is illustrated in
Fig.~\ref{fig:kinem_DeltaRap_2ttag}, which shows the dijet rapidity
difference for events passing the fully hadronic event selection. The
plot on the left is inclusive in \mttbar, while the plot on the right
shows events with $\mttbar > 2\TeV$.

\section{Estimation of the background}
\label{sec:background}

\subsection{Dilepton channel}
\label{sec:bkg_dilepton}

\begin{figure}[!htbp]
\centering
\includegraphics[width=0.495\textwidth]{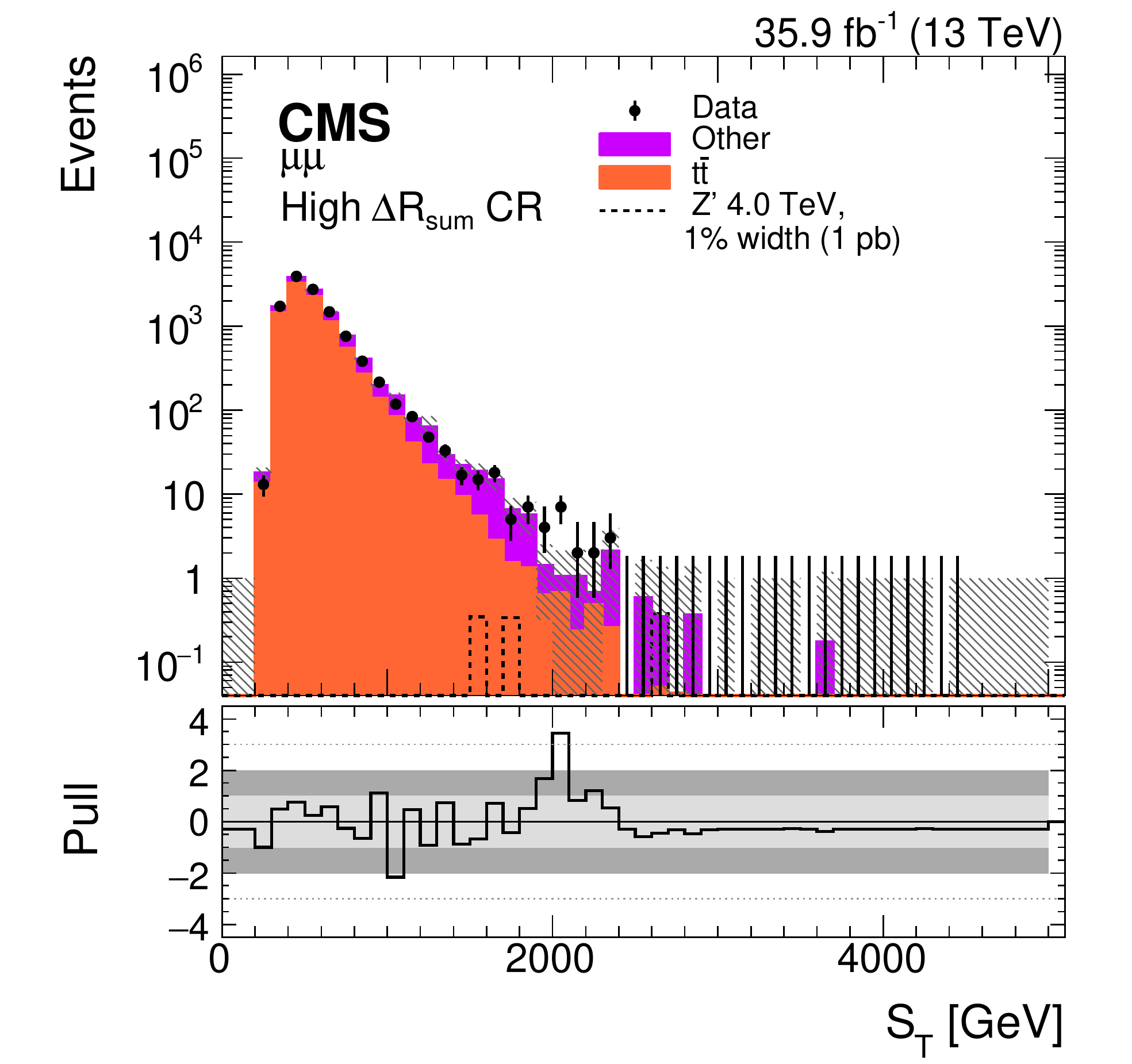}
\includegraphics[width=0.495\textwidth]{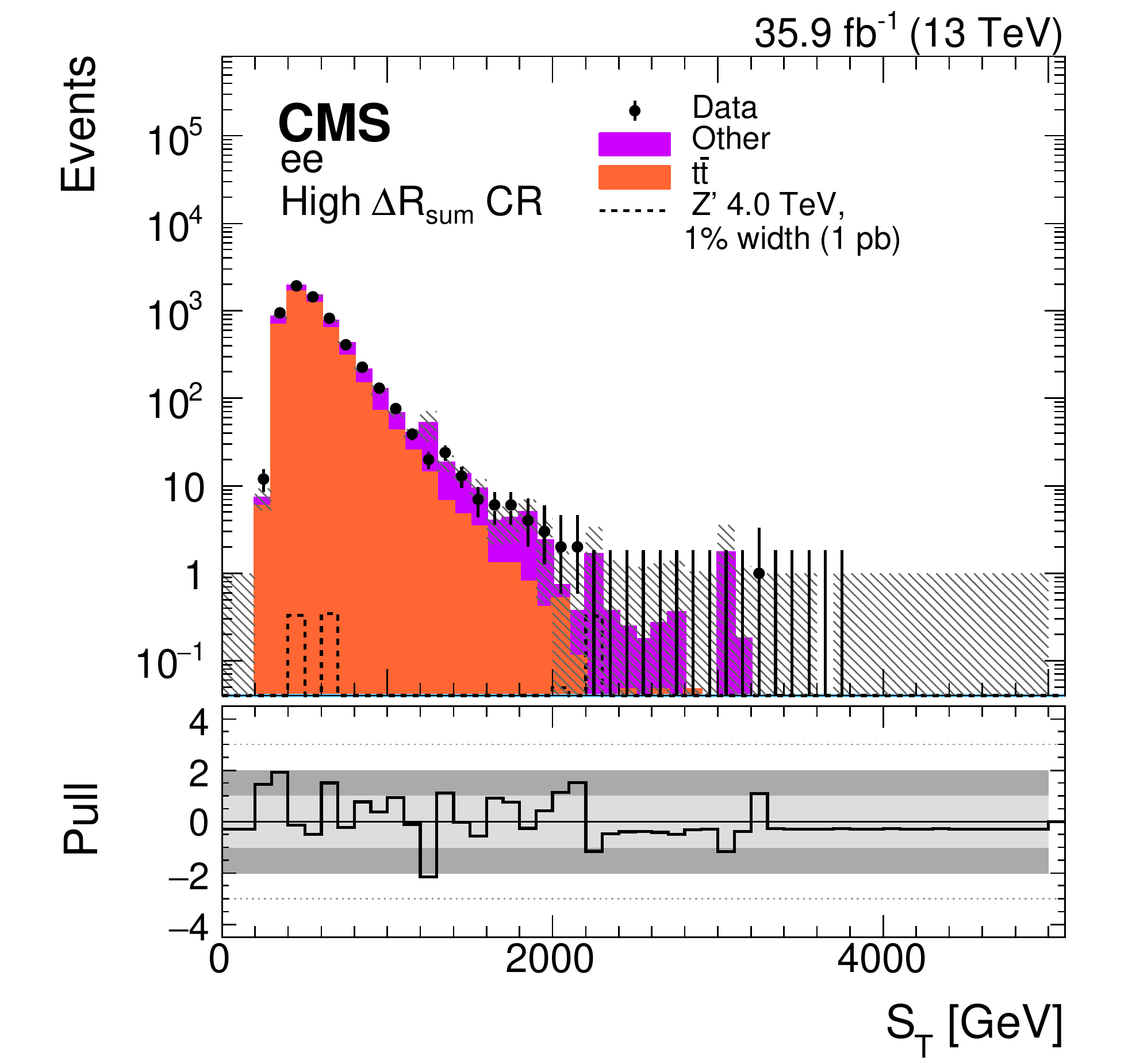}\\
\includegraphics[width=0.495\textwidth]{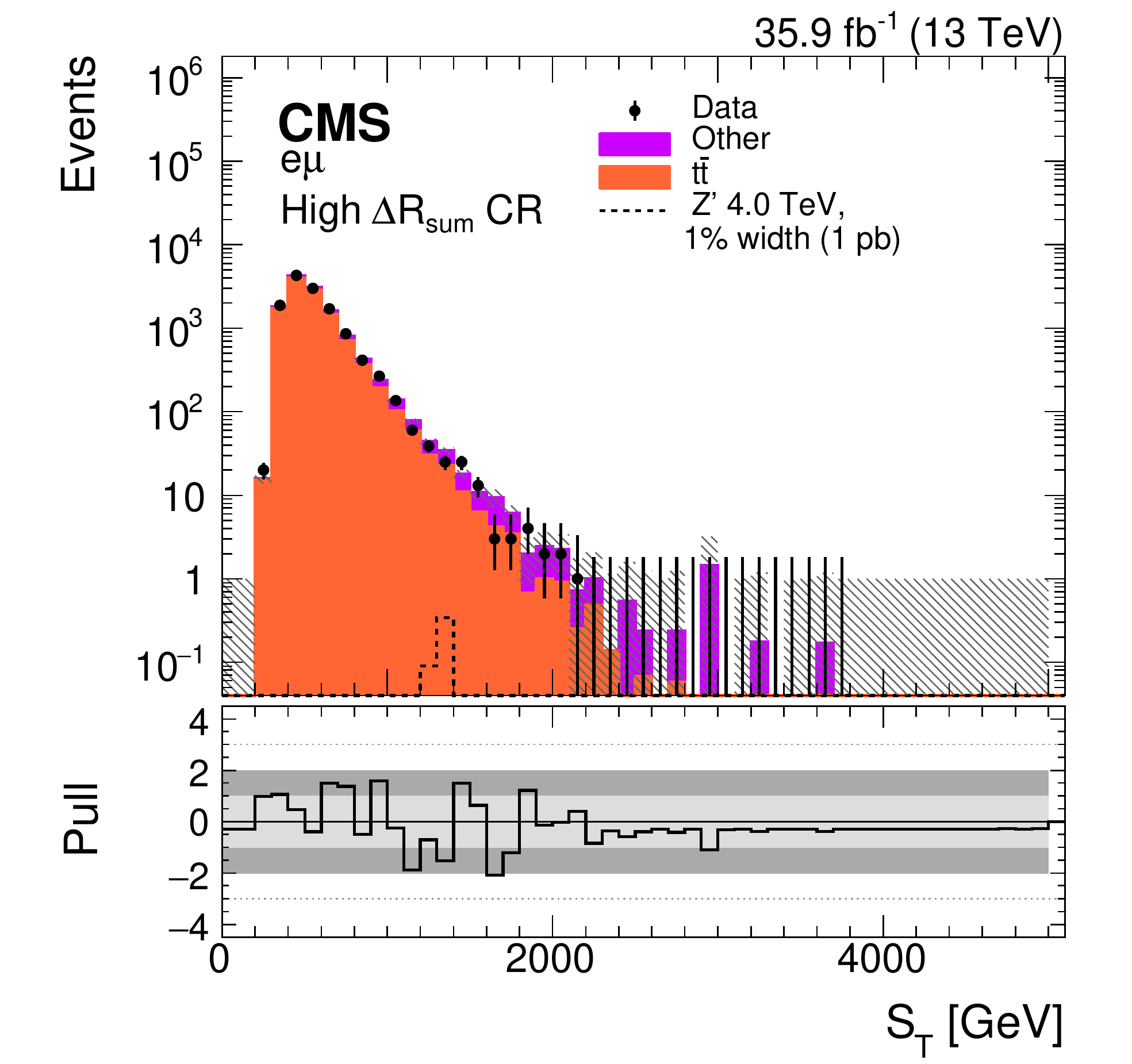}
\caption{Distributions of \ST in the background-enriched CR for \Pgm\Pgm\ (upper left), \Pe\Pe\ (upper right), and \Pe\Pgm\ (lower) subchannels. The contribution expected from a 4\TeV \PZpr\ boson, with a relative width of 1\%, is shown normalized to a cross section of 1\pb. The hatched band on the simulation represents the uncertainty in the background prediction. The lower panel shows the pull of each histogram bin from the SM prediction. The light (dark) gray band represents a pull of one (two) standard deviations (s.d.) from the predicted value.}
\label{fig:st_bkgd}
\end{figure}

The dominant irreducible background in the dilepton channel is
\ttbar production. Other secondary backgrounds arise from \zjets,
single top quark, and diboson processes. Simulated events are used to
model the shape of the kinematic distributions for the background
processes, including modeling the \ST variable used in the
statistical interpretation of the observations. The overall
normalization of the background processes is based on the
corresponding theoretical cross sections. The distributions are
allowed to vary within prior bounds of rate and shape uncertainties
during the statistical treatment, which employs six signal- and three
background-enriched regions, defined in
Section~\ref{sec:preselection_dilepton}. Modeling of the background is
separately checked in the background-enriched CR obtained with the
requirement $\Delta R_{\text{sum}} > 2$. Figure~\ref{fig:st_bkgd} shows the
distributions of \ST in the CR for \Pgm\Pgm, \Pe\Pe,
and \Pe\Pgm\ channels. The background simulation is in agreement with
data within the statistical and systematical uncertainties. The
quantity `pull', shown in Fig.~\ref{fig:st_bkgd} and subsequent
figures, is computed according to the following procedure. First, the
total uncertainty per bin is determined by adding the statistical and
all systematic uncertainties together in quadrature. Based on the
expected number of events and the total uncertainty in each bin,
pseudo-experiments are performed by sampling from a Gaussian
distribution with the mean equal to the expected number of events and
the standard deviation equal to the total uncertainty. For each
pseudo-experiment, a distribution of the number of expected events is
populated using Poisson statistics convolved with the Gaussian
distribution describing the variation in the expected number of events
in the bin. Finally, the number of events observed in data is used in
conjunction with the distribution of pseudo-experiments to calculate a
p-value, and the corresponding z-score is taken to be the pull.

\subsection{Single-lepton channel}
\label{sec:bkg_singlelepton}

Standard model \ttbar production is the main irreducible background in
the single-lepton channel. Other background processes include \wjets,
single top quark, \zjets, and diboson production. The QCD multijet
background is a minor contribution in the single muon channel
(${\approx}3$\%), and is suppressed to a negligible level in the
single-electron channel because of higher \pt and \ptmiss
requirements. All background processes in the single-lepton channel
are modeled from simulated events, and the normalization of each
background is based on its theoretical cross section. The rate and
shape of the backgrounds are allowed to vary in the statistical
analysis as described in Section~\ref{sec:stats}. Events that pass the
requirements in Section~\ref{sec:preselection_singlelepton} are
separated in two signal- and two background-enriched regions, defined
as follows.

\begin {enumerate}
\item Signal Region (SR1T): $\chi^{2} < 30$, $\wjetsbdt \geq 0.5$, 1 \cPqt-tagged AK8 jet.
\item Signal Region (SR0T): $\chi^{2} < 30$, $\wjetsbdt \geq 0.5$, no \cPqt-tagged AK8 jet.
\item Control Region (CR1): $\chi^{2} < 30$, $\wjetsbdt < -0.75$.
\item Control Region (CR2): $\chi^{2} < 30$, $0.0 < \wjetsbdt < 0.5.$
\end{enumerate}

\begin{figure}[!htbp]
\centering
\begin{tabular}{ccc}
\includegraphics[width=0.495\textwidth]{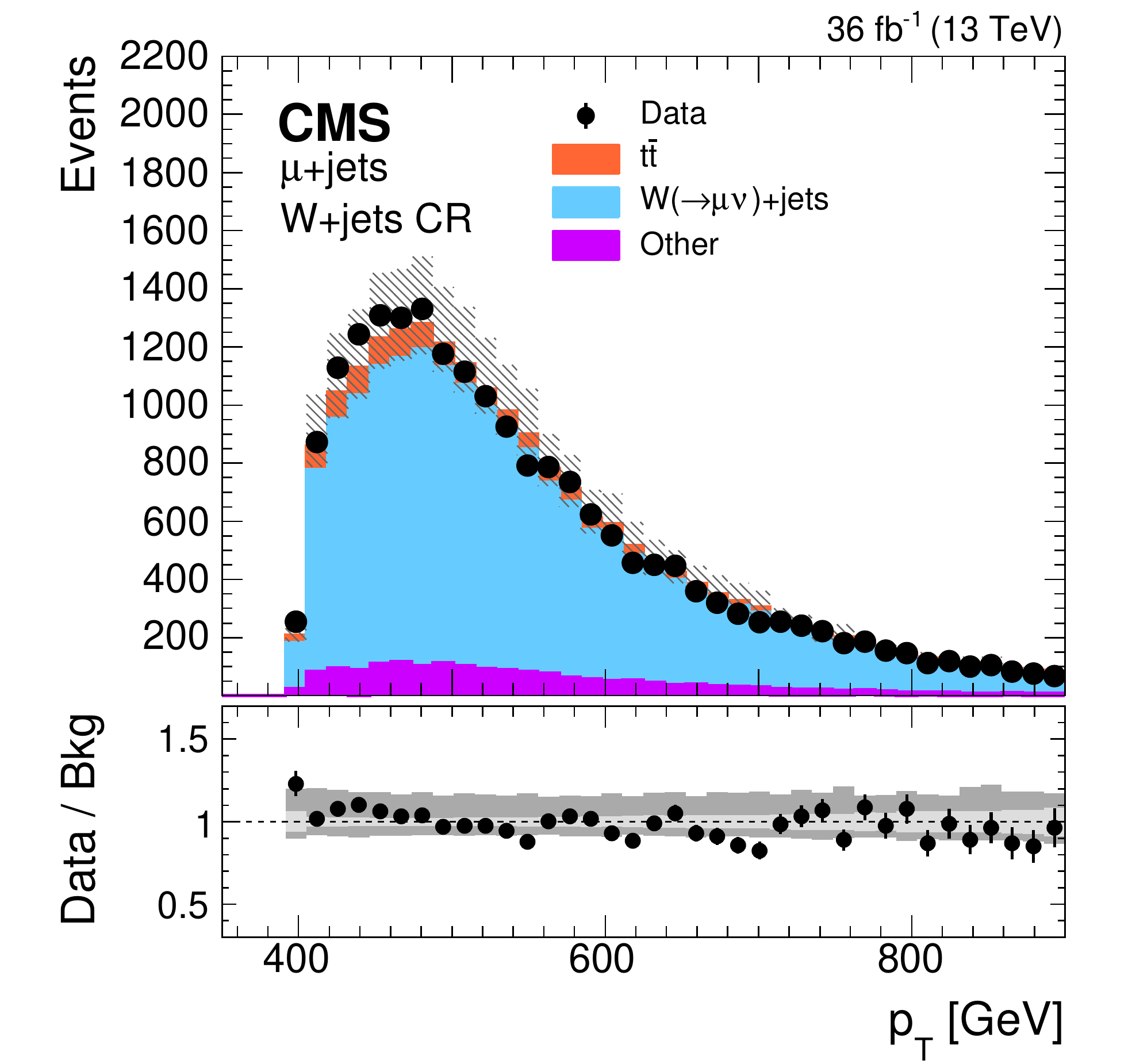}
\includegraphics[width=0.495\textwidth]{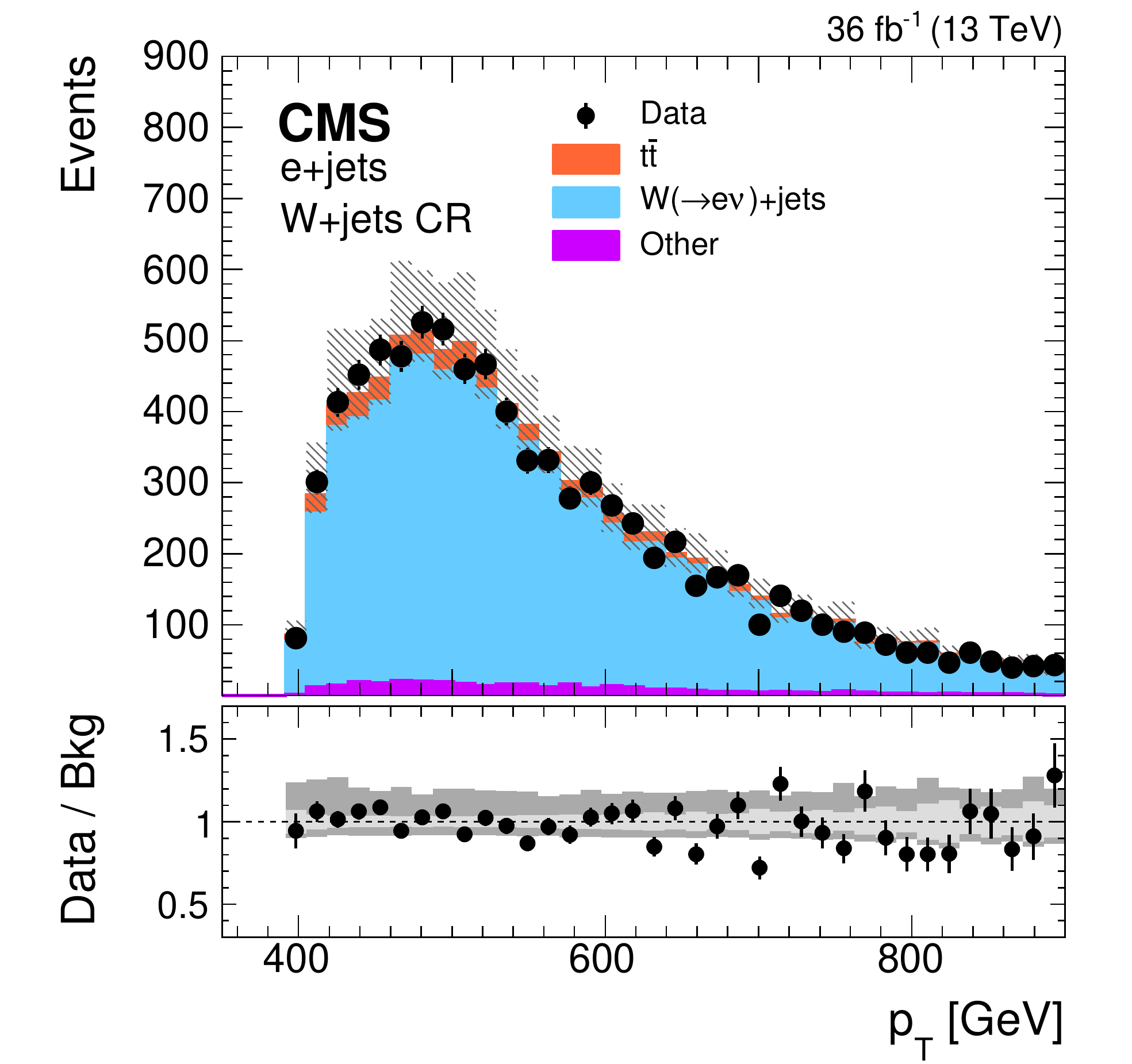}\\
\includegraphics[width=0.495\textwidth]{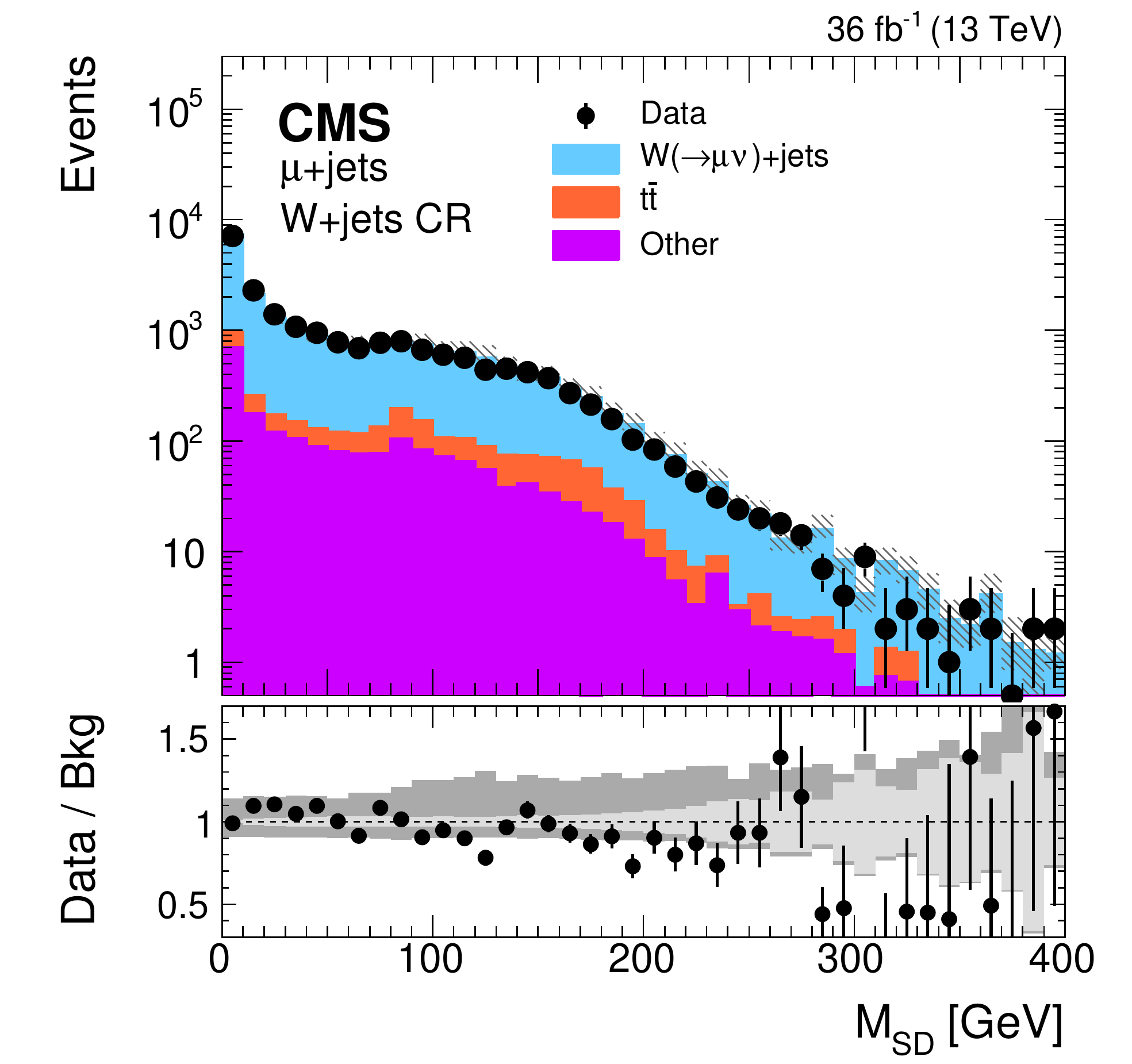}
\includegraphics[width=0.495\textwidth]{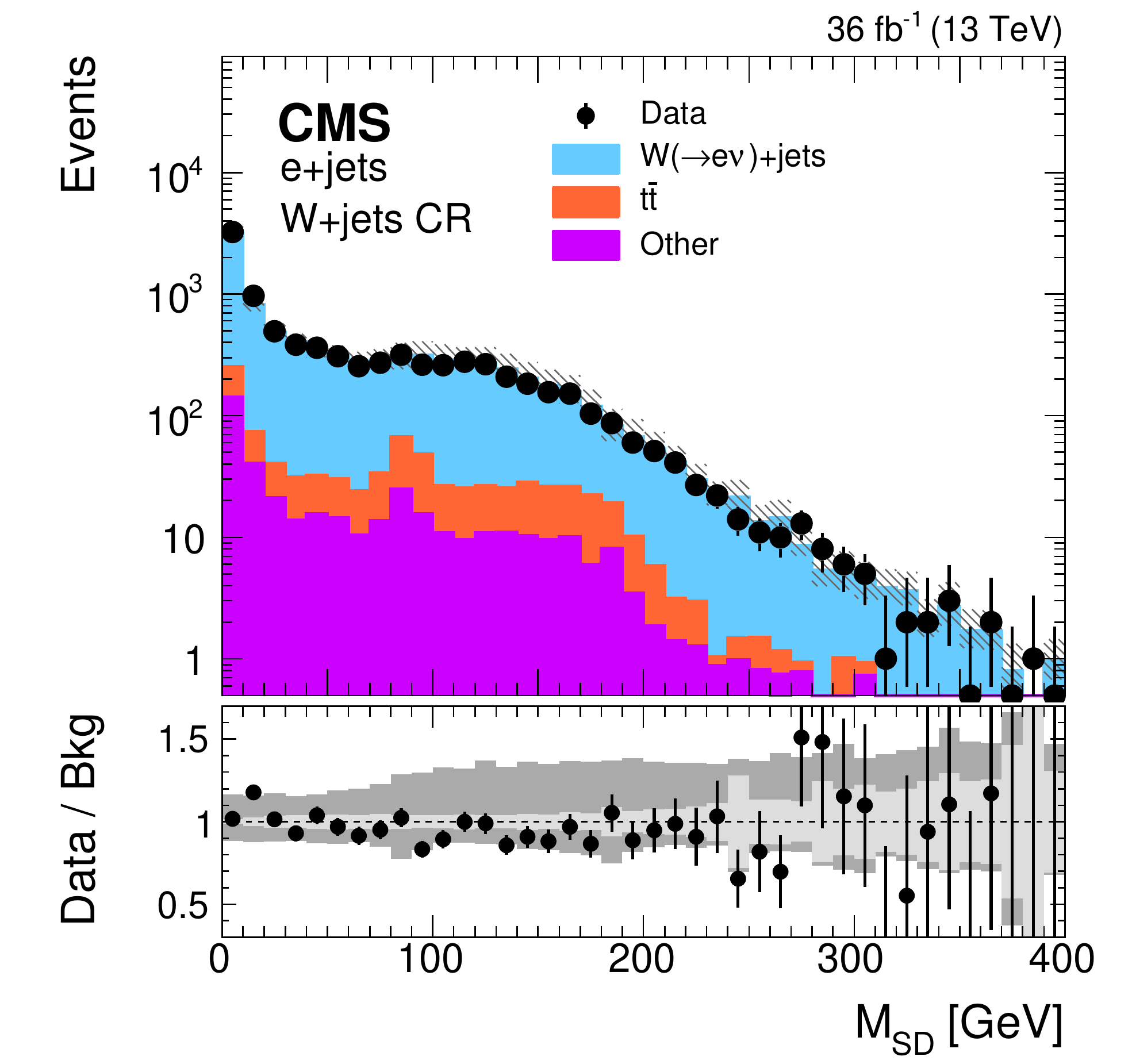}
\end{tabular}
\caption{Distributions of \pt (upper) and \msd (lower) for the \wjets background in the muon (left) and electron (right) channels using the \wjets mistag CR. The jet \pt information is taken from the CHS jets, while the \msd is take from the PUPPI jets. The hatched band on the simulation represents the uncertainty in the background prediction. The lower panels in each plot show the ratio of data to the SM background prediction and the light (dark) gray band represents statistical (systematic) uncertainty.}
\label{fig:sl_bkgd}
\end{figure}

The first control region (CR1) is dominated by \PW+jet events, while
CR2 is dominated by \ttbar events. For all regions, events are
separated based on the lepton flavor (\Pgm, \Pe), which results in
eight exclusive categories used in the binned maximum likelihood
fit. The rate at which light-flavor quarks and gluons are
misidentified as originating from top quarks (\tmistag) is measured in
data and simulation using a \wjets mistag CR with
$\chi^{2}_{\text{lep}} > 30$ and $\wjetsbdt < -0.5$. The \pt and
\msd distributions in the \wjets background can be seen in
Fig.~\ref{fig:sl_bkgd}.

\subsection{Fully hadronic channel}
\label{sec:bkg_allhad}

The two main sources of background in the fully hadronic channel are
QCD multijet and \ttbar production. For the latter background,
simulated events are used to model the shape of the \mttbar
distribution. This distribution is initially normalized to the
theoretical cross section, but it is allowed to vary within the bounds
of rate and shape uncertainties during the statistical treatment. The
final normalization and shape are determined by fitting the
distributions in the six SRs, defined in
Section~\ref{sec:preselection_allhad}.

\begin{figure}[!htbp]
\centering
\includegraphics[width=0.75\linewidth]{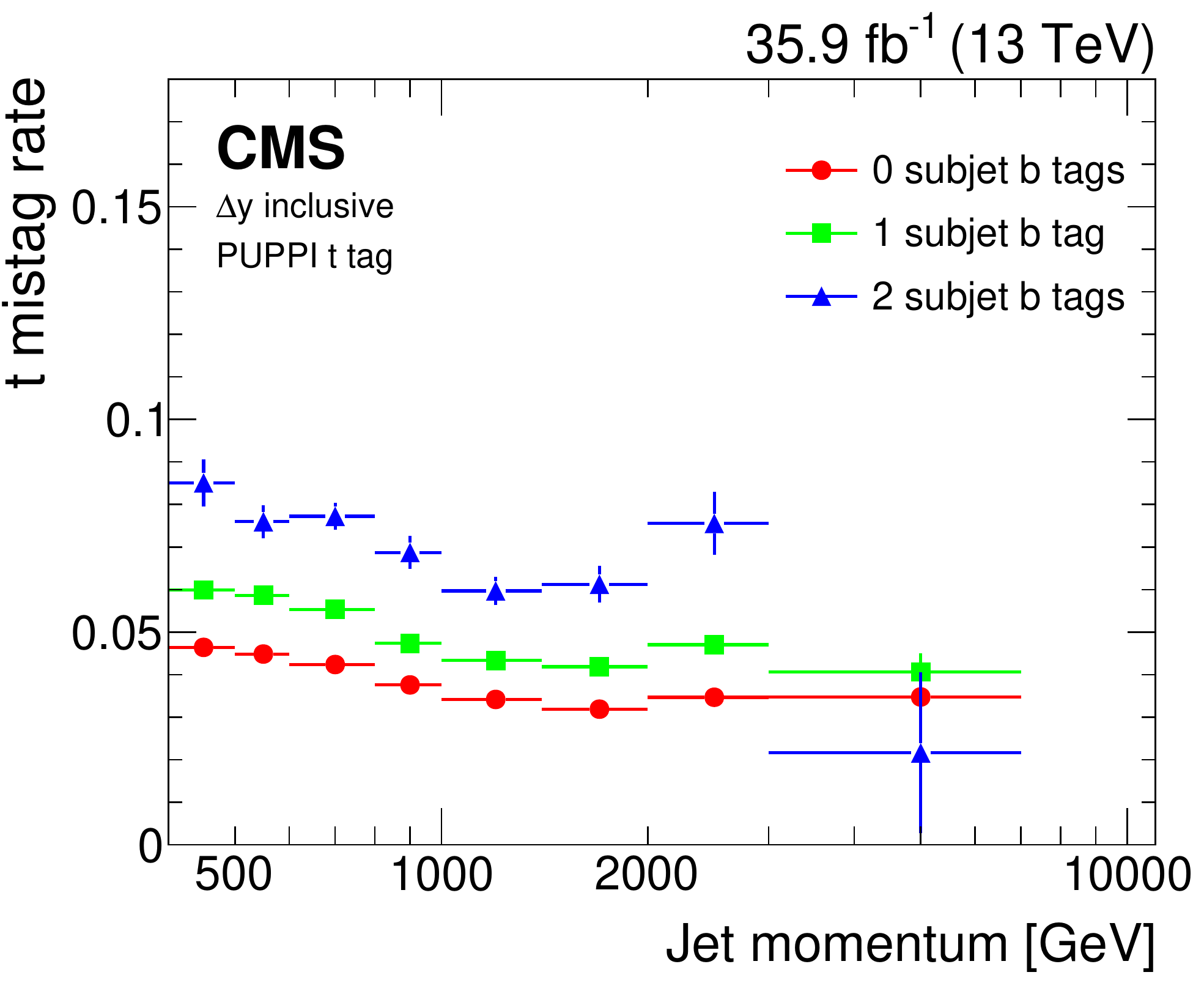}
\caption{The \tmistag rate as measured with an anti-tag and probe procedure separately for each \cPqb-tag category.}
\label{fig:mistag}
\end{figure}

The QCD multijet background is estimated from data, using a method similar to the techniques described in
Ref.~\cite{cms_ttbar_resonance4}. The preselection described in
Section~\ref{sec:preselection_allhad} is enforced in order to select a
back-to-back dijet event topology. In the first step of the background
estimate, the \tmistag rate in QCD multijet events is
measured. A QCD multijet enriched region is selected by requiring one
of the two jets to be ``anti-tagged,'' meaning it has a PUPPI soft
drop mass in the \cPqt-tag mass window $105 < \msd < 210\GeV$,
but the N-subjettiness requirement is inverted to
$\tau_{32} > 0.65$. The opposite ``probe'' jet is used to determine the
\tmistag rate. This rate is parametrized as a function of probe jet
momentum ($p$) and is measured for each of the three subjet \cPqb-tag
categories~(Fig.~\ref{fig:mistag}). This ``anti-tag and probe''
procedure is repeated for the \ttbar simulation, indicating that there is a small
(${\approx}2$\%) contribution from SM \ttbar events. The observed \ttbar contamination is
then subtracted from the anti-tag and probe data selection.

After the \tmistag rate has been measured in the QCD multijet
CR, it is used to estimate the \mttbar QCD multijet distribution in
the SR. First, a ``single-tagged'' region is selected, in which at
least one of the two jets is required to be \PQt\ tagged, meaning it has a PUPPI
\msd in the \cPqt-tag mass window $105 < \msd < 210\GeV$
and an N-subjettiness requirement of $\tau_{32} < 0.65$. One of the
two top quark jet candidates is randomly selected, in order to avoid
bias. If the selected jet is \PQt\ tagged, the event is included in the QCD
multijet estimate. The event is weighted by the previously measured
\tmistag rate, based on the momentum of the opposite jet and
the number of subjet \PQb\ tags in the event. Again, the procedure
is repeated for the \ttbar simulation, and the \ttbar contamination is
subtracted from the QCD multijet background estimate. This eliminates
double counting between the \ttbar and QCD multijet distributions.

Finally, a ``mass-modified'' procedure is employed in order to ensure
that the jets used in the QCD multijet estimate mimic the relevant
kinematics of the jets in the SR. If the mass of the second QCD
multijet jet is not in the top quark mass window, it is assigned a
random value within that window. This modified mass is randomly
selected from the distribution of simulated light-flavor jets, with
masses within the \cPqt-tag window, $105 < \msd <
210\GeV$. A check of the entire background estimation method
using simulated QCD multijet events is self-consistent.

\section{Systematic uncertainties}
\label{sec:systematics}

Several sources of uncertainty that impact the final results of this
search are considered. In all cases, the uncertainties in
reconstruction efficiency and event interpretation are propagated to
the distribution used for signal extraction. These uncertainties can
be broadly grouped into two categories: those uncertainties that
affect only the overall normalization of expected background events
and those uncertainties that can result in a different reconstruction
of the \ttbar system, and therefore change the shape of the \mttbar
distribution. Each source of systematic uncertainty is accounted for
through unique nuisance parameters applied to the likelihood described
in Section~\ref{sec:results}. For contributions that apply to multiple
analysis channels, the nuisance parameters are fully correlated,
allowing better constraints to be placed on sources of systematic
uncertainties. The individual sources of uncertainty are described in
detail below, and are summarized in Table~\ref{tab:systematics}.

\begin{table}[!htbp]
\centering
\topcaption{Sources of systematic uncertainty that affect the \mttbar and \ST distributions in each analysis channel. For uncertainty sources that apply to multiple channels, the corresponding nuisance parameter is fully correlated across these channels if the symbol \ck appears in the same row. For normalization uncertainties, the size of the effect on the prior distribution is indicated. Shape uncertainties have priors of $\pm1$ s.d., and the dependence on the kinematic quantities is shown.}
\begin{scotch}{lcccc}
\multicolumn{2}{c}{Uncertainty} & \multicolumn{3}{c}{Channel} \\
Source & Prior & Dilepton & Single-Lepton & Hadronic \\
\hline
\btagging efficiency & $\pm1\text{ s.d.}(\pt,\eta)$ & \ck & & \ck \\
\PQb\ mistag rate & $\pm1\text{ s.d.}(\pt,\eta)$ & \ck & & \ck \\
Parton distribution functions & $\pm1\text{ s.d.}$ & \ck & \ck & \ck \\
CSV discriminator shape & $\pm1\text{ s.d.}$ & & \ck & \\
Diboson cross section & 50\% & \ck & & \\
Electron trigger& $\pm1\text{ s.d.}(\pt,\eta)$ & \ck & \ck& \\
Electron identification& $\pm1\text{ s.d.}(\pt,\eta)$ & \ck & \ck & \\
Jet energy scale& $\pm1\text{ s.d.}(\pt,\eta)$ & \ck & \ck & \ck \\
Jet energy resolution& $\pm1\text{ s.d.}(\eta)$ & \ck & \ck & \ck \\
Integrated luminosity measurement& 2.5\% & \ck & \ck & \ck \\
QCD multijet modified mass procedure& $\pm1\text{ s.d.}$ & & & \ck \\
QCD multijet estimate closure test& $\pm1\text{ s.d.}$ & & & \ck \\
Muon trigger& $\pm1\text{ s.d.}(\pt,\eta)$ & \ck & \ck & \\
Muon identification& $\pm1\text{ s.d.}(\eta)$ & \ck & \ck & \\
Pileup reweighting& $\pm1\text{ s.d.}$ & \ck & \ck & \ck \\
Renorm/fact. scales (\ttbar production) & $\pm1\text{ s.d.}$ & \ck & \ck & \ck \\
Single top quark cross section & 50\% & \ck & & \\
\ttagging efficiency& unconstrained & & \ck & \ck \\
\tmistag rate (fully hadronic) & $\pm1\text{ s.d.}(p)$ & & & \ck \\
\tmistag rate (single-lepton) & $\pm1\text{ s.d.}$ & & \ck & \\
Top quark pair cross section & 20\% &\ck & \ck & \ck \\
Top quark \pt reweighting& $\pm1\text{ s.d.}$ & \ck & \ck & \ck \\
\wjets cross section & 25\% & \ck & \ck & \\
\zjets cross section & 50\% & \ck & \ck & \\
\end{scotch}
\label{tab:systematics}
\end{table}

Including all the systematic uncertainties degrades the final
cross section limits by 10\% for resonance masses above
2.5\TeV. Lower mass hypotheses are more sensitive to the
systematic effects, thus the limit on the cross section
degrades by up to 60\% for the lowest mass \PZpr\
resonance considered (500\GeV). The uncertainties in the
jet energy corrections, pileup distribution, and \ttbar cross
section are the most significant. They result in a reduction
of the excluded mass by 1.1, 1.0, and 1.0\%, respectively.
All other systematic uncertainties have less than a 1\%
effect. Per channel, the most significant systematic
uncertainties are the b tagging scale factor, the \ttbar
renormalization and factorization scales, and the standard
model \ttbar cross section for the dilepton, single-lepton,
and all hadronic channels, respectively. The most
constrained nuisance parameters are those associated
with the \ttbar renormalization and factorization scales
as well as the top tagging efficiency, which are
constrained to 8.5 and 9.2\% of their prior uncertainty.
The average nuisance parameter has a post-fit
uncertainty that is 75\% lower than its prior estimate.

\begin{enumerate}

\item \textit{Standard model cross sections}: Uncertainties in the cross sections used to normalize simulated background processes are obtained using the fitting procedure described in Section~\ref{sec:introduction}. For the \ttbar, \wjets, and \zjets backgrounds, a priori uncertainties of 20, 25, and 50\% are assigned, respectively. A cross section uncertainty of 50\% is used for the subdominant diboson and single top quark backgrounds. The values chosen reflect the relatively large uncertainties associated with modeling these backgrounds in the Lorentz-boosted phase space where the analysis is performed.

\item \textit{Integrated luminosity}: The uncertainty in the measurement of the integrated luminosity is 2.5\%~\cite{CMS:lumi}, and is applied to all simulated signal and background samples.

\item \textit{Pileup reweighting}: All simulated samples used in the analysis are reweighted to ensure that the distribution of the number of pileup interactions per event matches the corresponding distribution in data. This pileup distribution is obtained using a total inelastic cross section value of 69.2\unit{mb}~\cite{Sirunyan:2018nqx,Aaboud:2016mmw}. A systematic uncertainty in the distribution is obtained by varying the value by $\pm 4.6$\%, which is calculated using the method described in~\cite{Aaboud:2016mmw} using the cross sections from~\cite{Sirunyan:2018nqx}. The resulting uncertainty has both a normalization and shape component.

\item \textit{Lepton reconstruction and triggers}: Simulated events are corrected by scale factors to account for differences between data and simulation in the efficiencies in the identification criteria for muons and electrons. By applying the scale factors shifted up or down by their uncertainties, new templates are obtained that correspond to these uncertainties. These templates can be used as the nuisance parameters, which are correlated between channels as identical identification criteria are used. The scale factors are parametrized as functions of lepton \pt and $\eta$ to account for different detector response. In the same way, uncertainties in the trigger efficiency are also accounted for, in the muon and electron trigger selections for this analysis.

\item \textit{Jet energy scale and resolution}: Uncertainties in the energy corrections applied to jets are propagated to the final discriminating distributions by reconstructing events with the jet level corrections shifted within their corresponding uncertainties, which depend on the jet \pt and $\eta$.

\item \textit{Jet \btagging}: Simulated events are corrected with scale factors to account for differences in the efficiency for identifying a \PQb\ jet between data and simulation. There are two components to this process, each with an independent, uncorrelated nuisance parameter: one that accounts for the scale factor applied to the rate of identifying \cPqb-tagged jets (efficiency) and one that accounts for the scale factor applied to the rate of mistakenly identifying light-flavor jets as \PQb\ jets (\PQb\ mistag rate). In each case, the uncertainty is obtained by shifting these \pt-dependent scale factors within their uncertainties. The \btagging uncertainties are fully correlated between the dilepton and fully hadronic analyses, as they use the same \btagging criteria.

\item \textit{CSV discriminant shape}: The CSV tagger provides a continuous variable that can be used to identify \PQb\ jets. This continuous variable is used as an input to the \wjetsbdt described above. The \wjetsbdt is only used in the single-lepton analysis, therefore the CSV shape systematic uncertainty only applies to that analysis. Several sources of systematic uncertainties are evaluated, including jet energy scale, flavor effects, and statistical effects. Each of these effects contributes an additional uncertainty in the CSV value that is propagated to the final signal discrimination process.

\item \textit{Jet \ttagging}: It is not possible to define a CR that is capable of measuring the \ttagging scale factor without overlapping the \ttbar SR. The \ttagging efficiency scale factor is determined during the statistical analysis. This is done by including a nuisance parameter with a flat prior distribution that is unconstrained and correlated between the fully hadronic and single-lepton channels. Sources of misidentified \cPqt-tagged jets are different in the single-lepton channel, where they originate from \wjets processes, and in the fully hadronic channel, where they originate from QCD multijet processes. Therefore, the nuisance parameters corresponding to the uncertainty in the \tmistag rate are treated as uncorrelated between the channels, and are also uncorrelated with the nuisance parameter assigned to the \ttagging efficiency.

\item \textit{Parton distribution functions}: For the \ttbar simulated sample, the PDFs from the NNPDF3.0 set~\cite{Ball:2014uwa} are used to evaluate the systematic uncertainty in the choice of PDF, according to the procedure described in Ref.~\cite{Butterworth:2015oua}.

\item \textit{Scale uncertainties}: For the \ttbar sample, the matrix element renormalization and factorization scales were varied up and down independently by a factor of 2 to account for uncertainties in the choice of $Q^2$ used to generate the simulated sample.

\item \textit{Top quark \pt reweighting}: The simulated SM \ttbar process was corrected at parton-level using a function derived from the ratio of top quark \pt measured in data and next-to-NLO predictions from \POWHEG and \PYTHIA~\cite{Khachatryan:2016mnb}. The uncertainty in this process is estimated by taking the difference between the unweighted and weighted results applied symmetrically to the nominal value as a function of \pt. The top quark \pt reweighting does not significantly impact the \mttbar and \ST distributions, and would not obscure a resonance signal.

\item \textit{QCD multijet background estimation}: The `mass-modified' procedure described above to predict the shape of the background in the fully hadronic channel includes an uncertainty in the resulting distribution, equivalent to half of the difference between the uncorrected and `mass-modified' background shapes. This difference affects both the shape and normalization of the final distributions, and the corresponding nuisance parameter is independent from all other effects. The uncertainties in the \tmistag rates are propagated to the final distributions, and the corresponding uncertainty is handled via the \tmistag rate nuisance parameter described above. A closure test is performed with simulated QCD multijet events to test the accuracy of the method. An additional systematic uncertainty is included, equal to the magnitude of the discrepancy observed from the closure tests results, evaluated and applied on a bin-by-bin basis to the fully hadronic signal categories. This systematic most greatly affects the two \cPqb-tag, high-$\abs{\Delta y}$ category, for which the method only closes within 20\%. For the other categories, the method closes within ${\approx}4$\%.

\end{enumerate}

\section{Statistical analysis}
\label{sec:stats}

\begin{figure}
\centering
\includegraphics[width=0.455\textwidth]{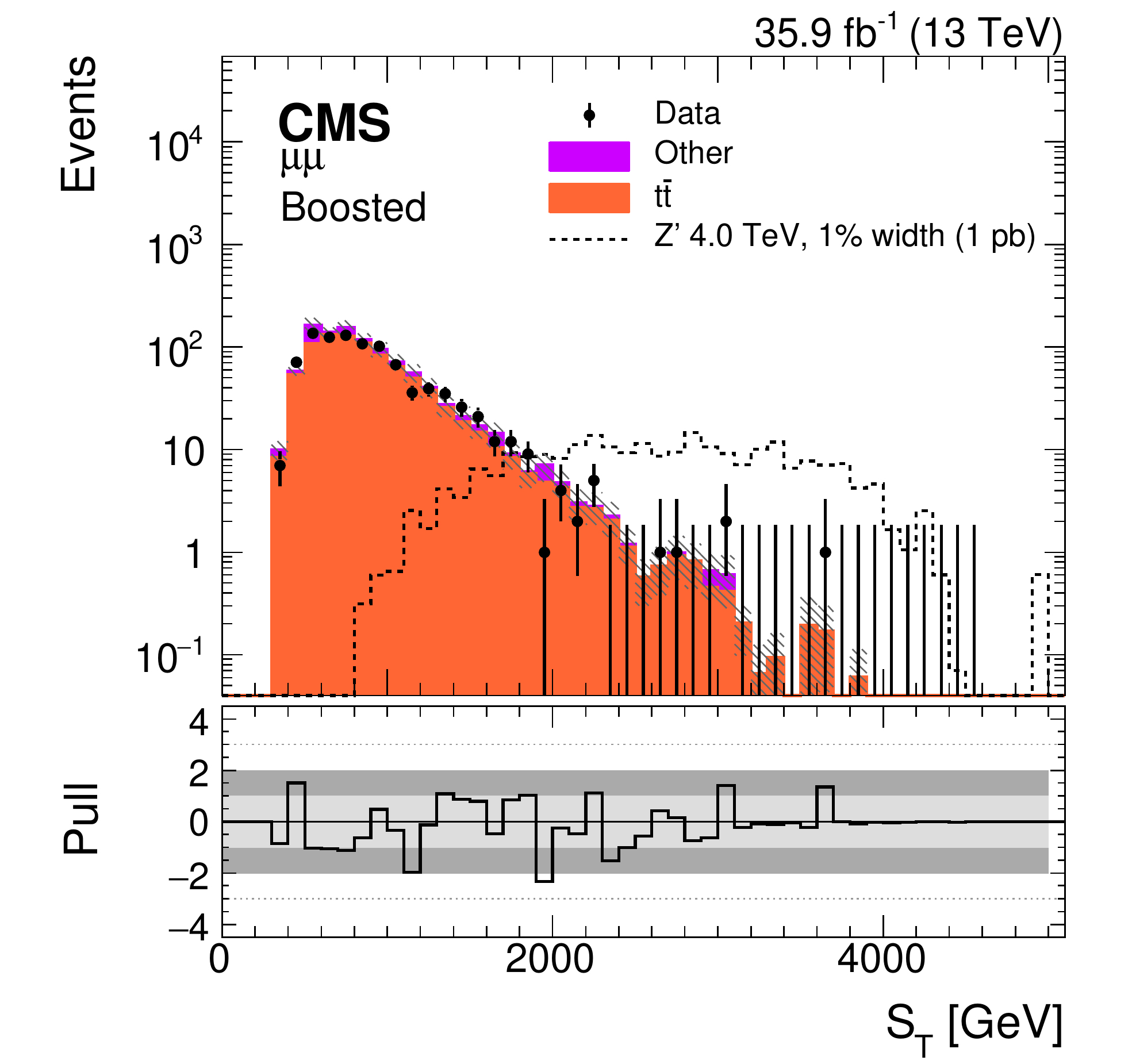}
\includegraphics[width=0.455\textwidth]{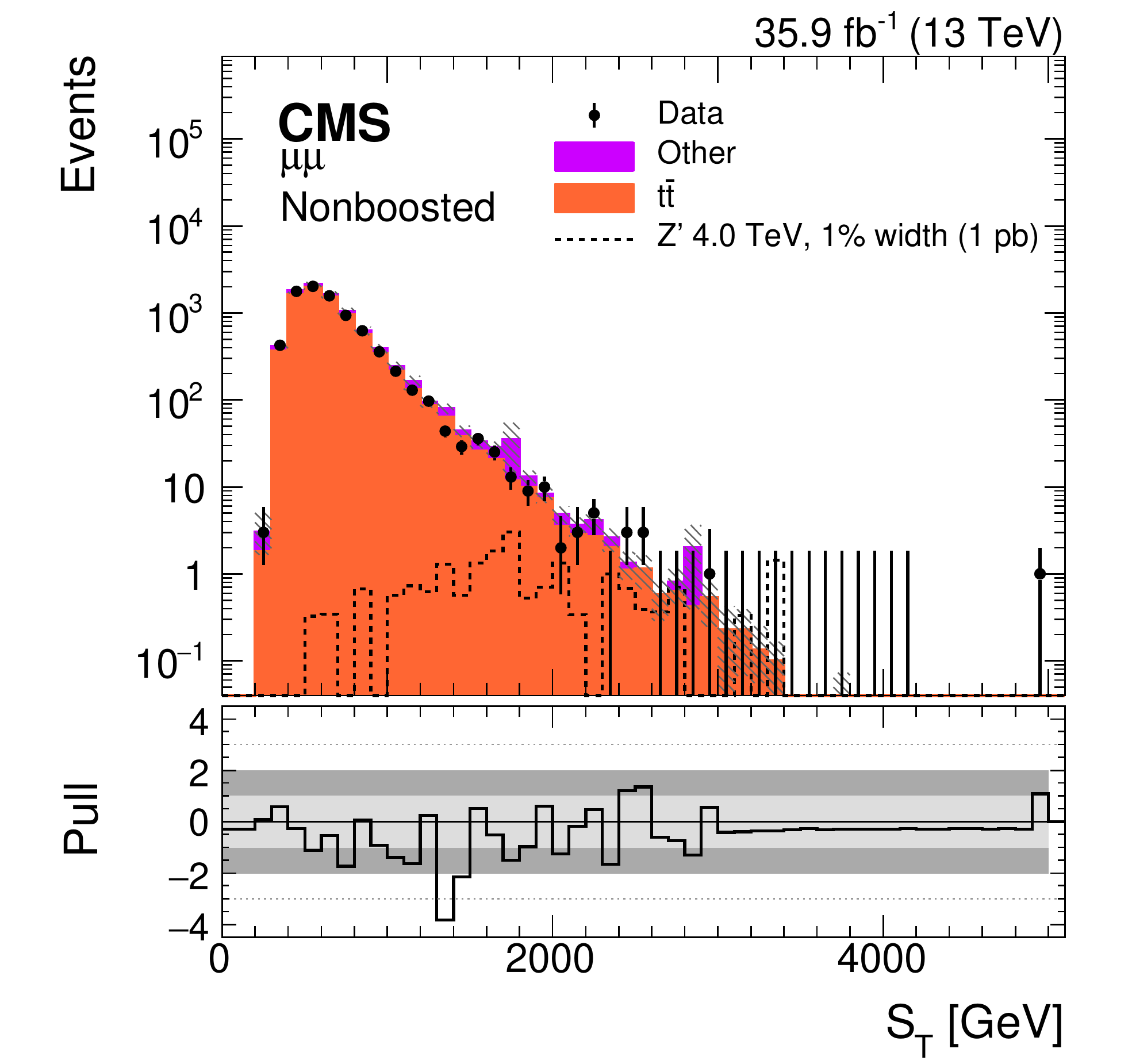}
\includegraphics[width=0.455\textwidth]{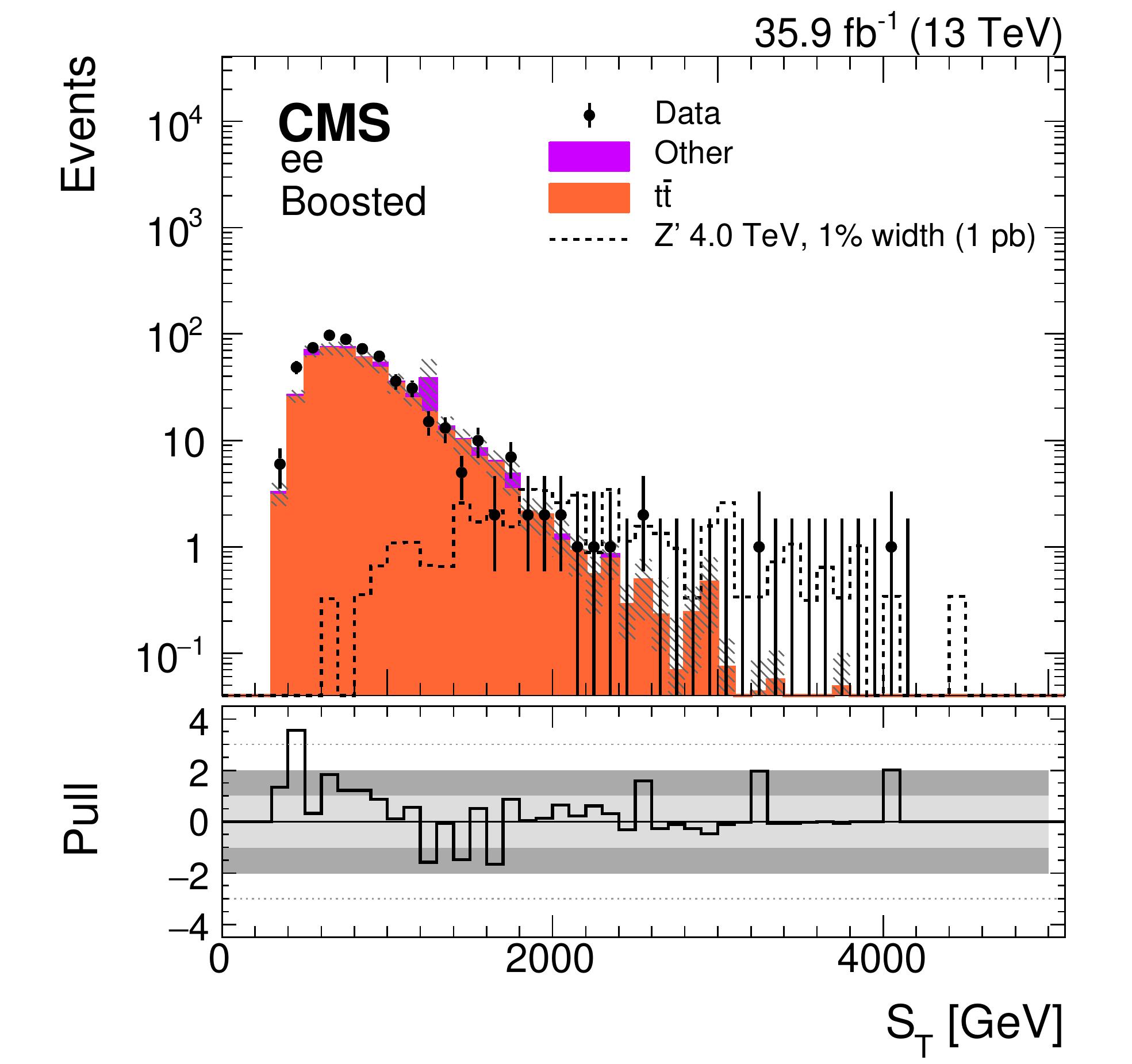}
\includegraphics[width=0.455\textwidth]{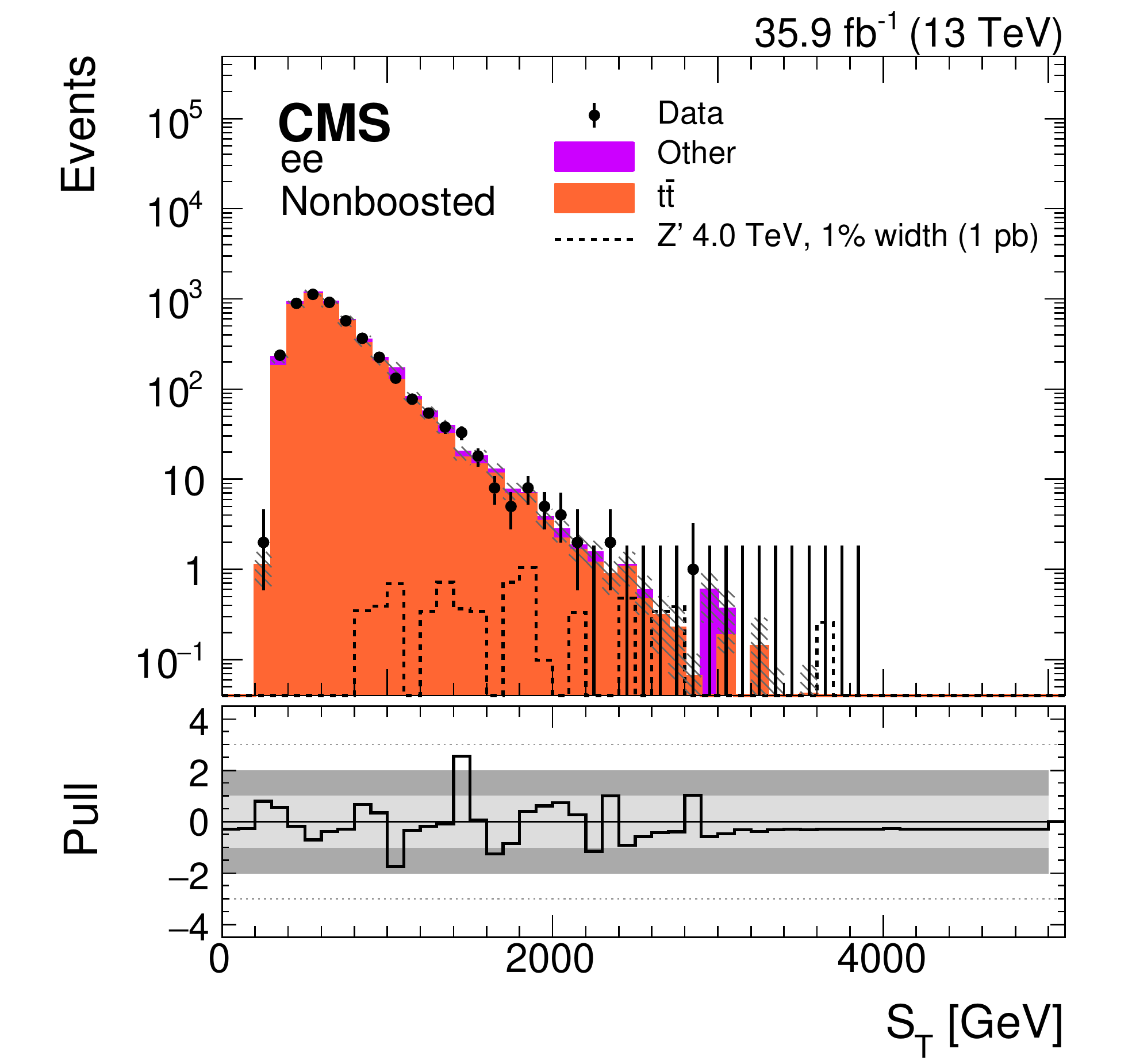}
\includegraphics[width=0.455\textwidth]{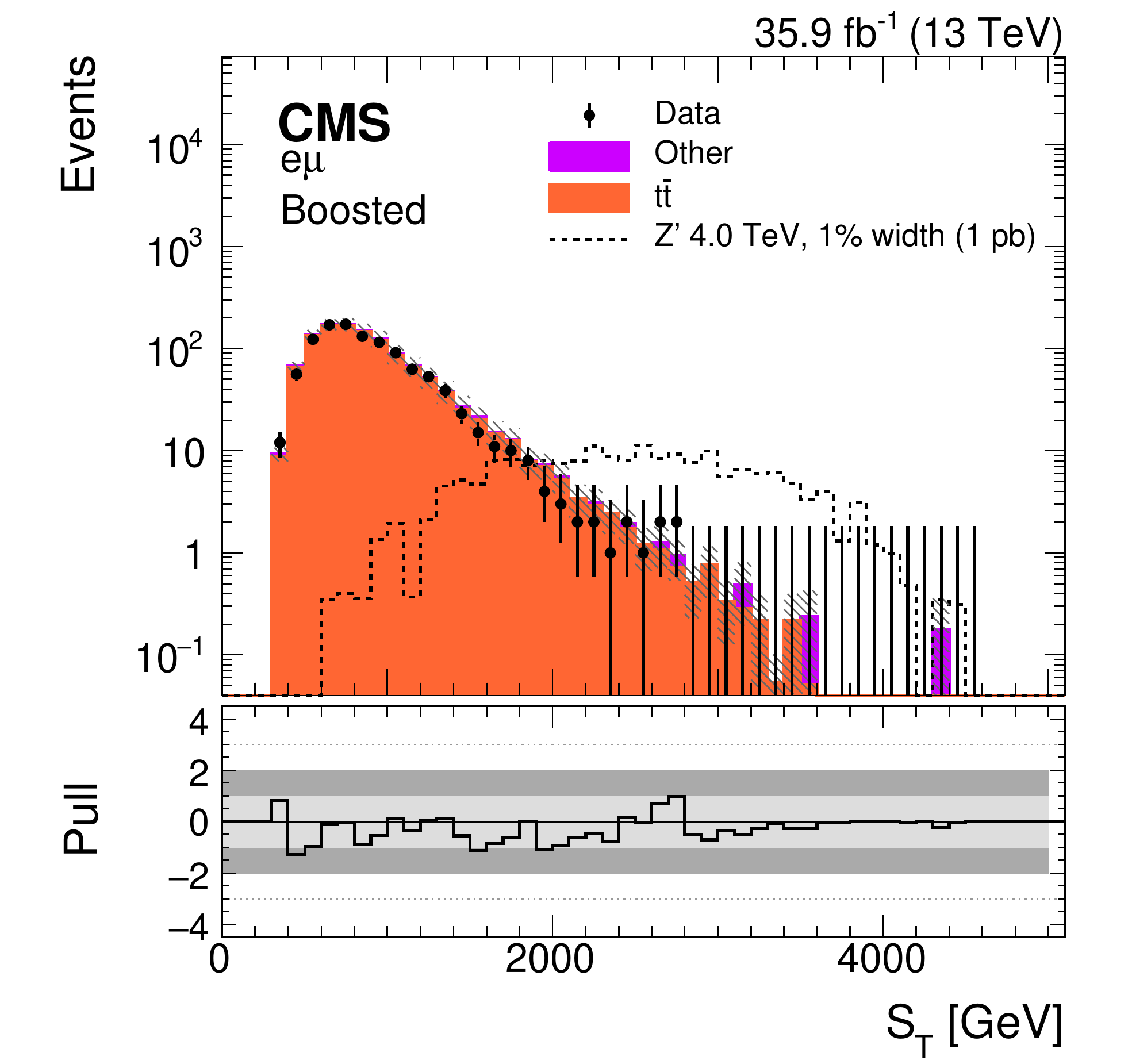}
\includegraphics[width=0.455\textwidth]{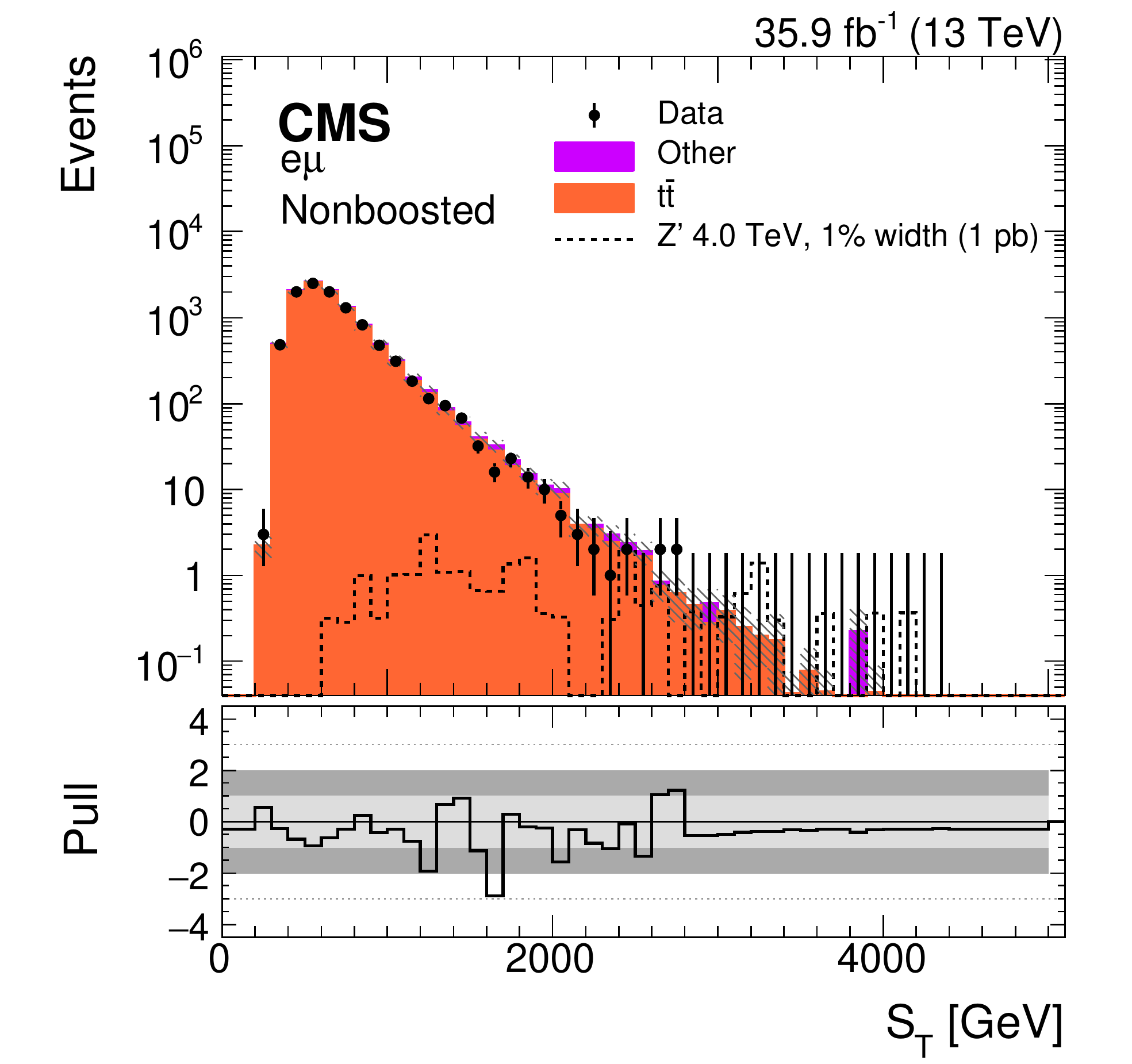}
\caption{Distributions of \ST for the \Pgm\Pgm\ (upper), \Pe\Pe\ (middle), and \Pe\Pgm\ (lower) SRs in the boosted (left) and nonboosted (right) regions, as defined in Section~\ref{sec:preselection_dilepton}. The contribution expected from a 4\TeV \PZpr\ boson, with a relative width of 1\%, is shown normalized to a cross section of 1\pb. The hatched band on the simulation represents the uncertainty in the background prediction. The lower panel in each plot shows the pull of each histogram bin from the SM prediction. The light (dark) gray band represents a pull of one (two) s.d. from the predicted value.}
\label{fig:mtt_dilep}
\end{figure}

\begin{figure}
\centering
\includegraphics[width=0.495\textwidth]{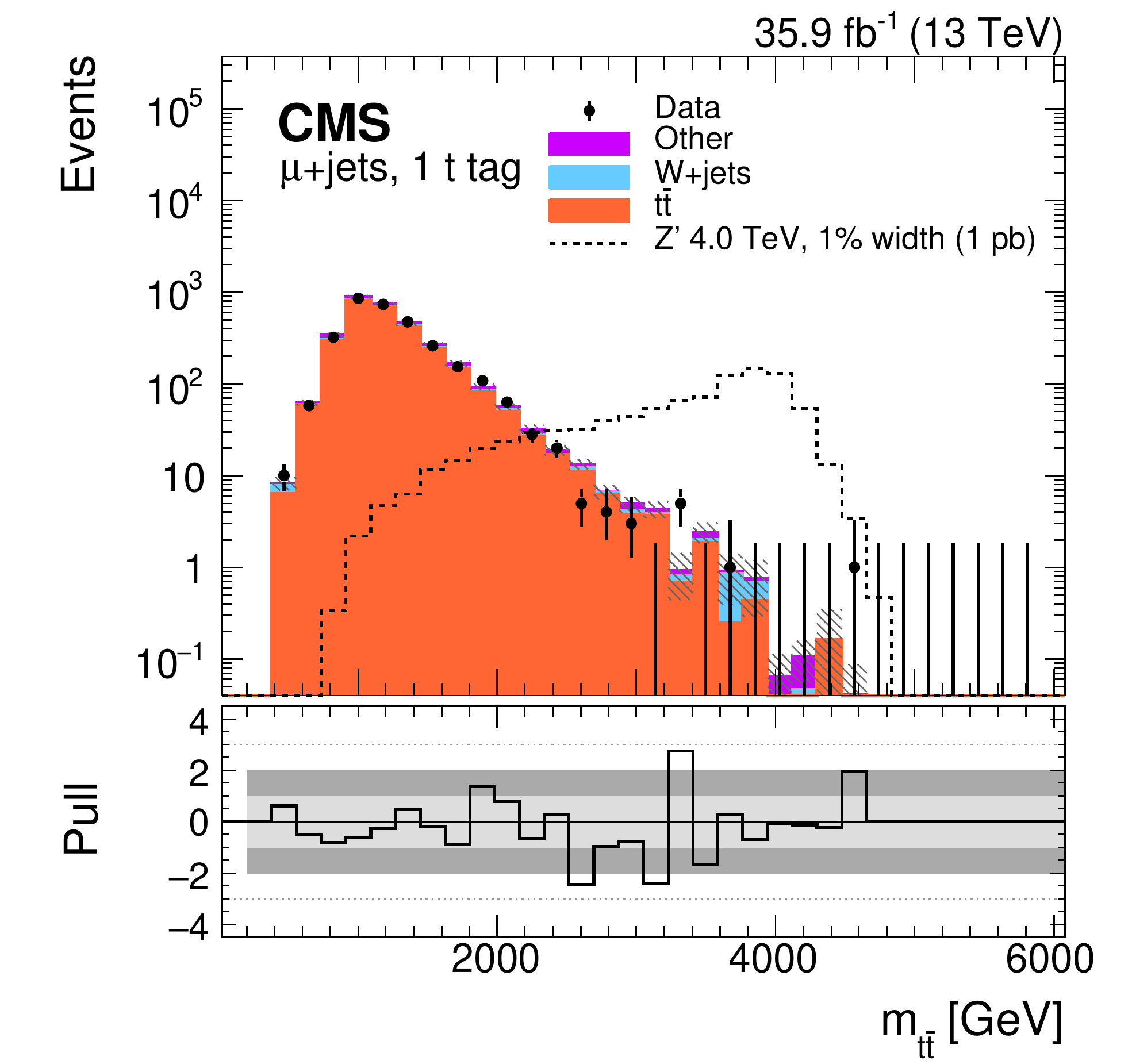}
\includegraphics[width=0.495\textwidth]{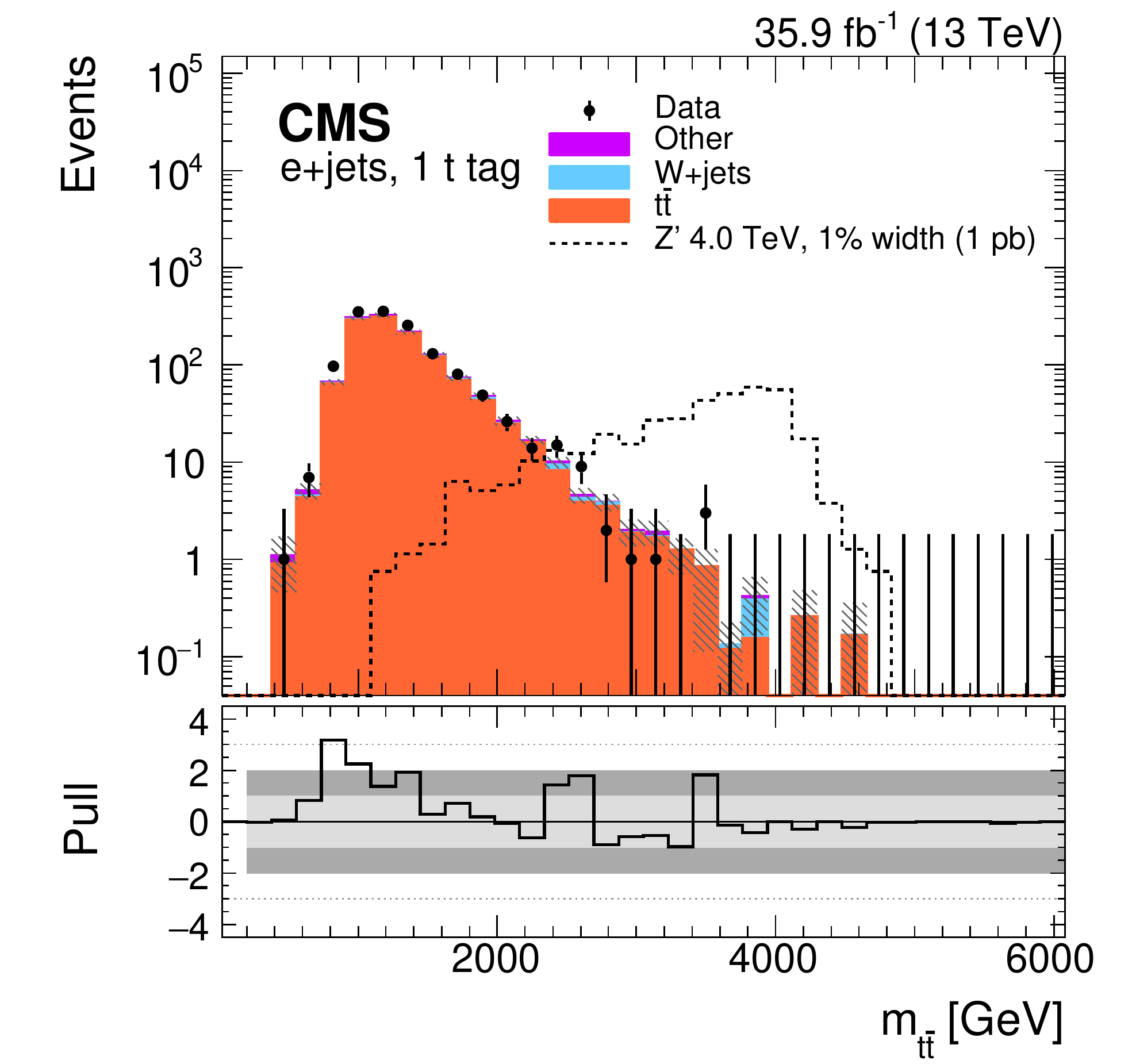}
\includegraphics[width=0.495\textwidth]{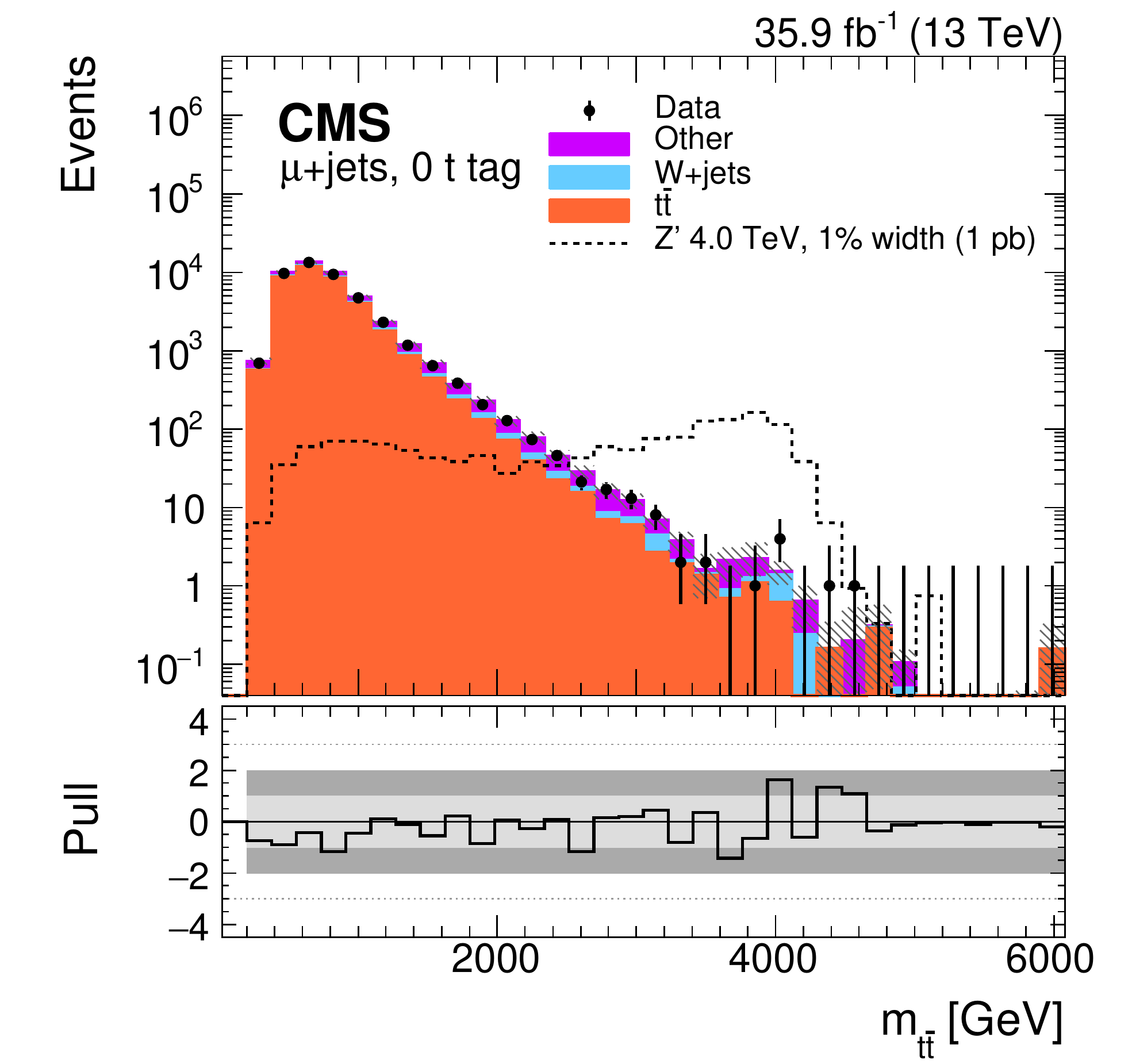}
\includegraphics[width=0.495\textwidth]{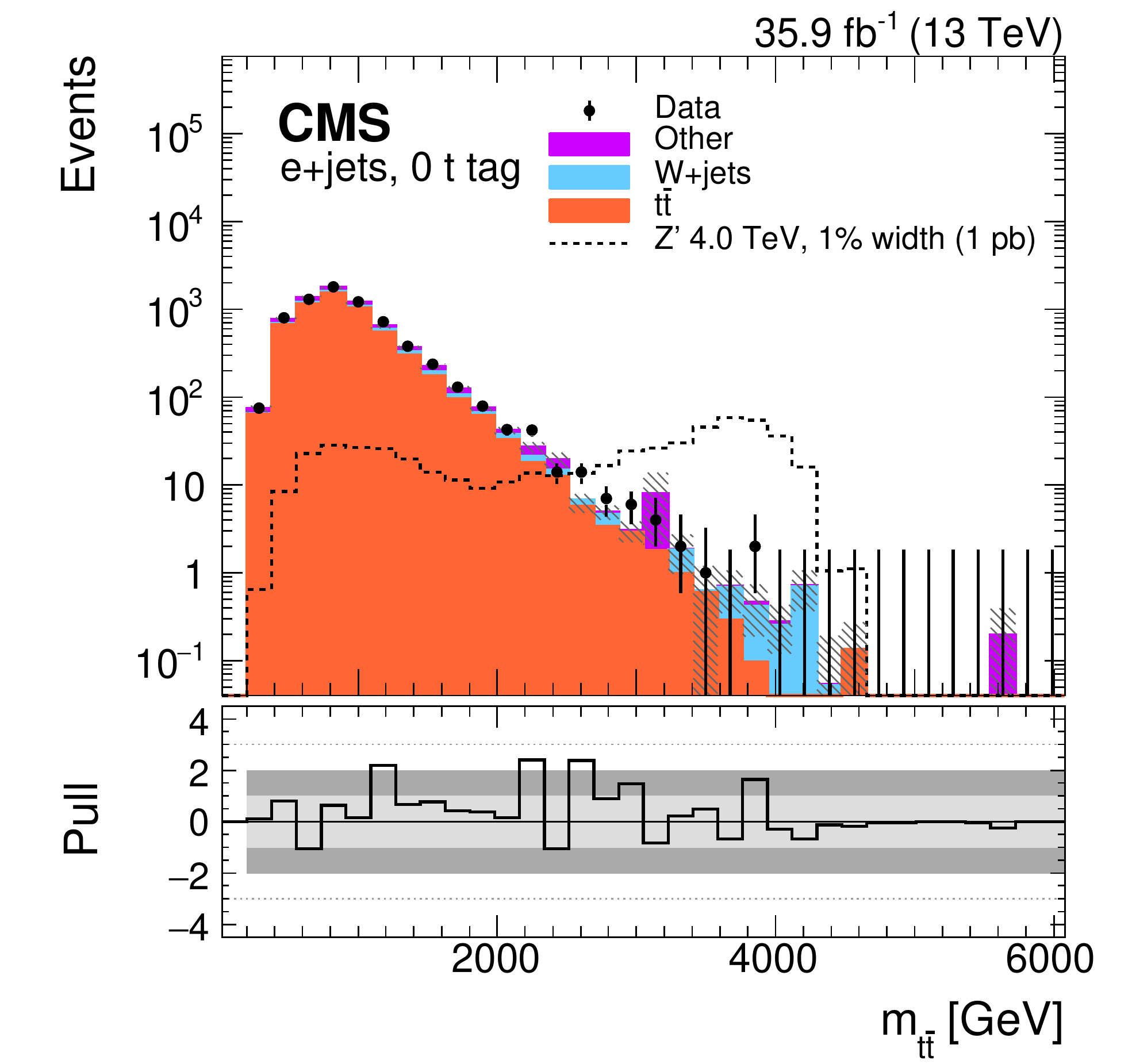}
\caption{Distributions of \mttbar for the single-lepton channel SRs for the muon (left) and electron (right) categories with (upper) and without (lower) \ttagging. The contribution expected from a 4\TeV \PZpr\ boson, with a relative width of 1\%, is shown normalized to a cross section of 1\pb. The hatched band on the simulation represents the uncertainty in the background prediction. The lower panel in each plot shows the pull of each histogram bin from the SM prediction. The light (dark) gray band represents a pull of one (two) s.d. from the predicted value.}
\label{fig:mtt_ljets_sr}
\end{figure}

\begin{figure}
\centering
\includegraphics[width=0.495\textwidth]{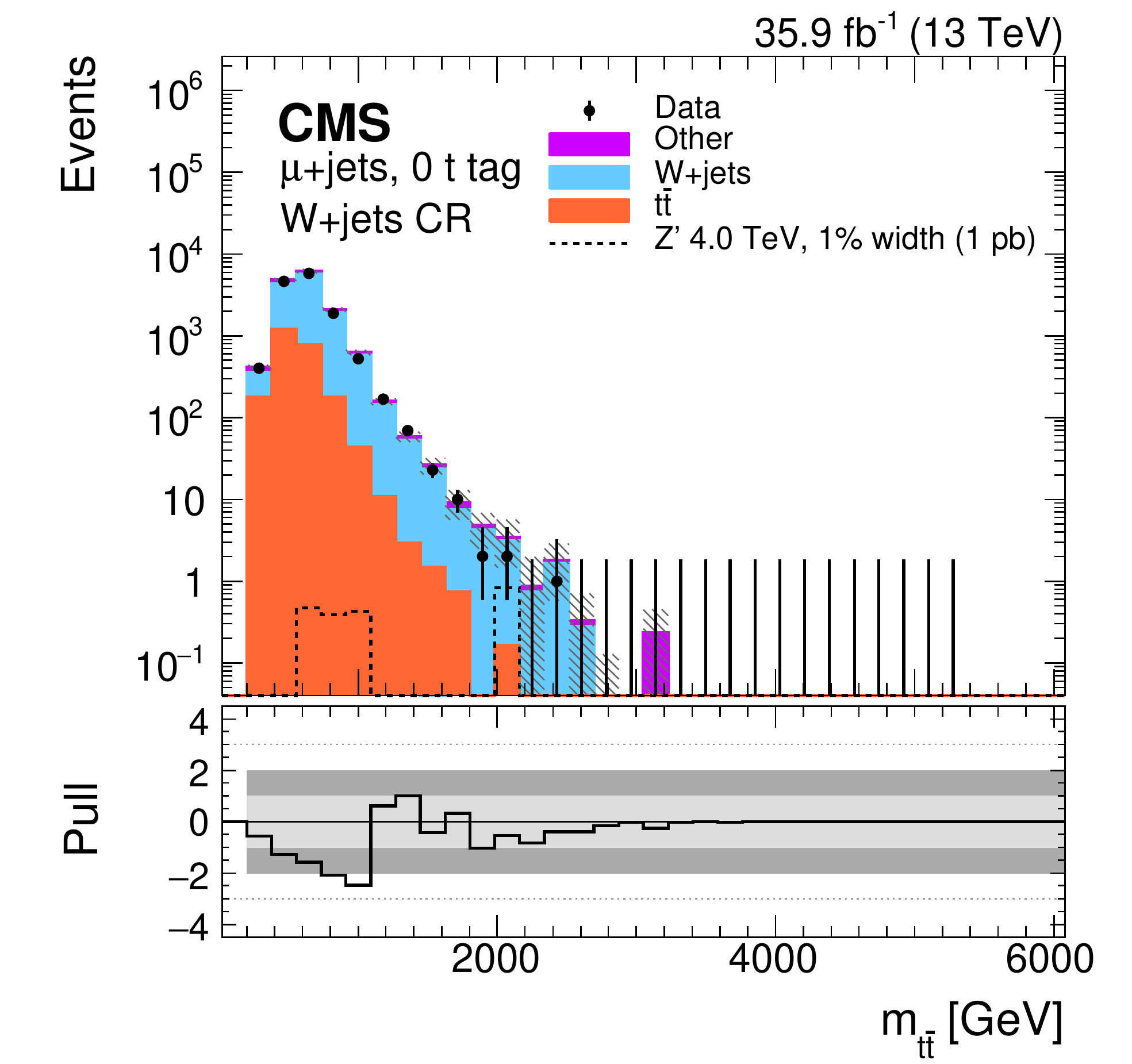}
\includegraphics[width=0.495\textwidth]{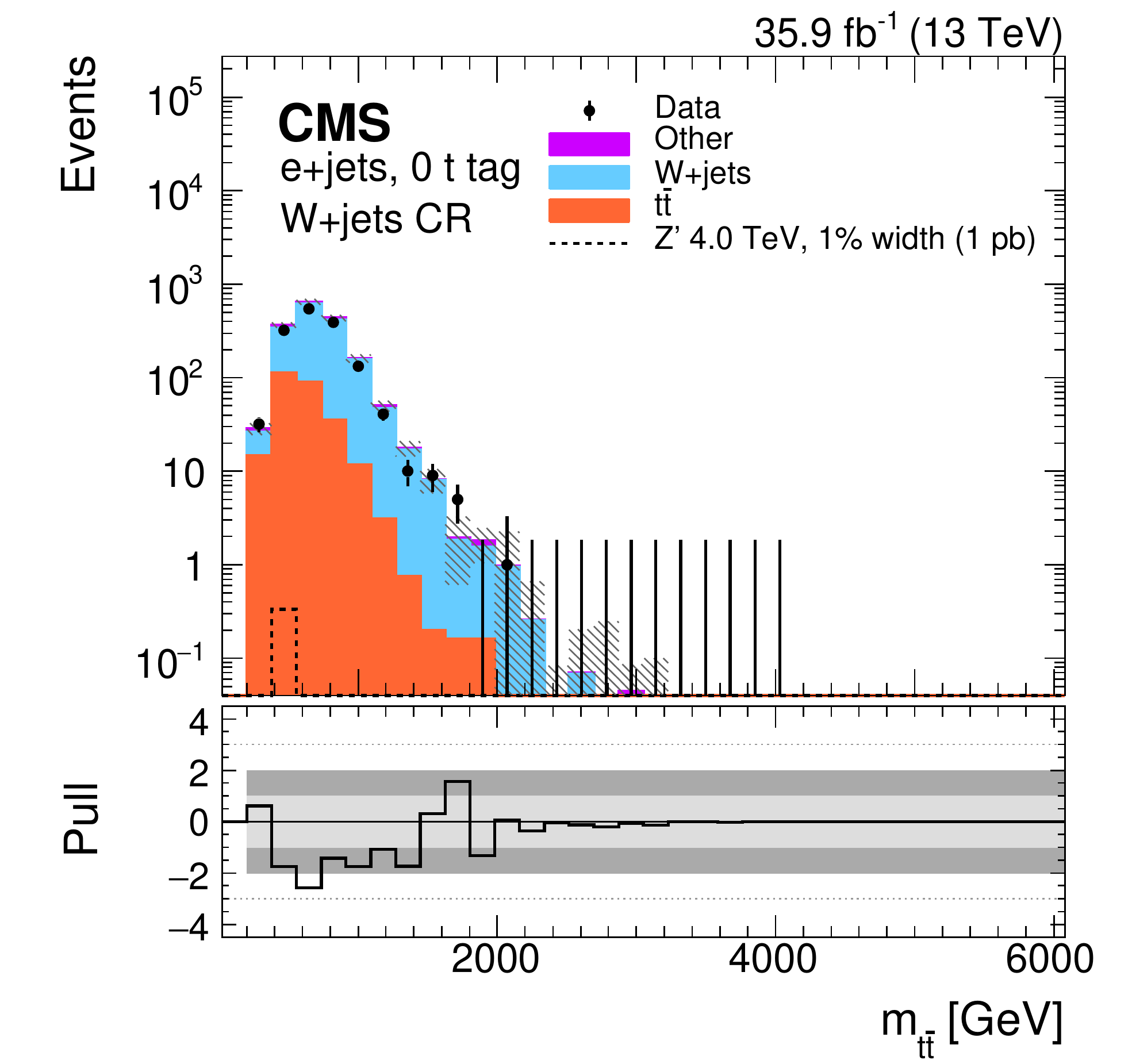}
\includegraphics[width=0.495\textwidth]{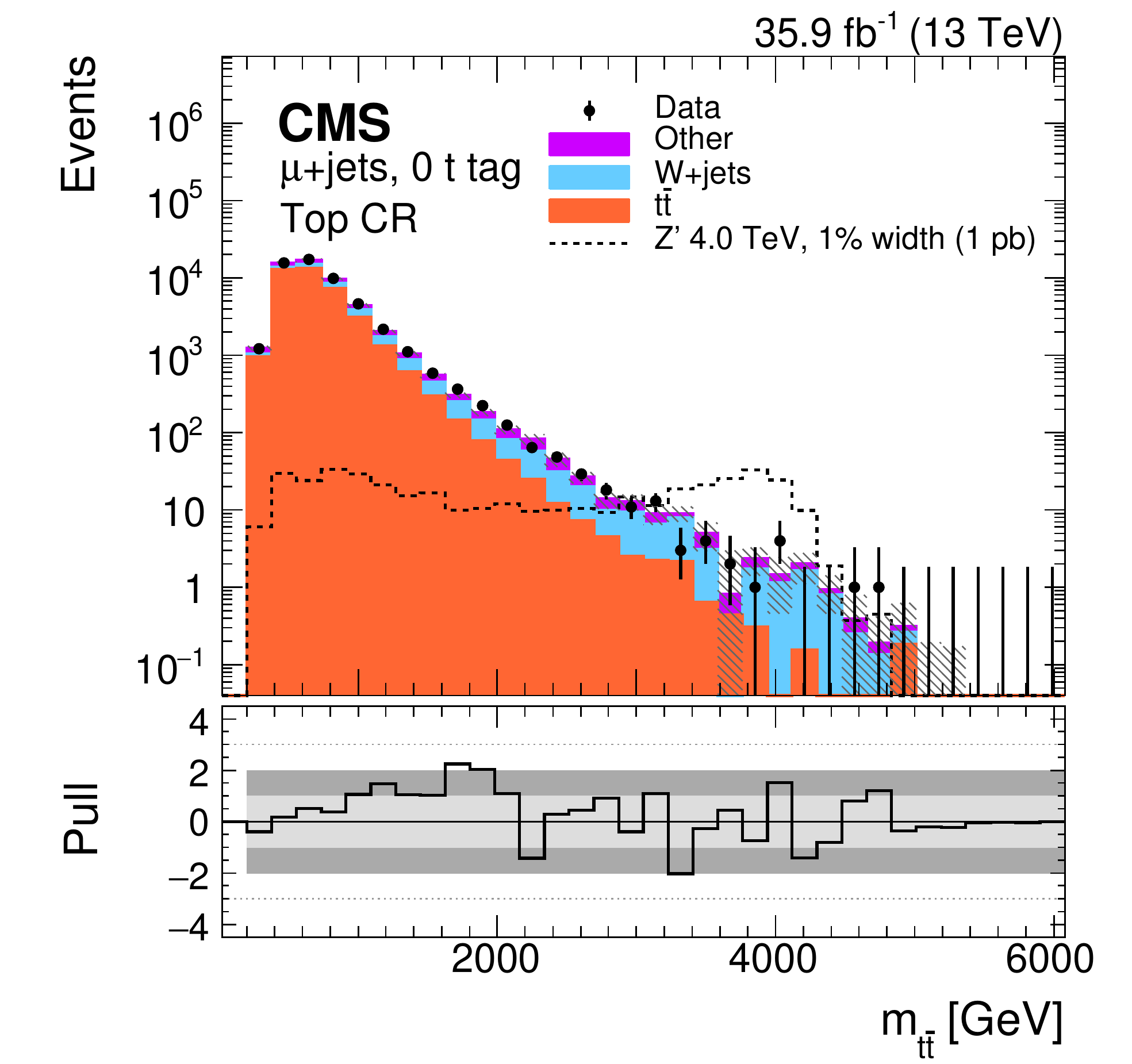}
\includegraphics[width=0.495\textwidth]{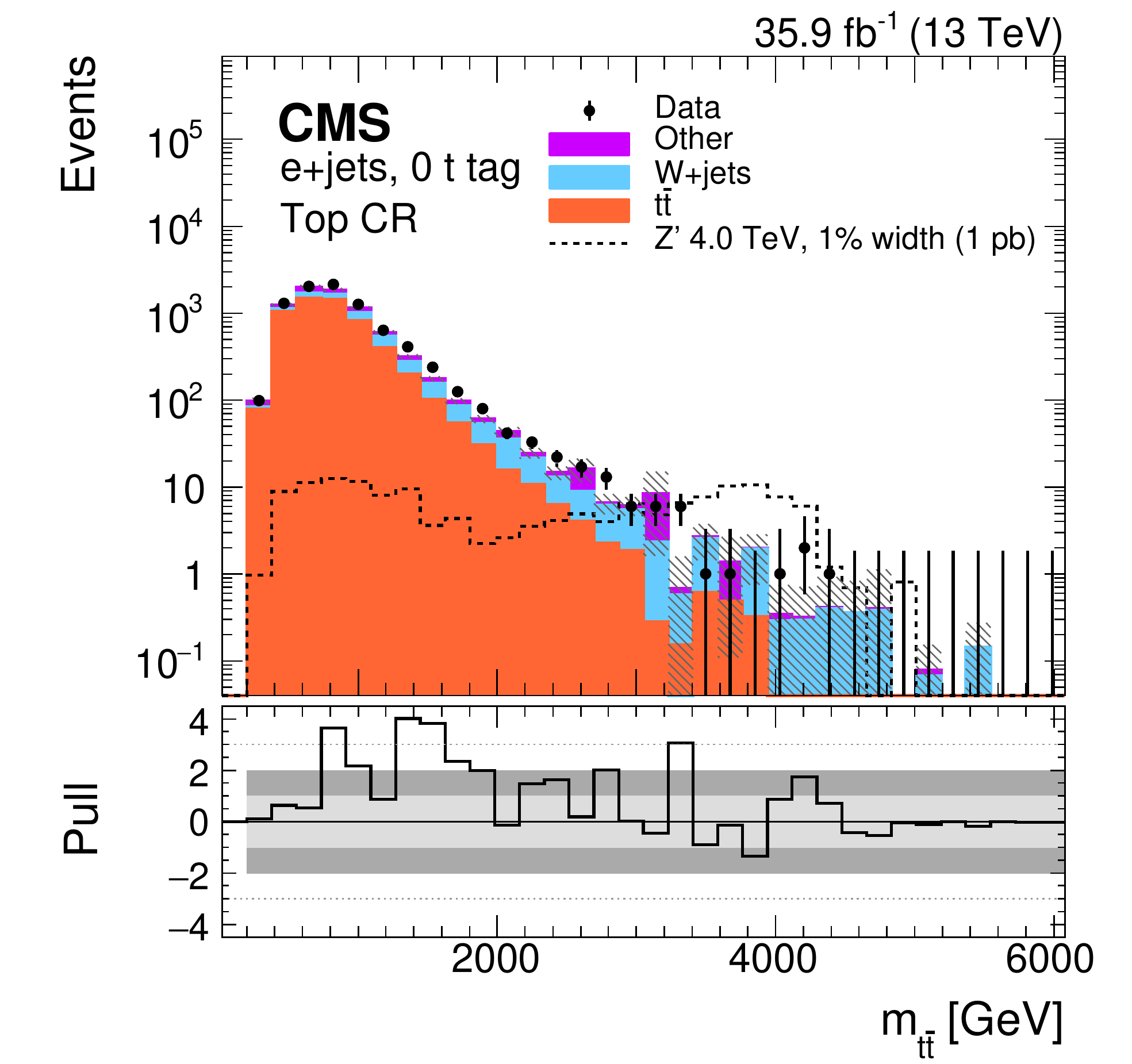}
\caption{Distributions of \mttbar for the single-lepton channel CR1 (upper) and CR2 (lower) for the muon (left) and electron (right) categories. The contribution expected from a 4\TeV \PZpr\ boson, with a relative width of 1\%, is shown normalized to a cross section of 1\pb. The hatched band on the simulation represents the uncertainty in the background prediction. The lower panel in each plot shows the pull of each histogram bin from the SM prediction. The light (dark) gray band represents a pull of one (two) s.d. from the predicted value.}
\label{fig:mtt_ljets_cr}
\end{figure}

\begin{figure}
\centering
\includegraphics[width=0.455\textwidth]{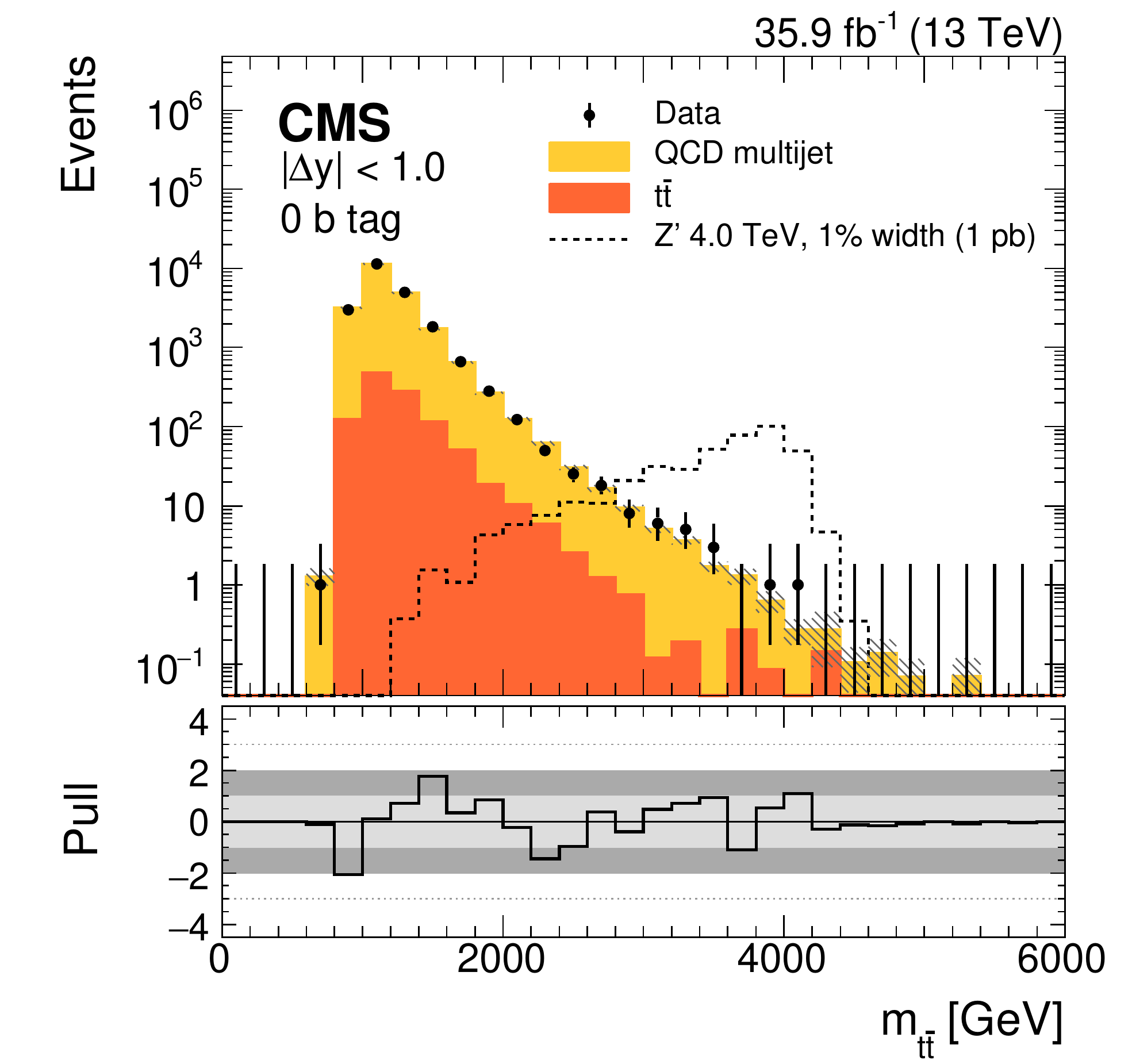}
\includegraphics[width=0.455\textwidth]{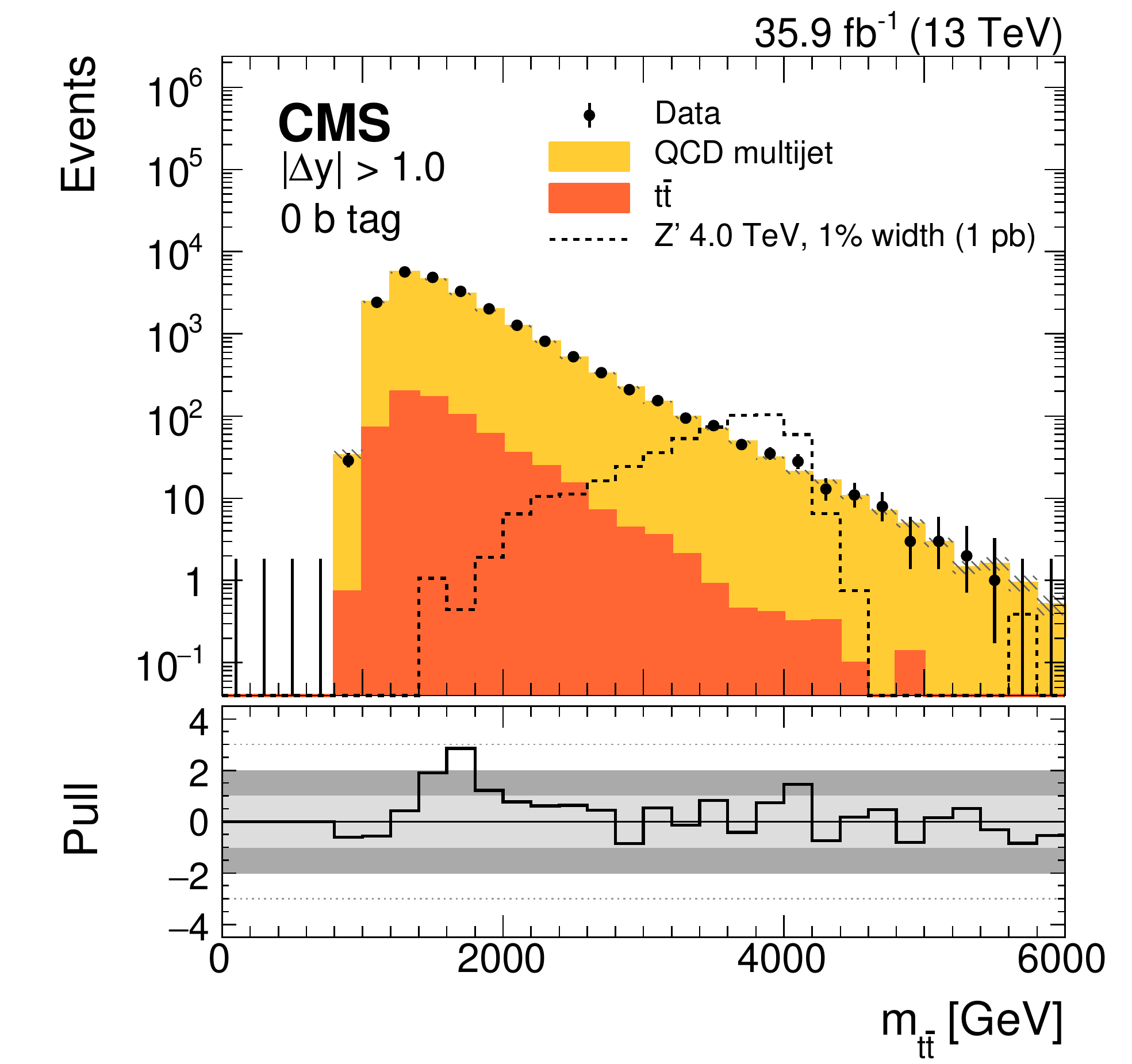}
\includegraphics[width=0.455\textwidth]{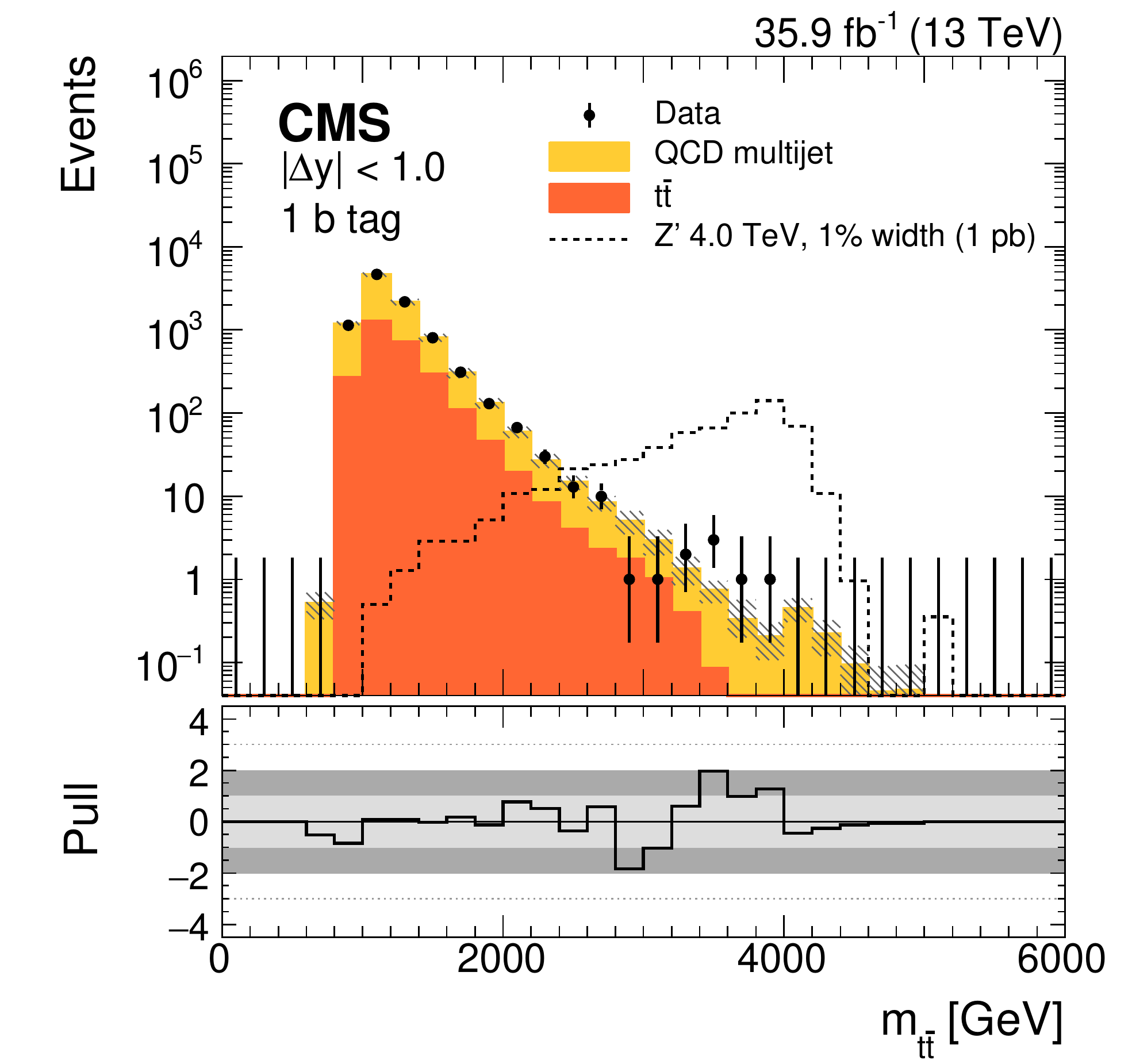}
\includegraphics[width=0.455\textwidth]{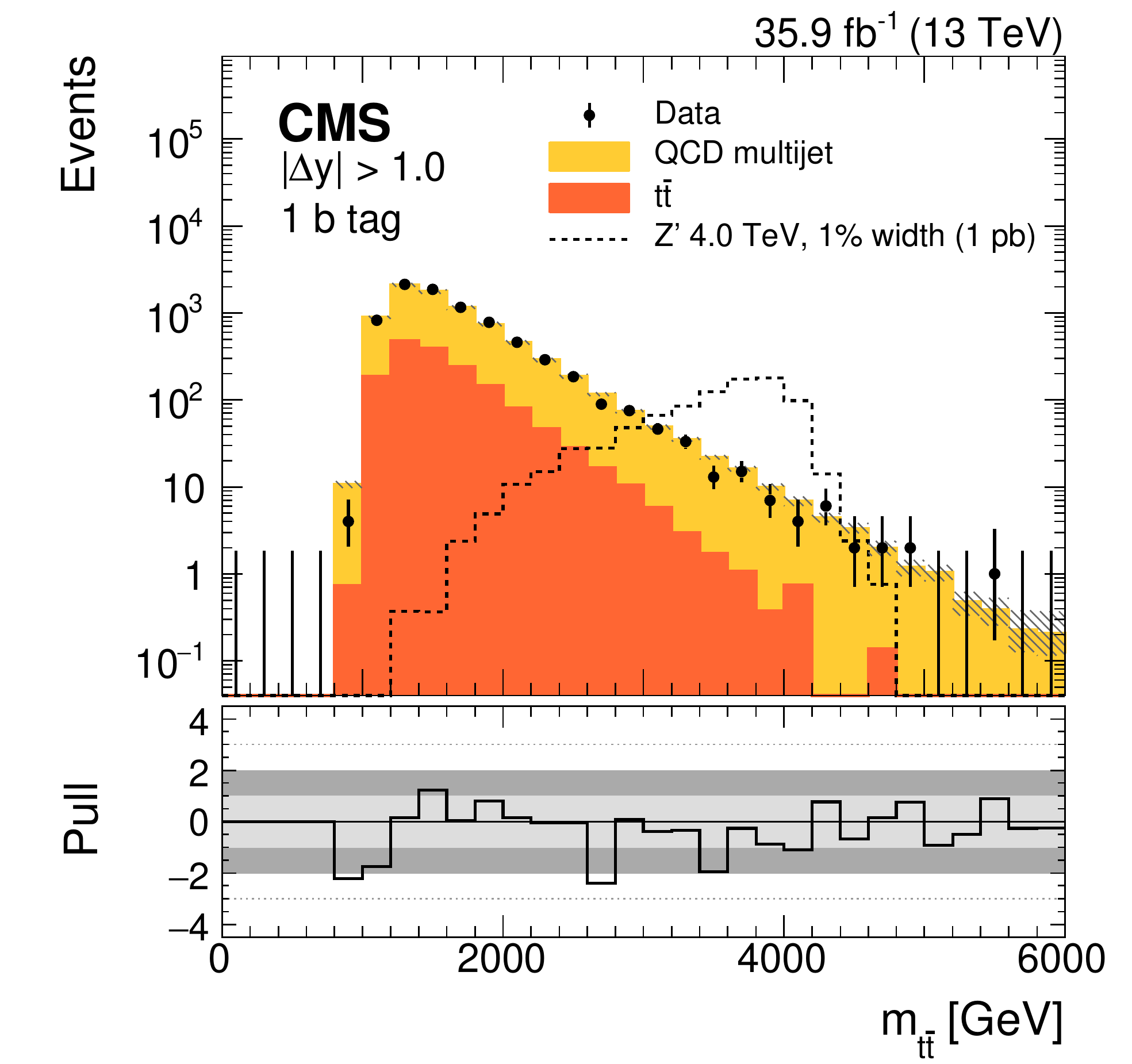}
\includegraphics[width=0.455\textwidth]{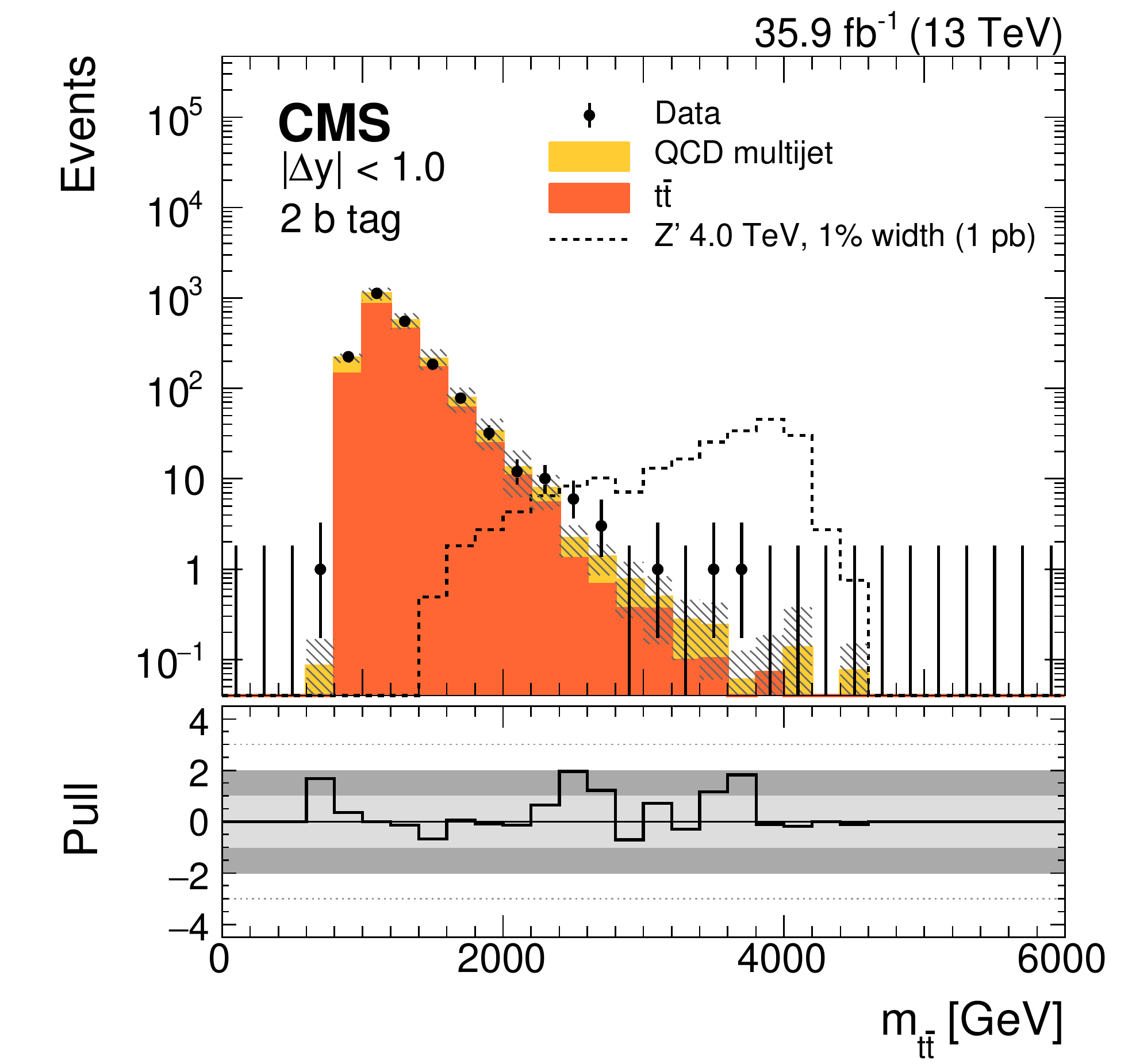}
\includegraphics[width=0.455\textwidth]{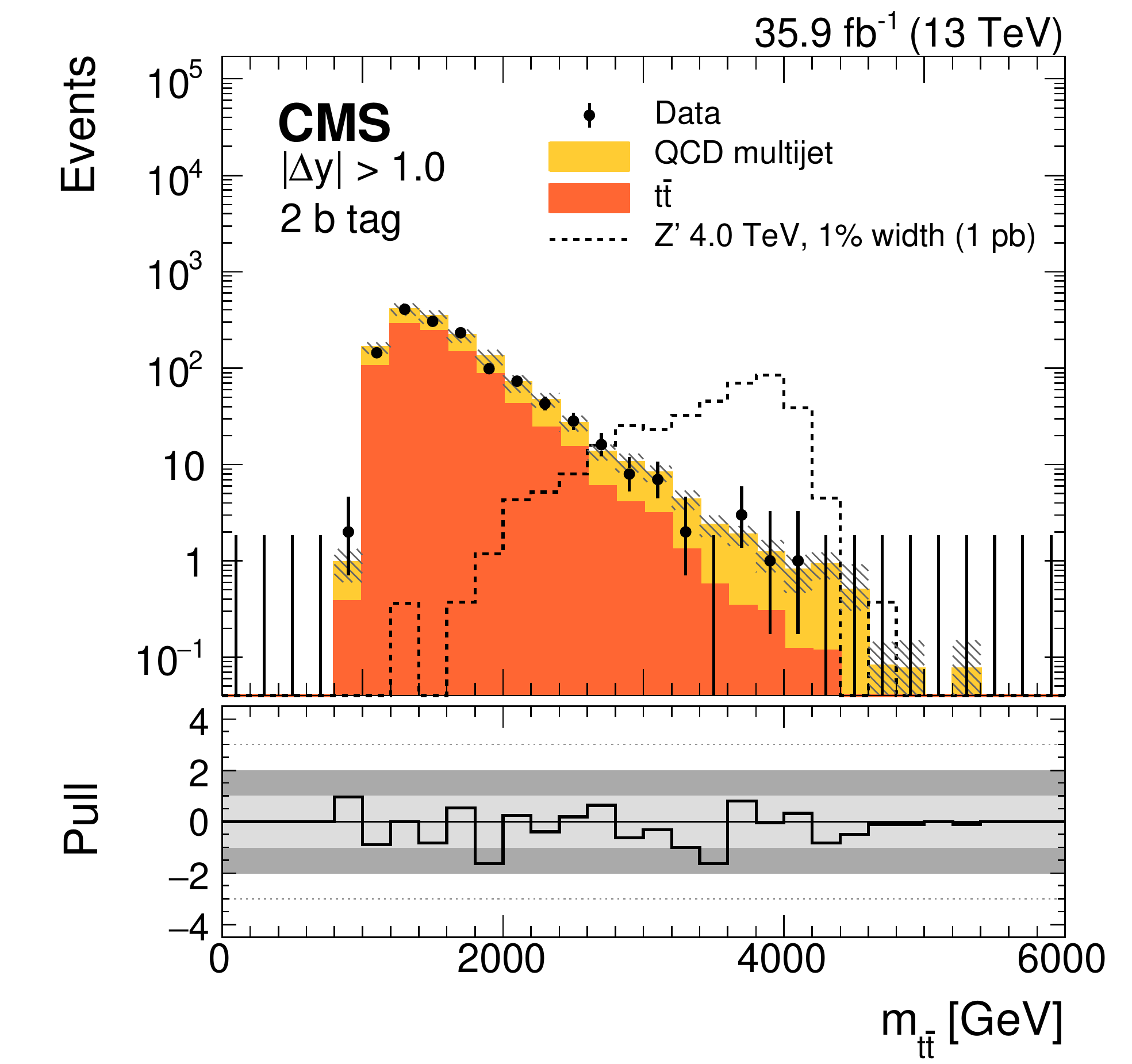}
\caption{Distributions of \mttbar for the fully hadronic channel SR categories, used to extract the final results. The contribution expected from a 4\TeV \PZpr\ boson, with a relative width of 1\%, is shown normalized to a cross section of 1\pb. The hatched band on the simulation represents the uncertainty in the background prediction. The lower panel in each plot shows the pull of each histogram bin from the SM prediction. The light (dark) gray band represents a pull of one (two) s.d. from the predicted value.}
\label{fig:mtt_had}
\end{figure}

Before extracting the final results of the analysis, a background-only binned maximum
likelihood fit is performed on the signal and control regions to determine the preferred values of the
background process normalizations and shapes, using constraints from
the sources of systematic uncertainty described above. Each source of
systematic uncertainty is included through a unique nuisance parameter
that is allowed to vary within the rate and shape constraints
described above, using a log-normal prior distribution. The post-fit
values of the nuisance parameters are used to correct the
normalization and shape of each background process. The \mttbar
and \ST distributions after the fitting procedure are shown in
Figs.~\ref{fig:mtt_dilep}, \ref{fig:mtt_ljets_sr}--\ref{fig:mtt_ljets_cr},
and \ref{fig:mtt_had}, for the dilepton, single-lepton, and fully
hadronic channels, respectively. The mild deficits at low \mttbar in
the two plots on the left in Fig.~\ref{fig:mtt_ljets_cr} do not
significantly impact the limit, because this region is used to evaluate
the \ttbar and \wjets cross sections and is not sensitive to the
resonance signal. The \ttagging efficiency is measured simultaneously
in signal and control regions during the maximum likelihood fit, as it is
not possible to select a CR that might not be contaminated by the
potential signal. The \ttagging efficiency scale factor is modeled as
a free nuisance parameter, with an unconstrained prior, in the binned
likelihood fit. The \ttagging efficiency scale factor measured by the fit is
$1.001 \pm 0.012$.

Data are found to be in agreement with expectations in each of the categories considered in
this analysis. Limits on the product of the production cross section
and branching fraction are calculated, $\sigma(\Pp\Pp\to
X)\,\mathcal{B}(X\to\ttbar)$, for heavy resonances decaying to a
pair of top quarks. A shape-based
analysis is performed using both the signal and control regions from the three
exclusive analysis channels. The \textsc{Theta} software
package~\cite{theta} simultaneously fits the \mttbar distributions from the
single-lepton and fully hadronic channels and the \ST
distributions from the dilepton channel. For the
limit calculation, a Bayesian likelihood-based method is
used~\cite{CowanPDGStat,CMS-NOTE-2011-005} with each bin of the
distributions combined statistically, along with the implementation of
unique nuisance parameters that correspond to the systematic
uncertainties described in Section~\ref{sec:systematics}. The signal
normalization is allowed to vary with a distinct unconstrained nuisance
parameter having a uniform prior, while the other nuisance parameters
have log-normal prior distributions. Finally, to account for the
limited number of simulated events, an additional statistical
uncertainty is included for each process relying on simulated events
through the ``Barlow--Beeston lite''
method~\cite{barlow_beeston}. Prior to the statistical analysis,
the \mttbar distributions are rebinned. For the fully hadronic and
dilepton channels, the total statistical uncertainty in the background
is required to be below 30\% in any given bin. In the single-lepton
channel, the total statistical uncertainty in the background
expectation for the sum of small backgrounds (single top quark,
multijet, \zjets, $\PW+\PQb$, or \PQc\ jets) is required to be below 10\%
in each bin. The tighter statistical uncertainty requirement is
needed for these backgrounds because the events are rejected
at a high rate, resulting in significantly fewer simulated events
that pass the final selection.

\begin{figure}
\centering
\includegraphics[width=0.75\textwidth]{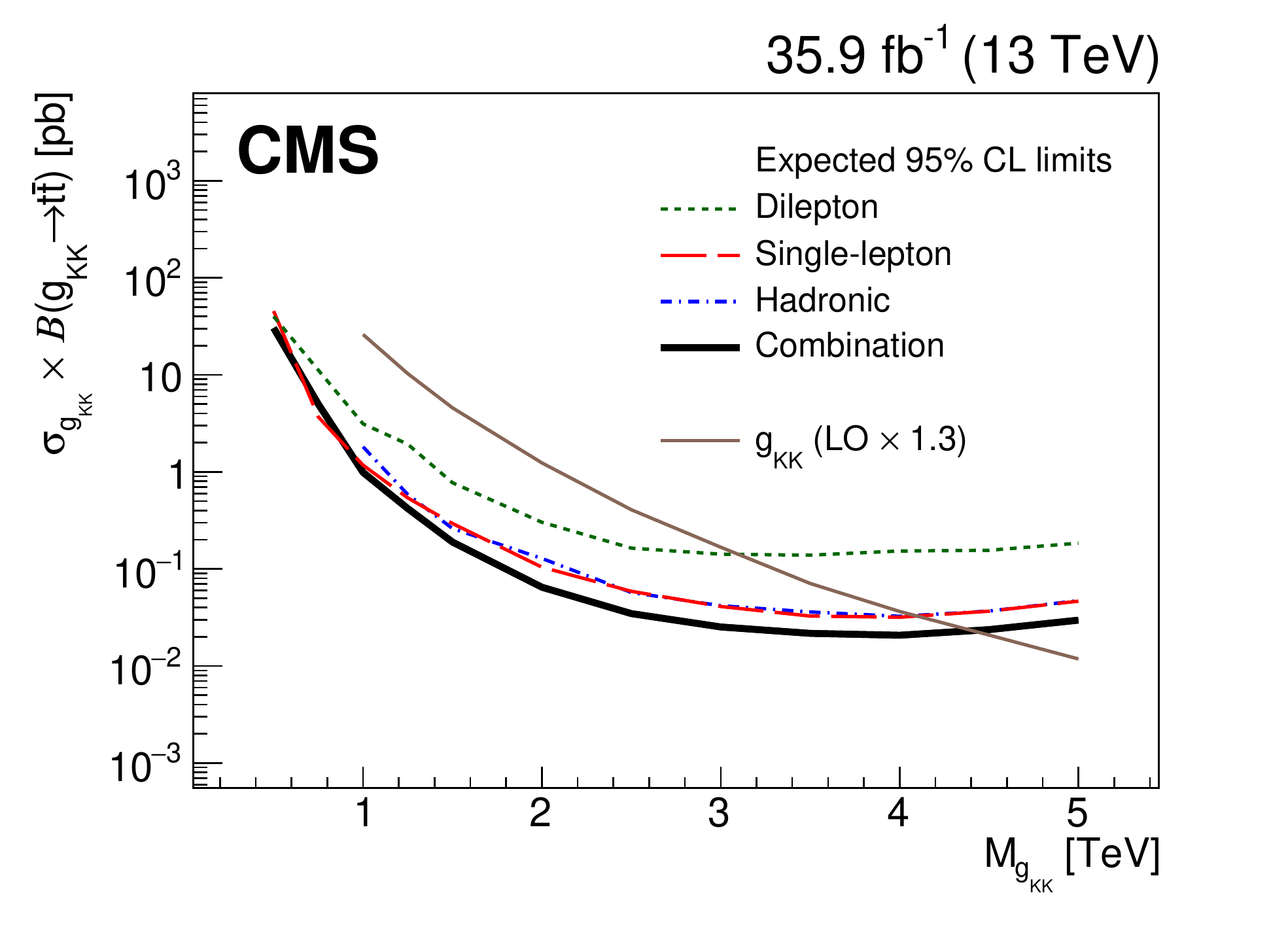}
\caption{Comparison of the sensitivities for each analysis channel contributing to the combination. The expected limits at 95\% \CL are shown for each channel with the narrow colored lines, while the combination result is shown with thick the black line. These results are shown specifically for the \gKK signal hypothesis, as this model has characteristics that are common to many \ttbar resonance searches. The multiplicative factor of 1.3 for the \gKK is the NLO $K$ factor.}
\label{fig:exp_comparison}
\end{figure}

Figure~\ref{fig:exp_comparison} shows a comparison of the expected
sensitivities in each of the three analysis channels in terms of the
expected limits for the \gKK signal model. The contributions from the
single-lepton and fully hadronic channels dominate the sensitivity
over most of the mass range, apart from the region of lowest masses,
where the dilepton channel makes a significant contribution.

\section{Results}
\label{sec:results}

The statistical analysis is performed for each of the signal models
considered in this analysis: three variations of a \PZpr\ boson having
a width-to-mass ratio of 1, 10, and 30\%, as well as a \gKK. In
each case, a 95\% \CL limit is obtained on the product of the
resonance production cross section and branching fraction. The observed and expected
limits and 1 and 2 s.d. bands are calculated for resonance masses
ranging from 0.5 to 5.0\TeV and are listed in
Tables~\ref{tab:limits_narrow}--\ref{tab:limits_kkg}.

\begin{table}[!htbp]
\centering
\topcaption{Limits at 95\% \CL on the product of the resonance production cross section and branching fraction for the narrow ($\Gamma/m = 1\%$) \PZpr\ boson resonance hypothesis.}
\begin{scotch}{ccccc}
Mass [TeV] & Obs. [pb] & Median exp. [pb] & 68\% Exp. [pb] & 95\% Exp. [pb]\\
\hline
0.50 & 29 & 28 & [13, 49] & [7.5, 78] \\
0.75 & 1.1 & 2.4 & [1.5, 3.7] & [1.0, 5.6] \\
1.00 & 0.37 & 0.54 & [0.37, 0.77] & [0.26, 1.1] \\
1.25 & 0.31 & 0.16 & [0.11, 0.24] & [0.080, 0.35] \\
1.50 & 0.091 & 0.076 & [0.051, 0.12] & [0.036, 0.17] \\
2.00 & 0.023 & 0.027 & [0.018, 0.041] & [0.012, 0.061] \\
2.50 & 0.018 & 0.012 & [0.0083, 0.019] & [0.0056, 0.029] \\
3.00 & 0.0042 & 0.0075 & [0.0051, 0.011] & [0.0035, 0.017] \\
3.50 & 0.0046 & 0.0052 & [0.0035, 0.0081] & [0.0025, 0.012] \\
4.00 & 0.0041 & 0.0042 & [0.0028, 0.0065] & [0.0020, 0.010] \\
4.50 & 0.0030 & 0.0035 & [0.0023, 0.0054] & [0.0016, 0.0082] \\
5.00 & 0.0023 & 0.0032 & [0.0021, 0.0049] & [0.0014, 0.0079] \\
6.00 & 0.0013 & 0.0027 & [0.0017, 0.0042] & [0.0011, 0.0069] \\
6.50 & 0.0012 & 0.0026 & [0.0016, 0.0040] & [0.0011, 0.0065] \\
7.00 & 0.0012 & 0.0024 & [0.0016, 0.0038] & [0.0011, 0.0063] \\
\end{scotch}
\label{tab:limits_narrow}
\end{table}

\begin{table}[!htbp]
\centering
\topcaption{Limits at 95\% \CL on the product of the resonance production cross section and branching fraction for the wide ($\Gamma/m = 10\%$) \PZpr\ boson resonance hypothesis.}
\begin{scotch}{ccccc}
Mass [TeV] & Obs. [pb] & Median exp. [pb] & 68\% Exp. [pb] & 95\% Exp. [pb]\\
\hline
0.50 & 31 & 22 & [9.8, 43] & [5.4, 70] \\
0.75 & 2.9 & 3.6 & [2.2, 6.1] & [1.3, 9.5] \\
1.00 & 0.93 & 0.72 & [0.48, 1.1] & [0.34, 1.5] \\
1.25 & 0.55 & 0.24 & [0.16, 0.37] & [0.11, 0.54] \\
1.50 & 0.17 & 0.12 & [0.073, 0.18] & [0.050, 0.29] \\
2.00 & 0.041 & 0.040 & [0.027, 0.063] & [0.018, 0.096] \\
2.50 & 0.027 & 0.020 & [0.013, 0.030] & [0.0088, 0.046] \\
3.00 & 0.0084 & 0.013 & [0.0088, 0.020] & [0.0061, 0.031] \\
3.50 & 0.0091 & 0.011 & [0.0073, 0.017] & [0.0051, 0.025] \\
4.00 & 0.0092 & 0.010 & [0.0064, 0.015] & [0.0044, 0.023] \\
4.50 & 0.0087 & 0.010 & [0.0067, 0.016] & [0.0046, 0.024] \\
5.00 & 0.0097 & 0.012 & [0.0078, 0.019] & [0.0056, 0.029] \\
6.00 & 0.015 & 0.021 & [0.014, 0.034] & [0.0095, 0.053] \\
6.50 & 0.016 & 0.025 & [0.017, 0.040] & [0.011, 0.062] \\
7.00 & 0.022 & 0.032 & [0.021, 0.050] & [0.014, 0.081] \\
\end{scotch}
\label{tab:limits_wide}
\end{table}

\begin{table}[!htbp]
\centering
\topcaption{Limits at 95\% \CL on the product of the resonance production cross section and branching fraction for the extra-wide ($\Gamma/m = 30\%$) \PZpr\ boson resonance hypothesis.}
\begin{scotch}{ccccc}
Mass [TeV] & Obs. [pb] & Median exp. [pb] & 68\% Exp. [pb] & 95\% Exp. [pb]\\
\hline
1.0 & 2.0 & 1.1 & [0.63, 1.8] & [0.41, 2.7] \\
2.0 & 0.078 & 0.066 & [0.041, 0.11] & [0.027, 0.18] \\
3.0 & 0.019 & 0.026 & [0.017, 0.040] & [0.012, 0.061] \\
4.0 & 0.019 & 0.023 & [0.015, 0.035] & [0.011, 0.053] \\
5.0 & 0.022 & 0.025 & [0.016, 0.039] & [0.011, 0.062] \\
6.0 & 0.029 & 0.035 & [0.023, 0.055] & [0.015, 0.086] \\
6.5 & 0.030 & 0.040 & [0.026, 0.061] & [0.018, 0.097] \\
7.0 & 0.035 & 0.044 & [0.029, 0.070] & [0.019, 0.11] \\
\end{scotch}
\label{tab:limits_extrawide}
\end{table}

\begin{table}[!htbp]
\centering
\topcaption{Limits at 95\% \CL on the product of the resonance production cross section and branching fraction for the \gKK gluon resonance hypothesis.}
\begin{scotch}{ccccc}
Mass [TeV] & Obs. [pb] & Median exp. [pb] & 68\% Exp. [pb] & 95\% Exp. [pb]\\
\hline
0.50 & 9.5 & 30 & [13, 55] & [6.1, 82] \\
0.75 & 4.6 & 5.0 & [2.6, 8.3] & [1.5, 13] \\
1.00 & 0.71 & 0.99 & [0.64, 1.5] & [0.44, 2.3] \\
1.25 & 0.77 & 0.42 & [0.26, 0.67] & [0.18, 1.0] \\
1.50 & 0.30 & 0.19 & [0.12, 0.32] & [0.081, 0.56] \\
2.00 & 0.090 & 0.065 & [0.042, 0.10] & [0.028, 0.17] \\
2.50 & 0.045 & 0.035 & [0.022, 0.055] & [0.015, 0.089] \\
3.00 & 0.021 & 0.025 & [0.017, 0.039] & [0.012, 0.061] \\
3.50 & 0.016 & 0.022 & [0.014, 0.033] & [0.0098, 0.051] \\
4.00 & 0.020 & 0.021 & [0.014, 0.032] & [0.0096, 0.050] \\
4.50 & 0.019 & 0.024 & [0.016, 0.038] & [0.011, 0.059] \\
5.00 & 0.025 & 0.030 & [0.020, 0.047] & [0.014, 0.074] \\
\end{scotch}
\label{tab:limits_kkg}
\end{table}

\begin{figure}
\centering
\includegraphics[width=0.495\textwidth]{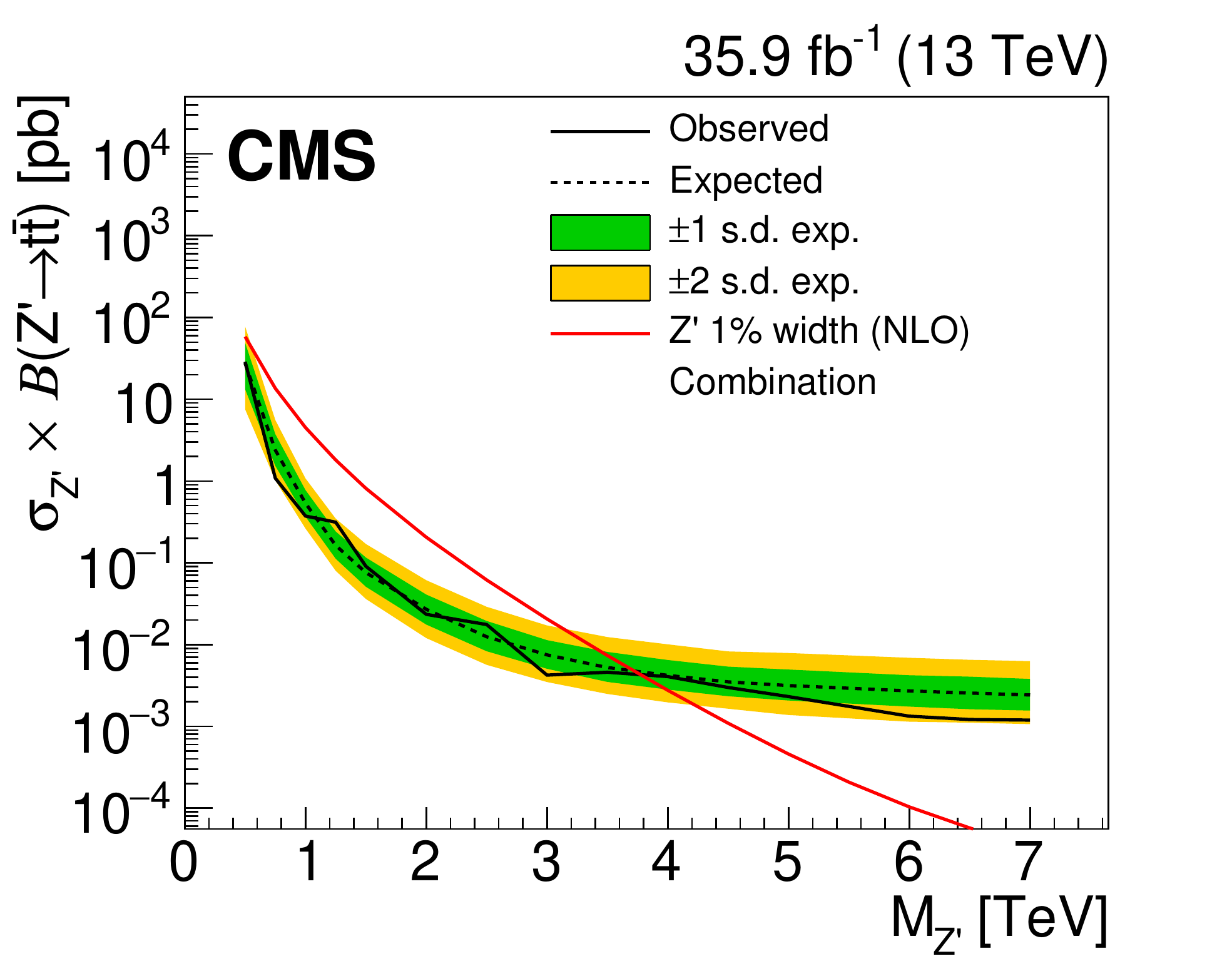}
\includegraphics[width=0.495\textwidth]{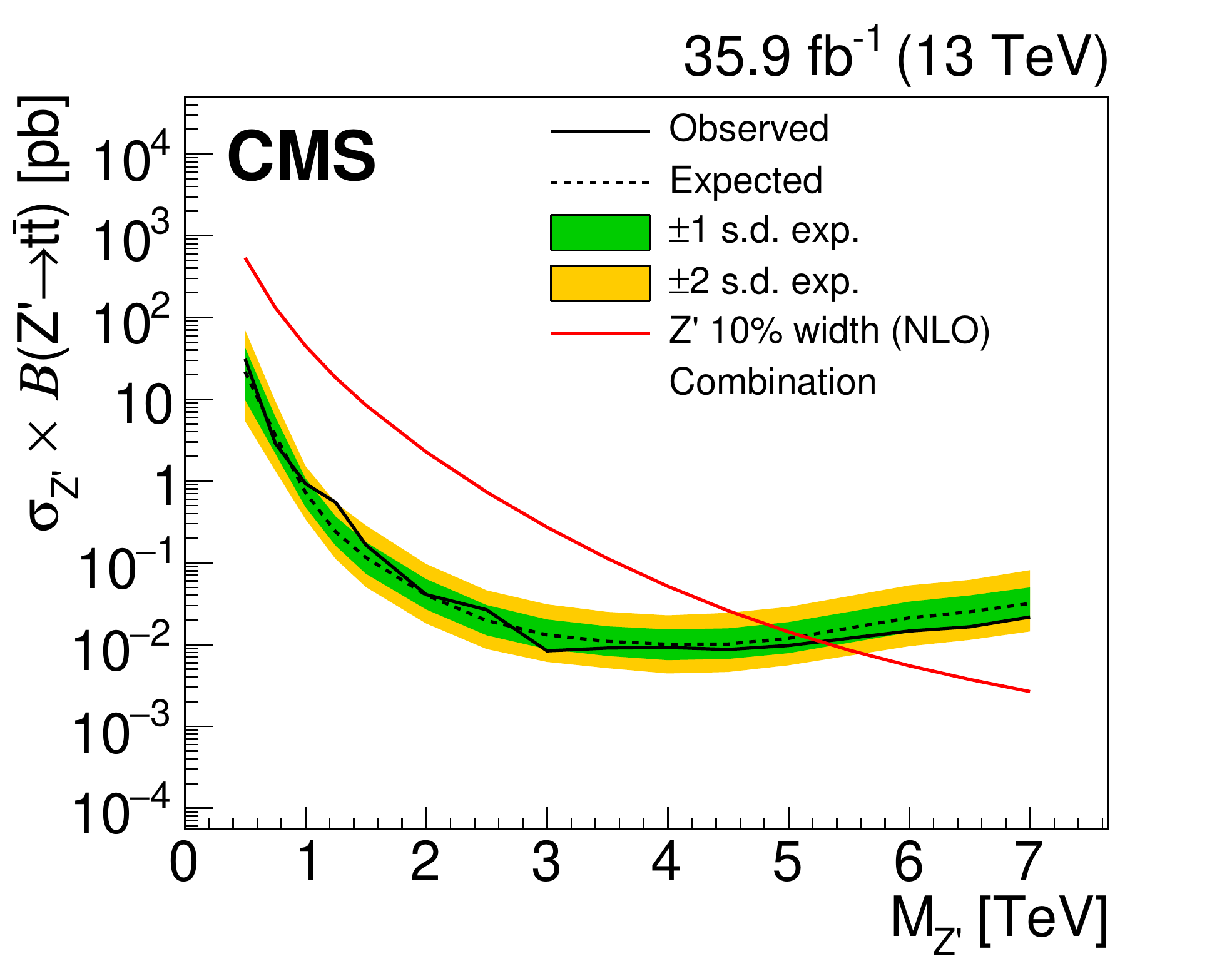}
\includegraphics[width=0.495\textwidth]{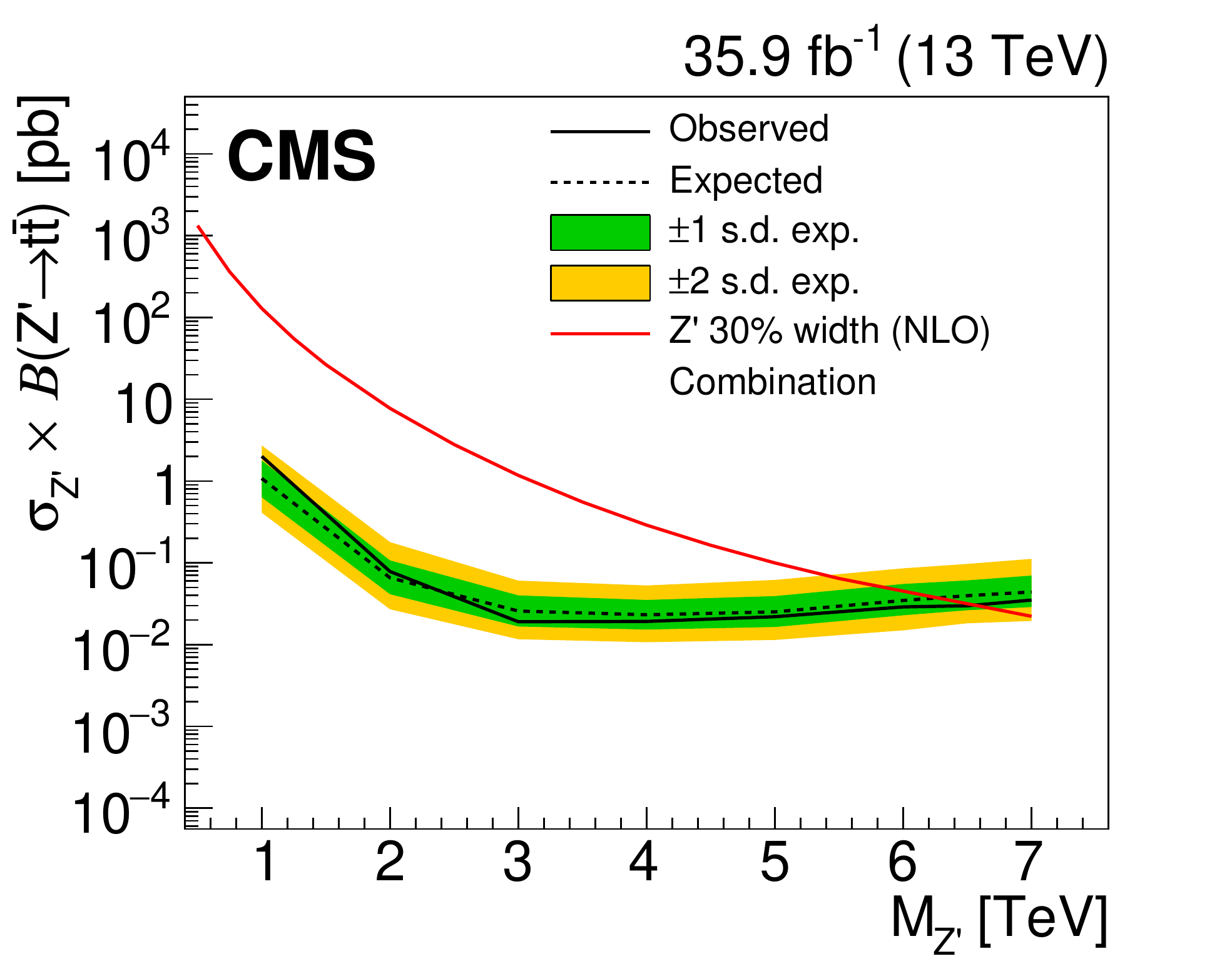}
\includegraphics[width=0.495\textwidth]{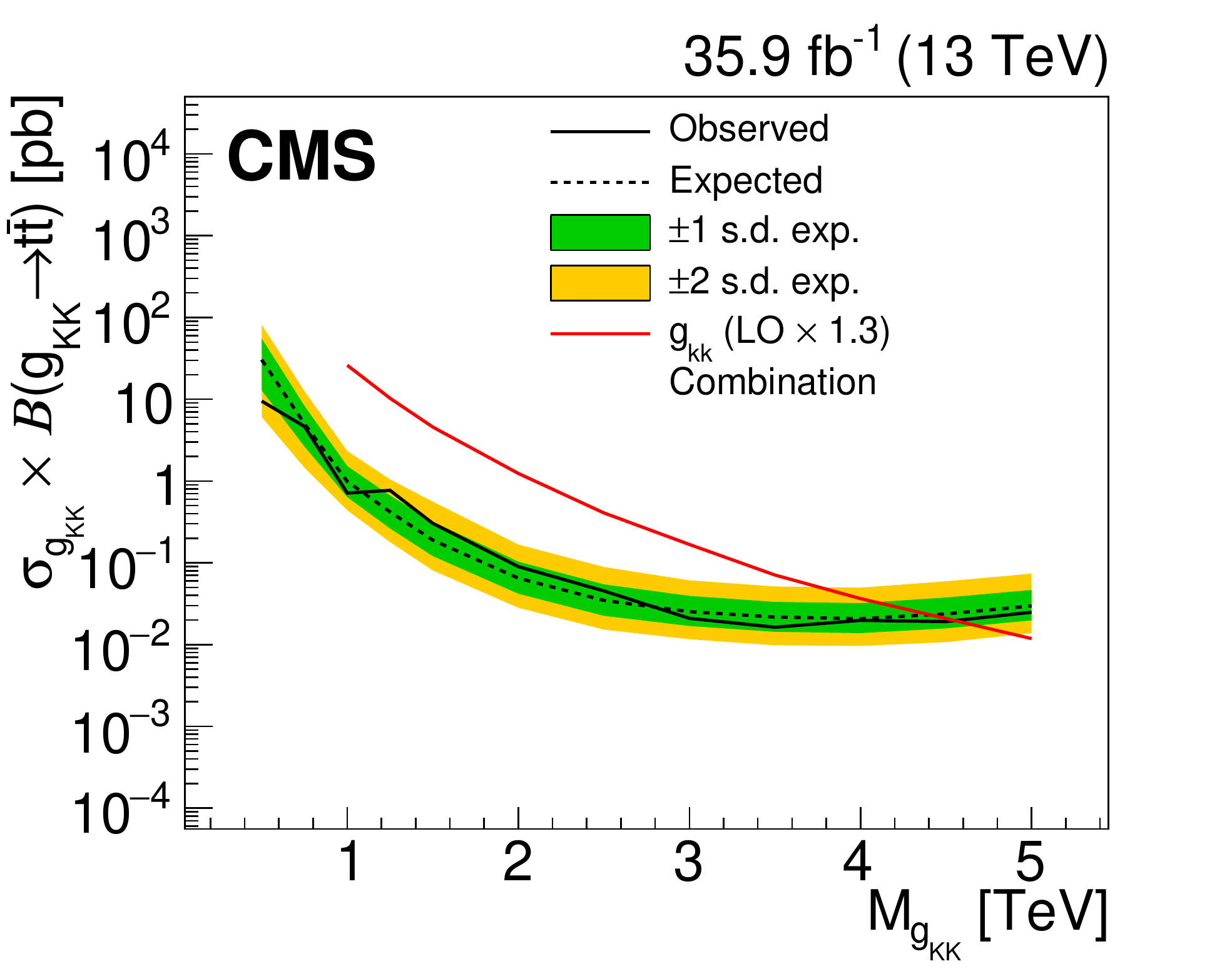}
\caption{Observed and expected limits at 95\% \CL for each of the four signal hypotheses considered in this analysis.}
\label{fig:limits}
\end{figure}

New exclusion limits on the mass of resonances decaying to \ttbar are set by
comparing the observed limit to the theoretical cross section, where the branching fraction
$\mathcal{B}(X\to\ttbar)$ is assumed to be 1. As
shown in Fig.~\ref{fig:limits}, the analysis excludes narrow
\PZpr\ bosons with masses up to 3.80\TeV (3.75\TeV expected),
wide \PZpr\ bosons with masses up to 5.25\TeV (5.10\TeV
expected), and extra-wide \PZpr\ bosons with masses up to
6.65\TeV (6.40\TeV expected). For the \gKK resonance hypothesis, the
analysis excludes masses up to 4.55\TeV (4.45\TeV expected). These
results represent a significant improvement on the previous results in
this channel from the 2015 data taking period, not only because of the
increase in integrated luminosity, but also the reduction in the
uncertainty in the multijet background estimate in the fully hadronic
channel, the improved \wjets rejection via the \wjetsbdt in the
single-lepton channel, and the inclusion of dilepton event categories
in the combination. The absolute cross section limits are 10--40\%
better, for \mttbar above 2\TeV, than the previous result released by
CMS~\cite{Sirunyan:2017uhk} scaled to an integrated luminosity of
35.9\fbinv. These results are the most stringent exclusion
limits on a \ttbar resonance to date.

\section{Summary}
\label{sec:summary}

A search for a generic massive top quark and antiquark (\ttbar)
resonance has been presented. The analysis was performed using data
collected by the CMS experiment in 2016 at the LHC at $\sqrt{s} =
13\TeV$, corresponding to an integrated luminosity of 35.9\fbinv.
The analysis is focused on searching for a \ttbar resonance above 2\TeV,
where the decay products of the top quark become collimated because of
its large Lorentz boost. The analysis performed a simultaneous
measurement of the backgrounds and the \ttagging efficiency from data. The
data are consistent with the background-only hypothesis, and no
evidence for a massive \ttbar resonance has been found. Limits at 95\%
confidence level are calculated for the production cross section for a
spin-1 resonance decaying to \ttbar pairs with a variety of decay widths.

Limits were calculated for two benchmark signal processes that decay to \ttbar
pairs. A topcolor \PZpr\ boson with relative widths of 1, 10, or
30\% is excluded in the mass ranges 0.50--3.80, 0.50--5.25, and
0.50--6.65\TeV, respectively. The first Kaluza--Klein excitation of the
gluon in the Randall--Sundrum scenario (\gKK) is excluded in the range
0.50--4.55\TeV. This is the first search by any experiment at $\sqrt{s}=13\TeV$
for \ttbar resonances that combines all three decay topologies of
the \ttbar system: dilepton, single-lepton, and fully hadronic.

The sensitivity of the analysis exceeds previous searches at $\sqrt{s}
= 8$ and 13\TeV, particularly at high \ttbar invariant mass.
Previous measurements have excluded a topcolor \PZpr\ up
to 3.0, 3.9, and 4.0\TeV, for relative widths of 1, 10, and 30\%, and
\gKK from 3.3 to 3.8\TeV, depending on
model~\cite{atlas_ttbar_resonance3,Sirunyan:2017uhk}. The presented
analysis improves upon those limits, extending the \PZpr\ exclusions to
3.80, 5.25, and 6.65\TeV and the \gKK exclusion to 4.55\TeV. These
are the most stringent limits on the topcolor \PZpr\ and \gKK
models to date.

\begin{acknowledgments}
\label{sec:acknowledgments}
We congratulate our colleagues in the CERN accelerator departments for the excellent performance of the LHC and thank the technical and administrative staffs at CERN and at other CMS institutes for their contributions to the success of the CMS effort. In addition, we gratefully acknowledge the computing centers and personnel of the Worldwide LHC Computing Grid for delivering so effectively the computing infrastructure essential to our analyses. Finally, we acknowledge the enduring support for the construction and operation of the LHC and the CMS detector provided by the following funding agencies: BMBWF and FWF (Austria); FNRS and FWO (Belgium); CNPq, CAPES, FAPERJ, FAPERGS, and FAPESP (Brazil); MES (Bulgaria); CERN; CAS, MoST, and NSFC (China); COLCIENCIAS (Colombia); MSES and CSF (Croatia); RPF (Cyprus); SENESCYT (Ecuador); MoER, ERC IUT, and ERDF (Estonia); Academy of Finland, MEC, and HIP (Finland); CEA and CNRS/IN2P3 (France); BMBF, DFG, and HGF (Germany); GSRT (Greece); NKFIA (Hungary); DAE and DST (India); IPM (Iran); SFI (Ireland); INFN (Italy); MSIP and NRF (Republic of Korea); MES (Latvia); LAS (Lithuania); MOE and UM (Malaysia); BUAP, CINVESTAV, CONACYT, LNS, SEP, and UASLP-FAI (Mexico); MOS (Montenegro); MBIE (New Zealand); PAEC (Pakistan); MSHE and NSC (Poland); FCT (Portugal); JINR (Dubna); MON, RosAtom, RAS, RFBR, and NRC KI (Russia); MESTD (Serbia); SEIDI, CPAN, PCTI, and FEDER (Spain); MOSTR (Sri Lanka); Swiss Funding Agencies (Switzerland); MST (Taipei); ThEPCenter, IPST, STAR, and NSTDA (Thailand); TUBITAK and TAEK (Turkey); NASU and SFFR (Ukraine); STFC (United Kingdom); DOE and NSF (USA).

\hyphenation{Rachada-pisek} Individuals have received support from the Marie-Curie program and the European Research Council and Horizon 2020 Grant, contract No. 675440 (European Union); the Leventis Foundation; the A. P. Sloan Foundation; the Alexander von Humboldt Foundation; the Belgian Federal Science Policy Office; the Fonds pour la Formation \`a la Recherche dans l'Industrie et dans l'Agriculture (FRIA-Belgium); the Agentschap voor Innovatie door Wetenschap en Technologie (IWT-Belgium); the F.R.S.-FNRS and FWO (Belgium) under the ``Excellence of Science - EOS" - be.h project n. 30820817; the Ministry of Education, Youth and Sports (MEYS) of the Czech Republic; the Lend\"ulet (``Momentum") Program and the J\'anos Bolyai Research Scholarship of the Hungarian Academy of Sciences, the New National Excellence Program \'UNKP, the NKFIA research grants 123842, 123959, 124845, 124850 and 125105 (Hungary); the Council of Science and Industrial Research, India; the HOMING PLUS program of the Foundation for Polish Science, cofinanced from European Union, Regional Development Fund, the Mobility Plus program of the Ministry of Science and Higher Education, the National Science Center (Poland), contracts Harmonia 2014/14/M/ST2/00428, Opus 2014/13/B/ST2/02543, 2014/15/B/ST2/03998, and 2015/19/B/ST2/02861, Sonata-bis 2012/07/E/ST2/01406; the National Priorities Research Program by Qatar National Research Fund; the Programa Estatal de Fomento de la Investigaci{\'o}n Cient{\'i}fica y T{\'e}cnica de Excelencia Mar\'{\i}a de Maeztu, grant MDM-2015-0509 and the Programa Severo Ochoa del Principado de Asturias; the Thalis and Aristeia programs cofinanced by EU-ESF and the Greek NSRF; the Rachadapisek Sompot Fund for Postdoctoral Fellowship, Chulalongkorn University and the Chulalongkorn Academic into Its 2nd Century Project Advancement Project (Thailand); the Welch Foundation, contract C-1845; and the Weston Havens Foundation (USA).
\end{acknowledgments}
\clearpage
\bibliography{auto_generated}
\cleardoublepage \appendix\section{The CMS Collaboration \label{app:collab}}\begin{sloppypar}\hyphenpenalty=5000\widowpenalty=500\clubpenalty=5000\vskip\cmsinstskip
\textbf{Yerevan Physics Institute, Yerevan, Armenia}\\*[0pt]
A.M.~Sirunyan, A.~Tumasyan
\vskip\cmsinstskip
\textbf{Institut f\"{u}r Hochenergiephysik, Wien, Austria}\\*[0pt]
W.~Adam, F.~Ambrogi, E.~Asilar, T.~Bergauer, J.~Brandstetter, M.~Dragicevic, J.~Er\"{o}, A.~Escalante~Del~Valle, M.~Flechl, R.~Fr\"{u}hwirth\cmsAuthorMark{1}, V.M.~Ghete, J.~Hrubec, M.~Jeitler\cmsAuthorMark{1}, N.~Krammer, I.~Kr\"{a}tschmer, D.~Liko, T.~Madlener, I.~Mikulec, N.~Rad, H.~Rohringer, J.~Schieck\cmsAuthorMark{1}, R.~Sch\"{o}fbeck, M.~Spanring, D.~Spitzbart, A.~Taurok, W.~Waltenberger, J.~Wittmann, C.-E.~Wulz\cmsAuthorMark{1}, M.~Zarucki
% ...
\vskip\cmsinstskip
% ...
1:  Also at Vienna University of Technology, Vienna, Austria\\
% ...
\end{sloppypar}
\end{document}